# The CODALEMA/EXTASIS experiment: Contributions to the 35th International Cosmic Ray Conference (ICRC 2017)

**Hervé Carduner[a], Didier Charrier[a,c], Richard Dallier[a,c], Laurent Denis[c], Antony Escudie[a], Daniel García-Fernández[a], Florian Gaté[a,e], Alain Lecacheux[b], Vincent Marin[a], Lilian Martin[a,c], Benoît Revenu[a,c], Matias Tueros[d]**

[a] *Subatech, Institut Mines-Télécom Atlantique, CNRS, Université de Nantes, Nantes, France*
[b] *CNRS-Observatoire de Paris, Meudon, France*
[c] *Unité Scientifique de Nançay, Observatoire de Paris, CNRS, PSL, UO/OSUC, Nançay, France*
[d] *Instituto de Física La Plata, CONICET, CCT-La Plata, La Plata, Argentina*
[e] *LAPP, CNRS, Université Savoie Mont-Blanc, Anneçy, France*

## Contents

# The CODALEMA/EXTASIS experiment: a multi-scale and multi-wavelength instrument for radio-detection of extensive air-showers

**B. Revenu**[1,3], **D. Charrier**[1,3], **R. Dallier**[1,3], **A. Escudie**[1], **D.García-Fernández**[1], **A. Lecacheux**[2], **and L. Martin**[1,3]

[1] *Subatech, Institut Mines-Telecom Atlantique, CNRS, Universite de Nantes, Nantes, France*
[2] *CNRS-Observatoire de Paris, Meudon, France*
[3] *Unité Scientifique de Nançay, Observatoire de Paris, CNRS, PSL, UO/OSUC, Nançay, France*
*E-mail:* revenu@in2p3.fr

Hosted by the Nançay Radio Observatory, the CODALEMA experiment is dedicated to radio detection of cosmic ray induced extensive air showers. It is composed of:

- 57 self-triggering radio detection stations working in the $20-200$ MHz band and spread over 1 km$^2$;
- an array of 13 scintillators as particle detector;
- a compact array made of 10 cabled antennas, triggered by the particle detector;
- a small array of 5 cabled antennas, and whose role is to figure out the capabilities of a phased antenna cluster to cleverly select air shower events;
- a 3D detector, measuring the electric field in three orthogonal polarizations.

Also supported by CODALEMA is the R&D EXTASIS project, aiming at detecting the low-frequency signal (below 9 MHz) produced at the sudden disappearance of the air shower particles hitting the ground. All these antenna arrays present different density and extent, and can be operated in a joint mode to record simultaneously the radio signal coming from an air shower. Therefore, the Nançay facilities may offer a complete description of the air shower induced electric field at small, medium and large scale, and over an unique and very wide frequency band (from 2 to 200 MHz). We describe the current instrumental set-up and the performances of CODALEMA/EXTASIS.

## 1. Radio signal and cosmic rays

Cosmic rays entering our atmosphere interact with the air molecules and produce atmospheric air showers. These showers are constituted by many secondary particles, in particular electrons and positrons radiating a detectable electric field. The emission mechanisms are now well understood [1, 2] and regularly observed in various experiments [3, 4, 5]... This radio signal can be compared to theoretical expectations through simulations [6, 7], in order to successfully compute the energy of the primary cosmic ray and its nature. The goal is to provide an accurate estimation of the cosmic ray composition above $10^{16}$ eV. This composition is still poorly known but we expect large progress in the next decade due to the use of the radio signal together with improved cosmic ray observatories such as the Pierre Auger Observatory [8].

It has been shown that the radio signal allows to estimate the nature of the primary cosmic ray through the measurement of the atmospheric depth of the maximum of the shower development $X_{\mathrm{max}}$ [9, 10]. In LOFAR for instance, which has a high density of antennas spread over a small area (0.2 km$^2$), the uncertainty on $X_{\mathrm{max}}$ is 17 g/cm$^2$ [11], similar to what is done using with the fluorescence technique ($\sim$ 20 g/cm$^2$ [12]) but with a duty cycle close to 100% compared to the 14% [13] of a fluorescence telescope.

The detection of this very rich radio signal is realized in various frequency bands: $20-200$ MHz for CODALEMA, $30-80$ MHz for AERA and Tunka-Rex, $30-80$ or (exclusive) $110-190$ MHz for LOFAR. This is not an easy task as the current experiments rely either on external triggering thanks to a joint shower detector (particles or fluorescence) sensitive only to air showers or on internal triggering with the strong drawback that a huge fraction of the data is generated by noise (of anthropic or natural origin). In Nançay, we aim at efficiently detecting air showers with a very large frequency band in order to fully characterize the primary cosmic ray independently of any other shower detector. This will allow us to contribute to the cosmic ray domain (spectrum and composition) in the domain $10^{16}-10^{18}$ eV.

## 2. The CODALEMA/EXTASIS experiment at Nançay

The study of air showers created by high energy cosmic rays interacting in our atmosphere is ongoing since 15 years in Nançay. The instruments were upgraded several times and today, we run a total of 5 arrays of various detectors. Only one of them is using the well-established surface detector technique using scintillators, already in use at the time of the second period of the CODALEMA experiment. The others are all observing the radio counterpart of the air shower phenomenon. In FIG. 1, we display the ground map of the current setups, together with the location of the FR606 LOFAR station and the NenuFAR (New Extension in Nançay Upgrading LOFAR) radiotelescope, as a SKA pathfinder.

## 3. The scintillator detector array (SD)

The SD provides the classical reconstruction of air showers: energy, arrival direction and core position. These informations are used to help/confirm the shower detection and reconstruction using the radio signal. The 13 scintillators are spread over an area of $340 \times 340$ m$^2$; its energy threshold is roughly around the knee ($10^{15}$ eV) and is fully efficient above $10^{16}$ eV. It has a very low background as scintillators are sensitive to high energy particles of the shower only. The DAQ of the scintillators is able to trigger in real time other instruments throughout the observatory: a time-tagging station holding a GPS antenna for accurate datation, the compact radio array (see section 5), the tripole antenna (see section 6), the EXTASIS detectors (see section 7).

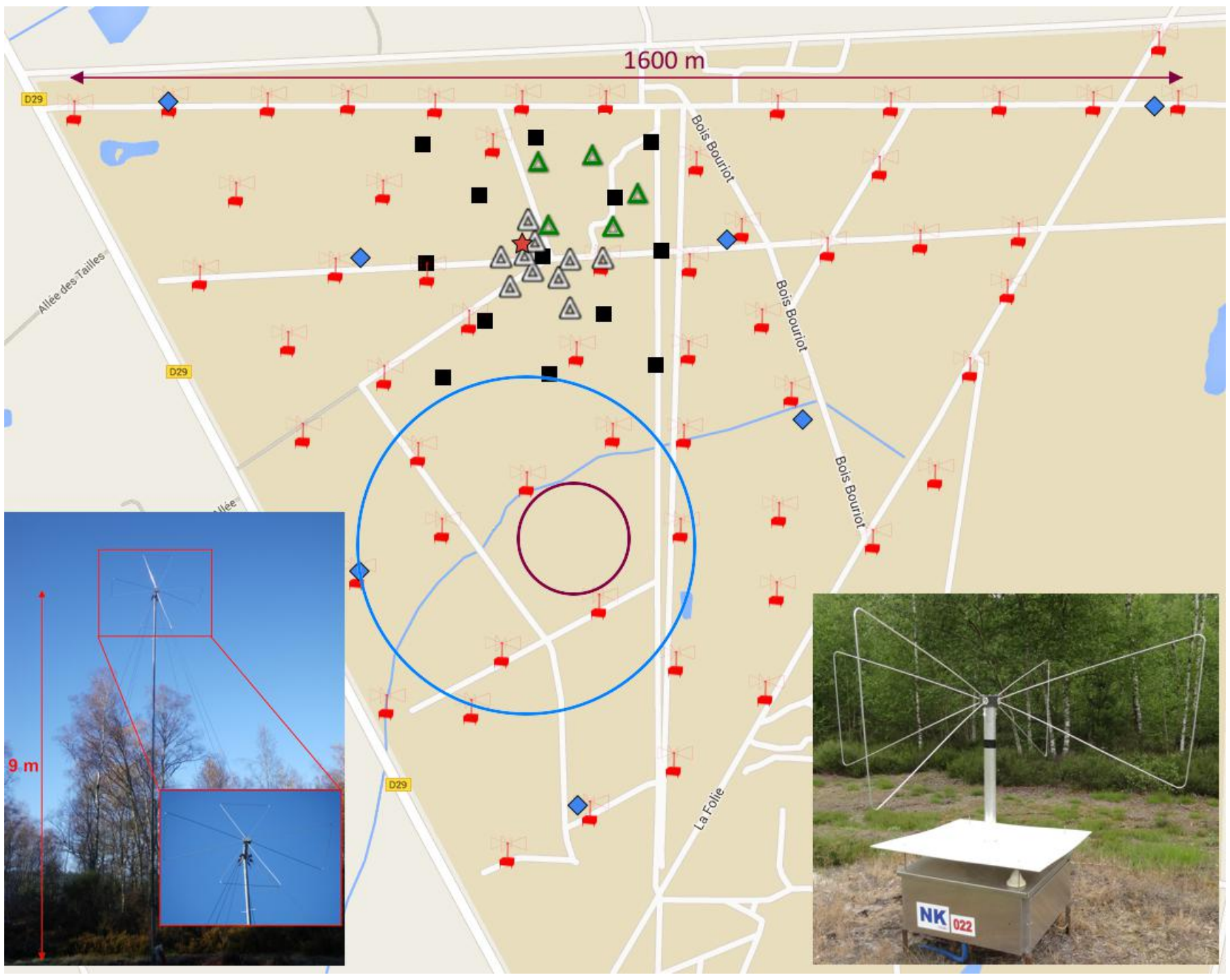

**Figure 1:** The CODALEMA/EXTASIS instruments cover an area of 1 km$^2$. Black squares are the scintillator positions of the particle array. The small red radio station symbols corresponds to the 57 autonomous stations (self-triggered). White triangles stand for the compact array (externally-triggered); green triangles are the antennas of the continuous array (self-triggered). The blue squares are the low frequency antennas of the EXTASIS experiment (externally triggered). The red star indicates the tripole antenna (externally-triggered). The picture of the bottom left is one of the EXTASIS detector. The picture of the bottom right is one of the 57 autonomous radio stations.

## 4. The autonomous stations radio array

The largest instrument dedicated to cosmic-ray detection in Nançay is the array of 57 autonomous stations, covering the full area of the observatory (1 km$^2$). The typical distance between two neighbouring stations is 150 m. These stations sample the incoming electric field in both horizontal east-west (EW) and north-south (NS) polarizations. We use the butterfly antenna [14], also in use in AERA. Both input channels pass into a splitter. One part is filtered in the 35 – 80 and 150 – 200 MHz bands and is compared to a voltage threshold. The other is continuously digitized at a sampling rate of 1 GHz. If the filtered analogic signal is larger than the threshold, this constitutes the level 1 trigger T1: we record the GPS time of the event together will the full-band signal in both channels, during 2.56 $\mu$s. We can choose the T1 conditions independently on each polarization (OR or AND). At the level of the trigger board, the maximum trigger rate is of the order of $10^3$ events per second. When selected by a second-level trigger T2, digitized data are read; this part of the process increases significantly the dead time and the maximum trigger rate decreases to roughly 28 events per second. The selection can be done at the station level by applying a cut on the rise time of the signal as the transients produced by air showers are known to last 10-20 ns only. A third-level trigger (T3) is currently in use in Nançay since more than a year: it is a selection on several stations having T2s with timestamps in agreement with the propagation of waves at the speed of light as it is the case for air shower radio signals. The resulting average T3 rate is around 1 Hz with some peaks at 20 Hz. The average number of radio transients corresponding to actual shower is 2 per day. They are identified through the time coincidence with the SD and also with the polarization pattern of the hit antennas. The energy threshold for the radio-detected cosmic rays is around $10^{17}$ eV. These autonomous stations are unique among those in use in other radio experiments on cosmic ray. The frequency band is very large 20-80 MHz and 110-200 MHz at once, associated with a fast sampling rate of 1 GHz. Measuring the electric field up to 200 MHz is important as the Cherenkov contribution clearly appears in the high frequency domain (above 100 MHz). Some results are presented in [3].

## 5. The Compact Radio Array

The compact array allows to study the small-scale variations of the electric field. We are using two separated arrays.

### 5.1 Externally triggered compact array

We are running a cluster of 10 antennas spread over an area of $150 \times 150$ m$^2$. These antennas detect the electric field in two linear horizontal polarizations. They are triggered by the SD; 20 channels are recorded over 6 $\mu$s, with a sampling rate of 400 MHz and a in the frequency band 10-200 MHz. Some results of this array are presented in [15].

### 5.2 Self triggered compact array

We also use a cluster of 5 antennas working in a self-trigger mode. Its main characteristics is to sample the sky in a continuous way: there is no trigger. This sampling is done on 8 channels in circular polarization after superposition of two linear polarizations added in quadrature. The sampling rate is 100 MHz with a bandwidth of 50 MHz, between 30 and 80 MHz. Data are stored in a ring buffer and analyzed in real time with a GPU (NVIDIA K20 at 5 Tflops). The data are first filtered, we merge the data into a circular polarized channel and the signal envelope is computed for each time window of 500 ns, corresponding to the time aperture of the array, We generate 2000 beams on the sky (by shifting the individual signals) with beamforming. Then we search for beams having an amplitude above a defined threshold. The beam forming technique increases the sensitivity by a factor $\sqrt{N}$, $N$ being the number of channels in use. A strong advantage of this continuous sampling is that the dead time is null. Work is in progress.

## 6. 3D electric field measurements

We measure at one location in Nançay the incoming electric field in coincidence with the SD. This detector is close to the center of the SD and contains three antennas allowing the electric field measurement in the three EW, NS and vertical polarizations. The goals of this 3D measurement are:

- to get the full electric field directly (i.e. avoid to estimate the vertical component from the measurement of two horizontal polarizations as it is done in most experiments);
- to determine wether or not the electric field emitted by air showers can be described in the far-field hypothesis.

The design of the 3D detector is based on three butterfly antennas with their LONAMOS LNA (Low Noise Amplifier) rotated twice around the NS and vertical axis. It is triggered by the SD and the three channels are recorded with a sampling rate of 100 MHz during 2.56 $\mu$s. A picture of the 3D detector is presented in FIG. 2. The results of this detector are presented in [16].

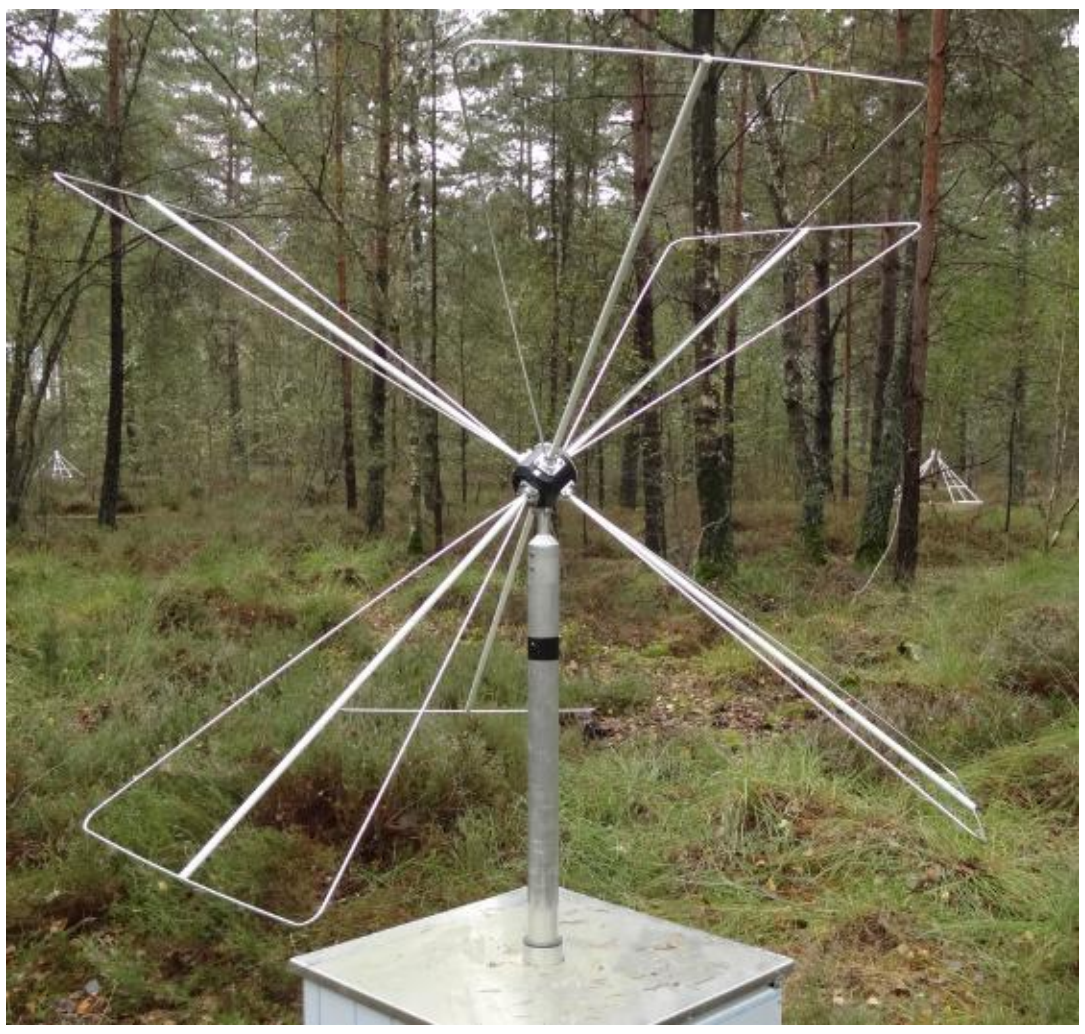

**Figure 2:** Picture of the 3D detector, whose location is indicated by the red star in FIG. 1.

## 7. EXTASIS: search for the low frequency radio signal

The most recent radio detector installed in Nançay is the EXTASIS experiment. EXTASIS stands for EXTinction of Air Shower Induced Signal and its main goal is to detect the electric field emitted by air showers when their front hits the ground. Secondary electrons and positrons are very quickly stopped after a few cm when reaching the ground. This phenomenon generates a large electric field as the time and space derivatives of the charge density and current in the shower are large. This "sudden death" signal is expected to peak at low frequency (1-10 MHz), as discussed in [17, 18]. Apart from this very specific signal, a low frequency detector should also detect the "regular" electric field emitted during the shower development while still in the air. The theoretical potential of the electric field from the sudden death of shower is huge: detection at large distances (i.e. possibility to have a sparse and cheap array), core position determination very easy and precise, direct $X_{\max}$ measurement. In reality, things are always more complex: the frequency band 1-10 MHz is overcrowded; many man-made emitters use this band and the atmospheric noise background is also very large, in particular during the night because of the ionosphere's daily variations. The current low frequency setup of EXTASIS consists today in 7 butterfly antennas installed at 9 m above the ground (as indicated by simulations of the antenna response at these frequencies), in both EW and vertical polarizations. These antennas are triggered by the SD and their data are digitized by oscilloscopes with a sampling rate of 500 MHz during 2 ms over 8 bits. The LNA associated to the antennas is a modified version of the LONAMOS LNA. One of the low frequency detector can be seen in FIG. 3. The first results are presented in [19].

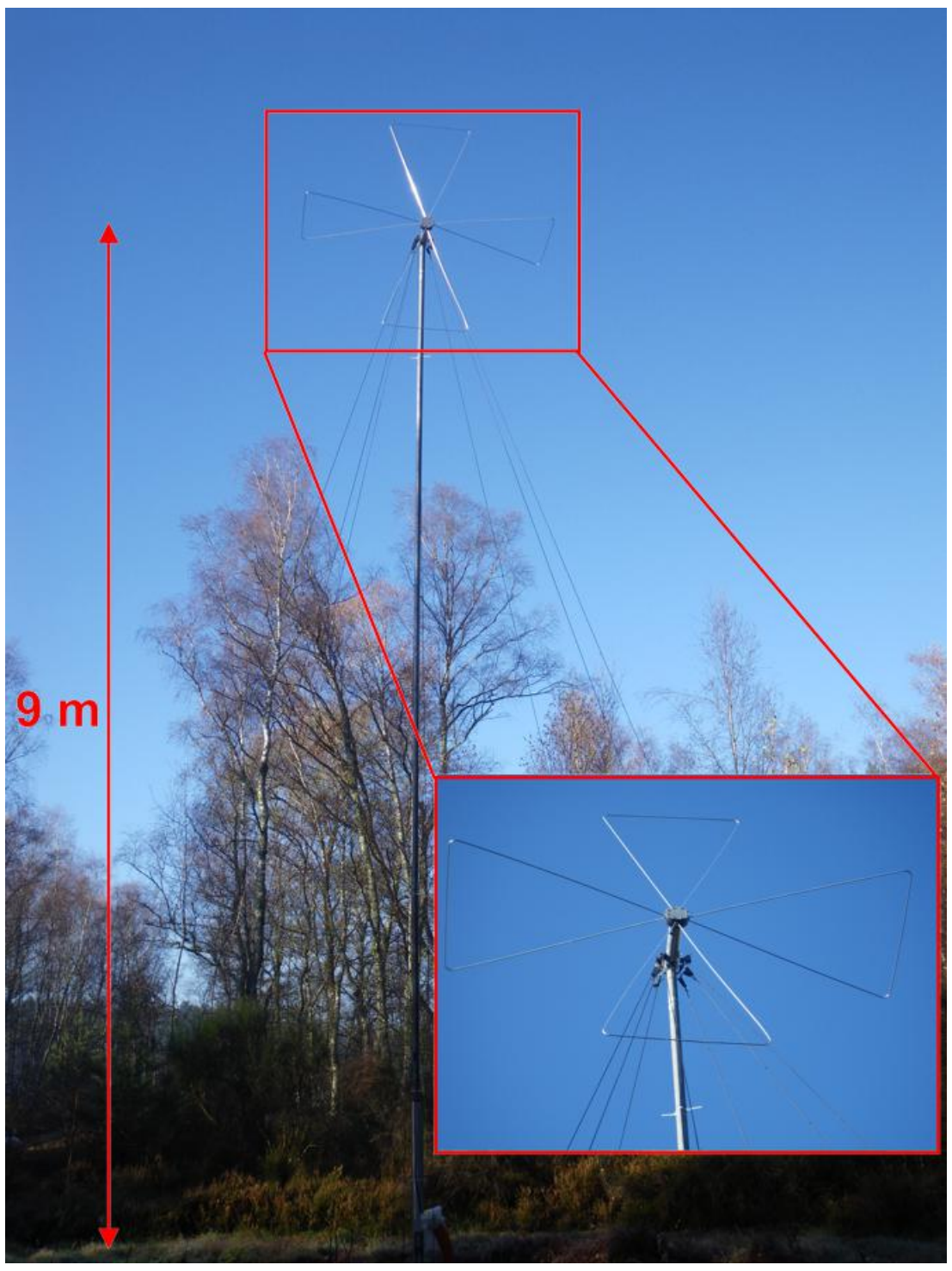

**Figure 3:** Picture of one of the low frequency detector of EXTASIS.

## 8. Conclusion

The CODALEMA/EXTASIS experiment in Nançay allows to detect air showers initiated by high-energy cosmic rays in a very wide frequency band: 1-200 MHz. This permits to have the full picture of the radio emission of showers and the various emission mechanisms in play: the geomagnetic effect, the charge-excess effect and the sudden death signal. This will allow a better reconstruction of the primary cosmic rays characteristics, in particular its nature. The first low-frequency events associated with cosmic rays are detected since a year and some results are shown in [19].

## Acknowledgements

We thank the Région Pays de la Loire for its financial support of the Astroparticle group of Subatech and in particular for its contribution to the EXTASIS experiment.

# Main features of cosmic ray induced air showers measured by the CODALEMA experiment

**Lilian Martin[*,1,3], R. Dallier[1,3] , A. Escudie[1], D. García-Fernández[1], F. Gaté[1], A. Lecacheux[2] and B. Revenu[1,3]**

[1]*Subatech, CNRS/IN2P3, IMT Atlantique, Université de Nantes, Nantes, France*
[2]*LESIA, Observatoire de Paris, CNRS, PSL, UPMC/SU, UPD, Meudon, France.*
[3]*Unité Scientifique de Nançay, Observatoire de Paris, CNRS, PSL, UO/OSUC, Nançay, France.*
*E-mail:* lilian.martin@subatech.in2p3.fr

The radio signals produced by extensive air showers initiated in the atmosphere by high energy cosmic rays are routinely observed and registered by the various instruments of the CODALEMA experiment located at the Nançay radio observatory and notably by the large array of self-triggering stations equipped with wide band and dual polarization antennas. Precise comparisons between observed radio signals and simulations performed with the SELFAS code allow most of the main features of the primary cosmic ray to be determined : arrival direction, energy and $X_{max}$ estimates from which the composition in the energy range covered by CODALEMA may be derived. After a presentation of the analysis methods, its sensitivity will be discussed and the results obtained over a significant set of experimental events will be detailed.

[*]Speaker

# 1. Introduction

Initiated at the earliest years of the modern period of the wide international R&D effort for the radio detection of extensive air showers (EAS), the CODALEMA experiment has been constantly improving its detection capabilities and sensitivity [1]. The AERA engineering array at the Pierre Auger Observatory is focusing on combining radio electric field measurement with other available detection techniques [2] while the LOFAR radio-telescope is relying on very high number of antennas on a small area to oversample the radio footprint on the ground and to deduce precise EAS information [3]. CODALEMA is still pursuing its original goal of developing both a sensitive antenna and triggering system toward an efficient and fully autonomous apparatus dedicated to high energy cosmic ray detection. This paper reports on the latest results obtained with a sample set of data and especially on the derivation of the shower $X_{max}$ and the cosmic ray energy estimates.

# 2.The experimental setup and the acquisition system

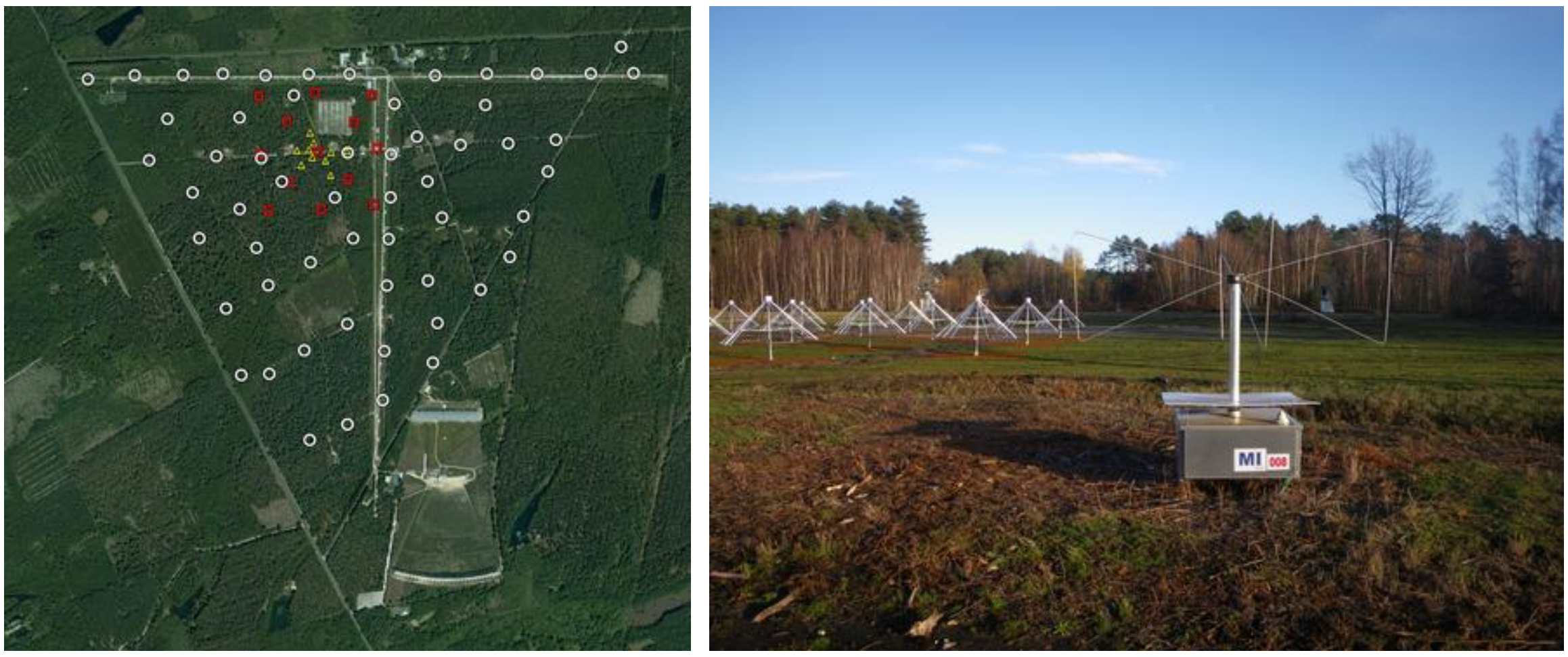

*Figure 1: The autonomous stations (white circle), the scintillators (red square) and the compact array (yellow triangle) superimposed on an areal view of the Nançay radioastronomy station (left). An autonomous station nearby a NenuFAR mini-array (right).*

3.Installed in the early 2000, at the radioastronomy station of Nançay, France, the CODALEMA experiment is now a set of various mature apparatus and prototyping instruments [1],[4]. Since few years, a set of 57 self-triggering autonomous stations (SA) scattered over roughly 1 km$^2$ is the main CODALEMA detection system (Figure 1). The stations feature dual horizontal polarization antennas equipped with the LONAMOS low noise amplifier and optimized to measure transients over a wide frequency band (10-200 MHz) and over a large dynamic range. Electronic boards have been developed to fulfill dedicated function (trigger, dating, digitization...) and installed in a double layer EMC compliant mechanical box (Figure 1). A multi-level triggering system has been developed to select narrow transients specific to air shower induced electric field and to reject signals produced by anthropic noise sources. A simple first trigger level compares band filtered analog signals to a programmable voltage level and is used to initiate on-board dating and digitization of the events. Further noise rejection is achieved using a second trigger level involving more sophisticated quantities such as the signal rise time. Finally, accepted events are sent by the stations to a central acquisition system which aggregates events in time coincidences, reconstructs arrival directions, excludes coincidences pointing to identified sources and finally requests the full information for the accepted event from the local stations.

Surviving the early stage of the experiment a set of 13 scintillators covering a limited surface of 340 by 340 $m^2$ allows to unambiguously detect high energy cosmic ray events (Figure 1). It provides a particle trigger sent to a dedicated GPS station for accurate time recording and off-line time coincidence with the autonomous stations. It requires that 5 scintillators out of 13 see a signal above 15 mV corresponding to a detection energy threshold of the order of $10^{15}$ eV (considering the array extension). This trigger is also used to record signals from dedicated cabled antenna arrays and to provide software triggers sent over the network to distant antennas [4].

CODALEMA is taking data in this mode steadily since several months. On a typical period of a month, the scintillator array is producing of the order of 40 000 events while the central daq system retains of the order of 1 200 000 events from the SA array. This figure reduces to 60 events per month when requiring for a time coincidence between both arrays.

## 4. Event reconstruction in the scintillator array

A meticulous procedure [1] is applied to the scintillator signals in order to deduce informations from the cosmic ray. At first an analysis on the raw signal is performed to quantify and subtract parasitic oscillations and transients produced by LW perturbations and mismatched electronic elements. For each channel, timing, amplitude and integrated charge are evaluated. Individual timings are combined using a plane front approximation to calculate the cosmic-ray arrival direction. For each event, the particle density in each station is extracted from the corresponding deposited charge and converted in VEM units. A correction factor is used to take into account the transition effect, wherein arriving gamma particles are converted into charged particles inside the scintillator. Then, event by event, station positions are projected into the plane perpendicular to the arrival direction and containing the central detector (shower plane). The reconstruction of the ground particle density lateral distribution function (LDF) as a function of distance from the shower axis is performed assuming a NKG theoretical shape. With a minimizing procedure, we obtain the shower size and the coordinates of the shower core in the shower plane, which are then converted back to the array plane. The primary energy is then calculated as a function of the reconstructed shower size parameter.

## 5. Event reconstruction in the autonomous station array

Event reconstruction using the data from the autonomous stations is rather straightforward. Once recorded on disk the data are transferred into a Firebird database for an easy data mining and data selection. As a matter of fact, retrieving coincidences consists simply in an SQL request requiring to group database entries by identical event time. Once aggregated, reconstructed data from the different stations are combined and treated. Similarly to the scintillator data, the arrival direction of the coincidences is deduced using a plane front approximation and using only the timing information (the time jitter induced by the station self triggering system is negligible).

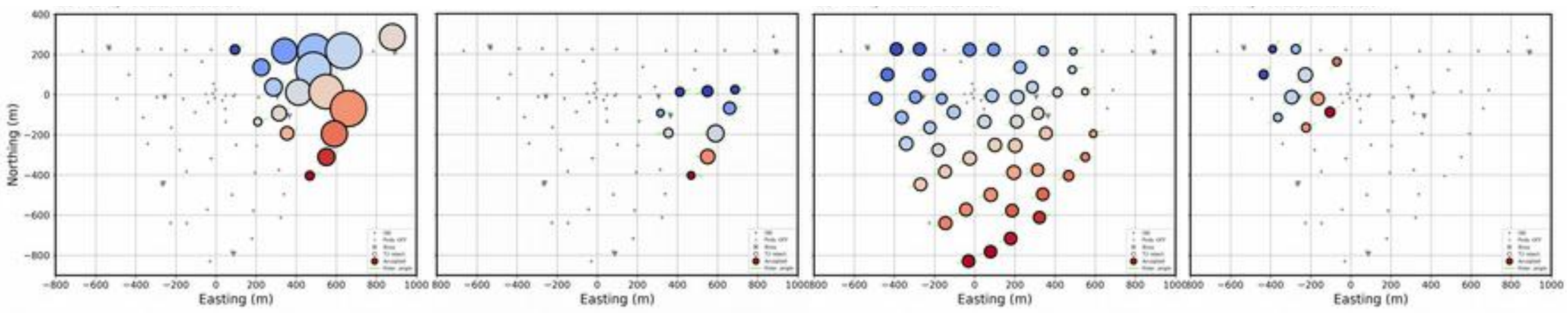

*Figure 2: Examples of radio signal amplitude footprints on the ground. The size of the circle surface is proportional to the sum of the square of the two polarization amplitudes while its color (early in blue, late in red) reflects the detection timing relative to the average timing. The gray dots represented the station locations.*

Few examples of the radio signal footprint on the ground are shown in Figure 2 reflecting the variety of arrival directions and energies of the cosmic rays observed by the SA array and exhibiting low or high multiplicities, uniform or very different amplitudes.

Analysis of the transients (Figure 3) gives further information on the air shower features and its unambiguous identification. Thanks to its very wide frequency band, the spectral content derived from the short transients recorded provides very useful information (Figure 3) : strongly influenced by the antenna response, the high frequency patterns are directly correlated to the arrival direction; a quickly varying spectral index is also typical of an increase of the distance between the antenna distance relative and the shower core. Finally, polarization patterns, simply expressed as the hodograph of the electric field measured in this EW-NS plane (Figure 7) can be used to infer the shower properties : polarization pattern relatively uniform indicates an arrival direction far from the geomagnetic field and/or antennas far from the shower core. On the opposite, a large dispersion of the observed patterns reveals antennas surrounding the shower core and/or a dominant charge excess effect (symmetrical around the shower axis).

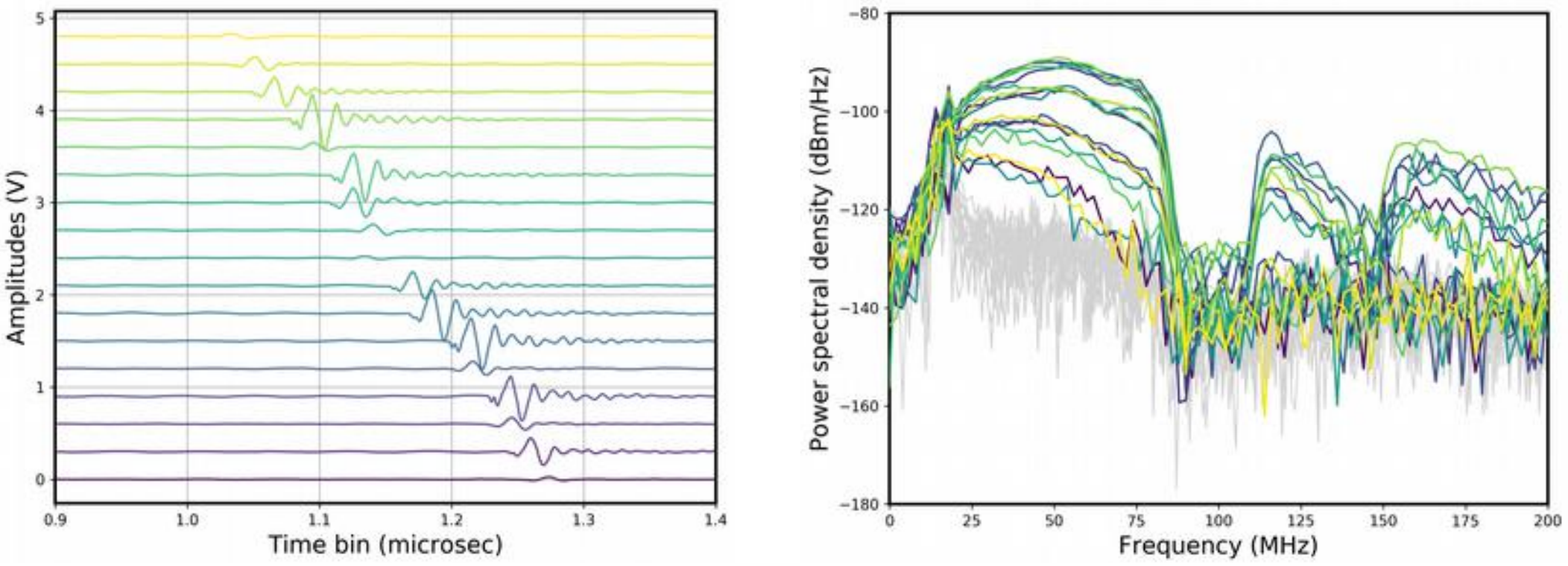

*Figure 3: Waveforms (left) observed in the 17 EW antennas of the left-most event of Figure 2. The waveforms are arbitrarily shifted in time and in amplitude for readability. The corresponding PSD (right) are calculated in a time window centered around the transients and compared to noise reference levels (light gray lines).*

## 6. Inferring the EAS features using SELFAS

Earliest methodology applied to infer the EAS and cosmic ray properties from the radio signal was modeled on the technique used for particle detector data. A symmetry was assumed around the shower axis in order to extract the shower location while an exponential lateral distribution function was employed in order to derive an energy estimate of the cosmic ray. The varying interplay between the geomagnetic effect and the charge excess effect clearly compromises a perfect symmetry of the electric field around the shower axis and leads for instance to a clear offset between the cores reconstructed respectively from the radio signals and from the particle density at ground [5]. In addition, an angular correction must be introduced for the energy calculation in order to take into account for the dependence of the geomagnetic effect to the angle between the local geomagnetic field and the cosmic-ray arrival direction. On the other hand, modern theoretical approaches simulating the radio signal induced by EAS have made significant progress and are now in relatively good agreement demonstrating that the radio emission mechanisms are understood. Therefore we have chosen in the analysis to proceed to careful and systematic comparisons between experimental data and simulations to infer the cosmic-ray properties.

### 6.1 SELFAS simulations

We use the SELFAS model [6] to calculate the electric field induced on a virtual array of antennas by a cascade of secondary particles initiated by ultra-high energy cosmic rays in the atmosphere. This approach is based on a microscopic description of the shower. Using relevant distributions of secondary electrons and positrons, individual contributions all along their tracks are calculated and summed up to compute the final electric field at given locations.

Basically, the SELFAS simulations produced for this analysis were performed with a fixed cosmic-ray energy of $10^{17}$eV, a random first interaction point (and thus random $X_{max}$) with a thinning factor resulting in 20 million of generated particles from CONEX distributions [7]. The shower axis was set to the azimuthal and zenith directions taken from the experimental event considered and its core set at the center of a virtual array of 320 antennas. In the shower frame, these antennas were arranged on 16 branches with an antenna density decreasing with increasing distance to the shower core (Figure 7). 50 proton and 10 iron induced EAS were simulated per experimental coincidence.

In order to do realistic comparisons, the full-band perfect radio-electric field must be convoluted with the full detection chain response and notably with the antenna response. This convolution is expressed in terms of a global transfer function which can be decomposed into the antenna transfer function, the analog chain transfer function and the digitizer transfer function allowing to convert a given three component electric field into digital values on specific ADC channels. The angular and frequency dependence of these transfer functions have been carefully measured using a network analyzer and calculated using the NEC antenna modeling code. An example of the convolution of SELFAS transients is illustrated in Figure 4. The very good agreement between the experimental data and the convoluted SELFAS transients is clearly obvious on the PSD.

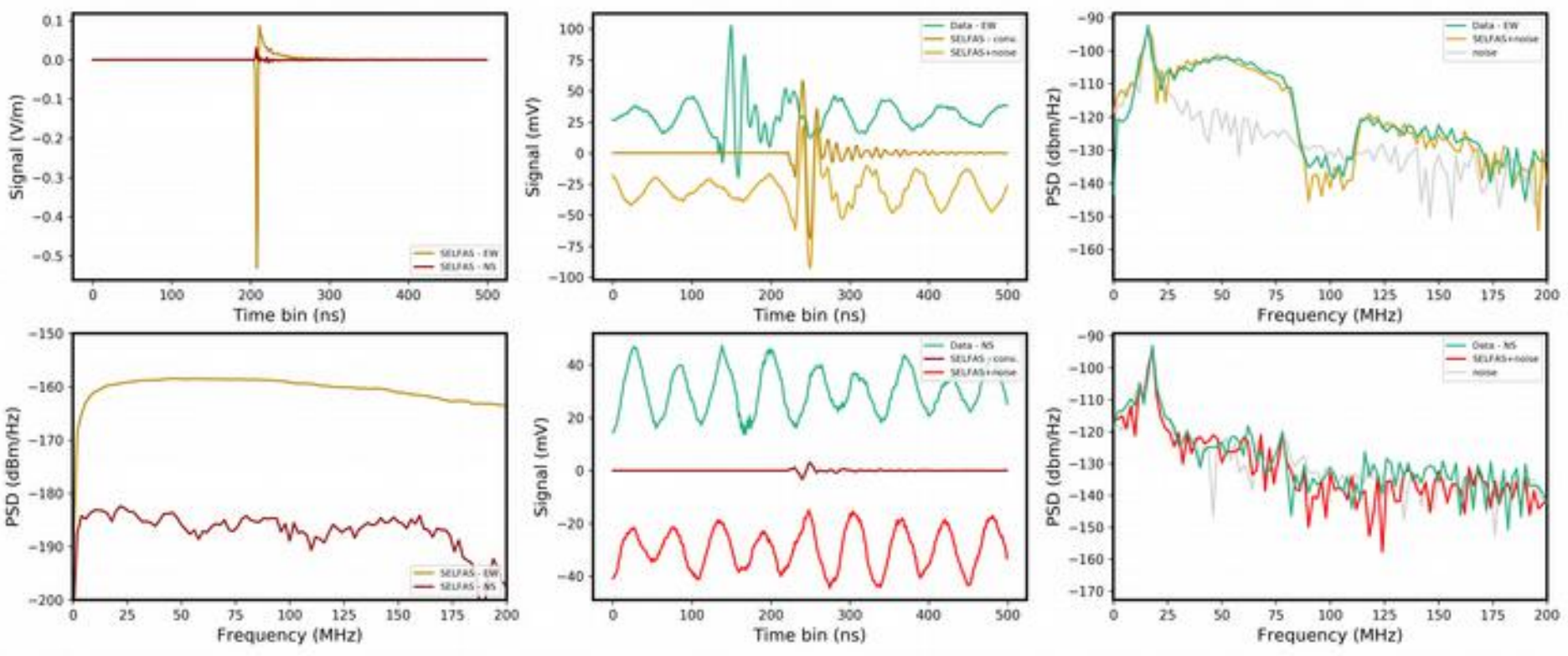

*Figure 4: Example of the convolution of SELFAS signals. Left: the EW (dark gold) and NS (dark red) transients and their power spectrum densities. Center: The convoluted waveforms (dark gold and dark red) with added noise (gold and red) are compared to real data (green). Right: the corresponding power spectrum density (gold and red) are compared to the experimental data (green).*

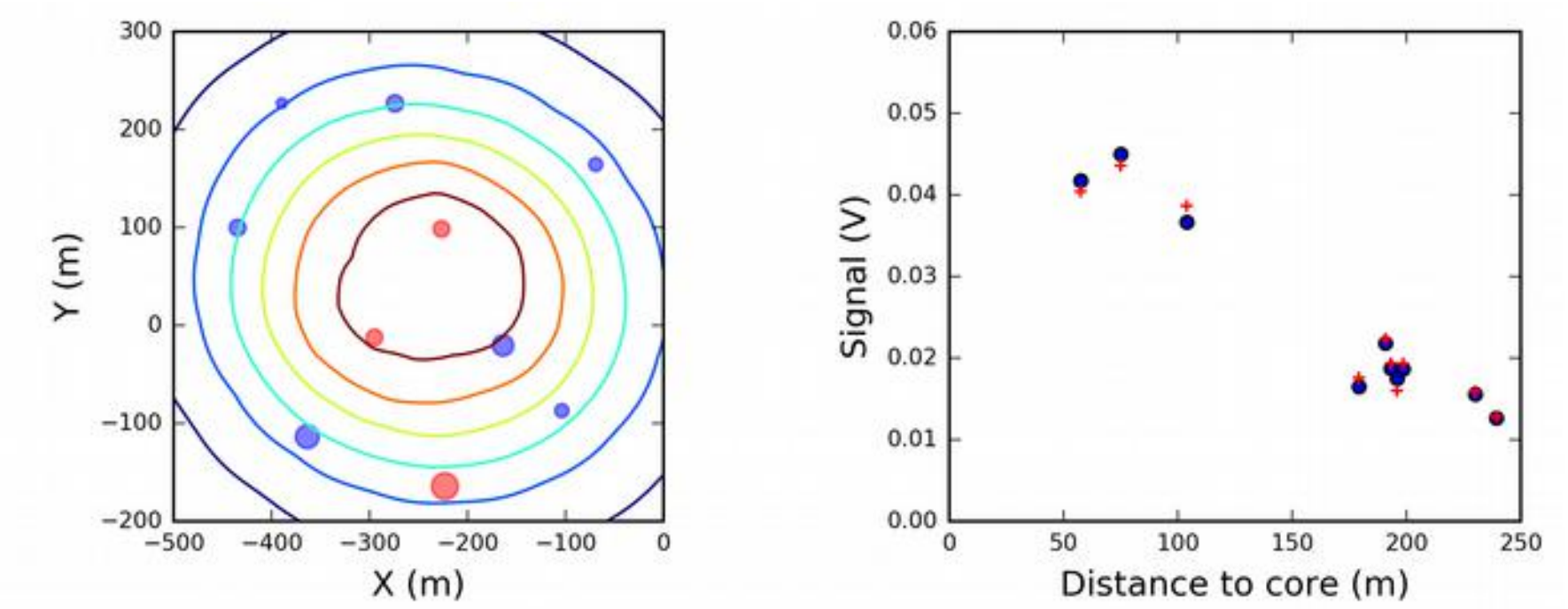

*Figure 5: Example of the SELFAS simulation producing the best match to the data (from the event on the right side of Figure 2). The SELFAS convoluted amplitudes are depicted as contour lines. Relative differences (in %) between the experimental data and the SELFAS values are shown as colored circles (left). Amplitudes are shown as a function of the distance to the core location (right) for experimental data (blue circles) and SELFAS simulation (red crosses).*

### 6.2 Minimization procedure

The procedure used to deduce the cosmic-ray features from the simulations is inspired from the methodology developed in [8]. For each experimental event, the quadratic sum of the EW and NS amplitudes are calculated for all the real antenna in a given frequency band (typically 30-80 MHz or 30-200 MHz). For each simulation, a range of core locations and a range of multiplication factor (to cope for unknown energy) are tested. Convoluted amplitudes are calculated at the real antenna locations and a $\chi^2$ is built by summing up the square of the amplitude differences. The best core location and best multiplication factor is determined from the smallest $\chi^2$. The results of the minimization for a particular event are illustrated on Figure 5 and Figure 6.

For each simulation, a core location, a multiplication factor and a $\chi^2_{min}$ are determined. By plotting the $\chi^2_{min}$ versus the $X_{max}$, the simulation best matching the experimental event is searched for. The distribution usually exhibits a minimum more or less pronounced that can be interpreted as the most probable value of $X_{max}$ for the considered experimental event (Figure 6).

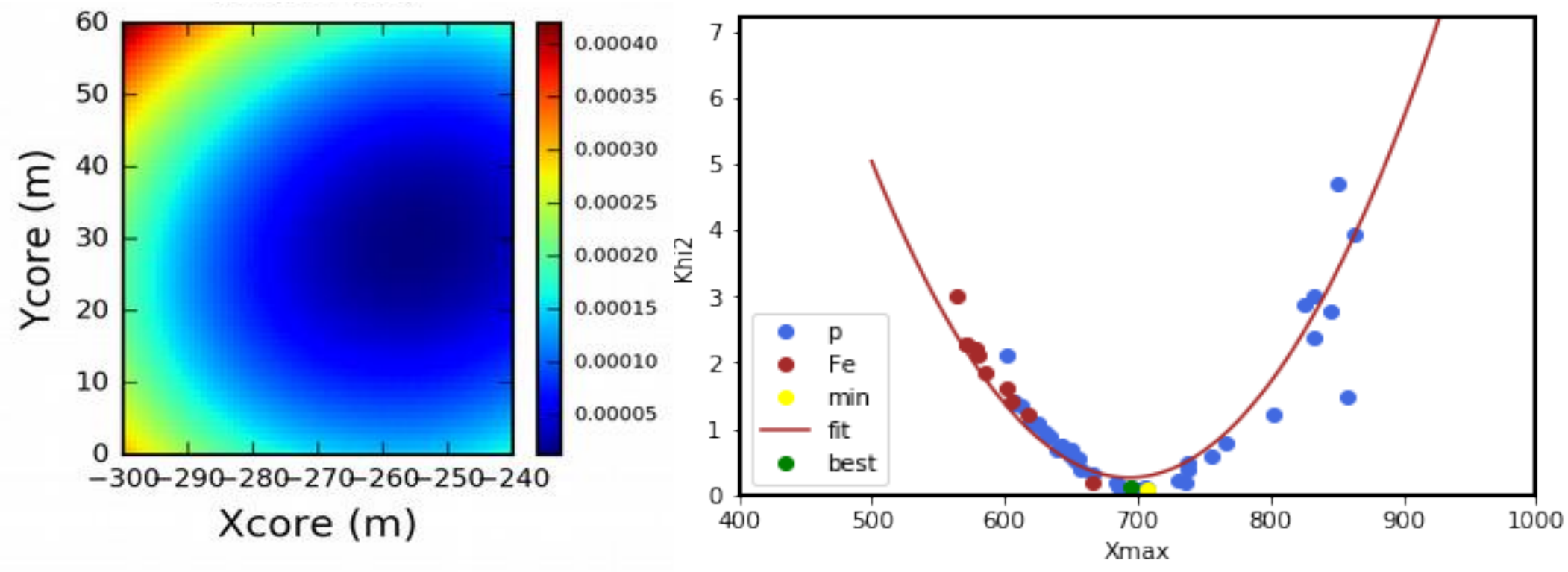

*Figure 6: Left: $\chi^2$ map used to deduce the core location (here around x=-250m and y=30m) for a simulation of the same experimental event (on the right of Figure 2) . Right: distribution of $\chi^2_{min}$ versus $X_{max}$ for the entire set of simulations.*

The overall features of the cosmic-ray are then accessible : the core location, the energy by combining the simulated energy and the multiplication factor, the $X_{max}$ and a particle identification probability.

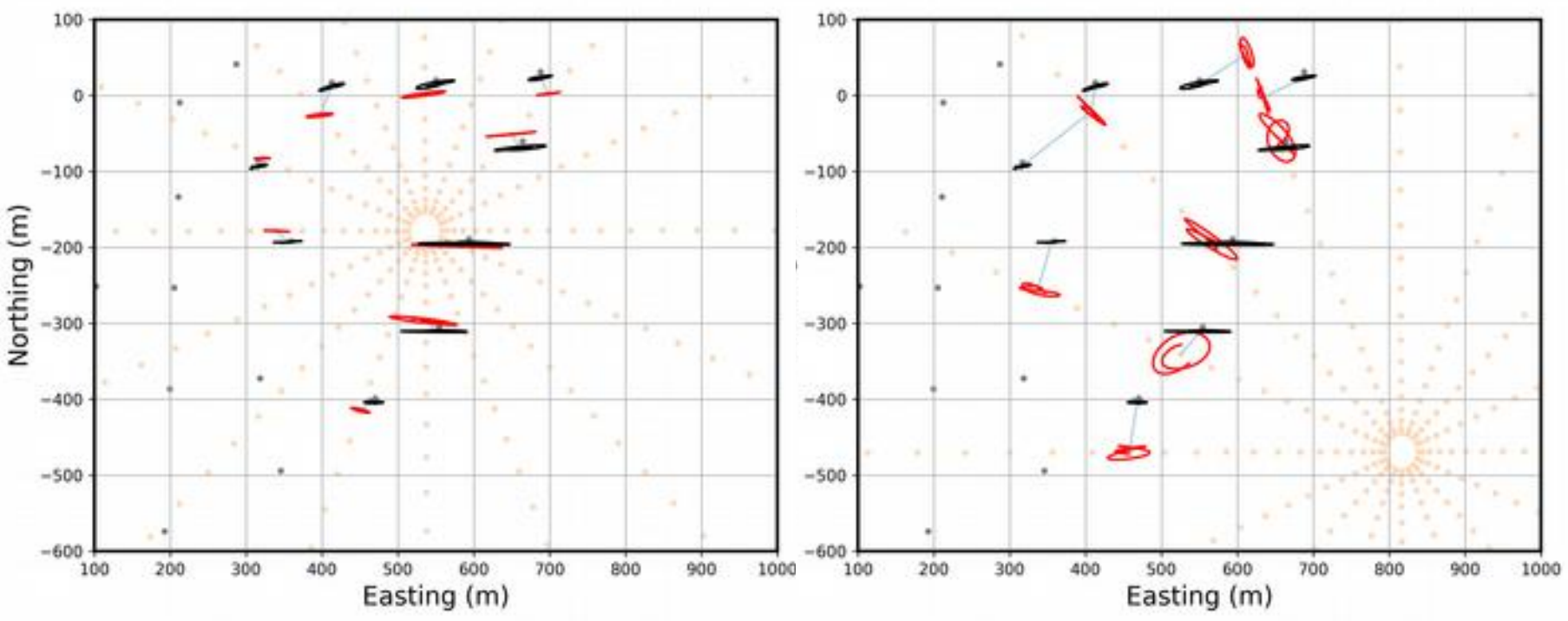

*Figure 7: Polarization patterns represented for the experimental data (black lines) and for SELFAS predictions (red line) obtained for a core location close to the best match (left) and a core location producing a local minimal $\chi^2$ (right).The full virtual array is shown (orange dots).*

The minimum in the distribution of $\chi^2_{min}$ versus $X_{max}$ is rather well formed for vertical showers and becomes less evident for more horizontal showers. In some rare cases, no minimum emerges (besides abnormal and extreme $X_{max}$ values). In such cases, a detailed inspection of the coincidence features usually reveals that the event is fortuitous or has very few chances to be a real EAS. Finally, a local minimum can eventually appears in the distribution which is associated to very different core location. Usually, in that case, polarization patterns exhibit very different variations and allow to select the correct core position (Figure 7). In addition, polarization pattern can be clearly a tool to distinguish between real cosmic ray events and parasitic noise events induced for instance by local static sources or aircrafts.

## 7. Comparison with the reconstruction in the particle detectors

The quality of the cosmic ray features deduced from the reconstruction combining the radio signals and the SELFAS simulations can be estimated by comparing them with the reconstructions in the scintillators. The energies deduced from the antennas $E_{radio}$ are plotted versus the energies derived from the scintillators $E_{scint}$ on the left panel of the Figure 8. The $E_{scint}$ are systematically lower than the radio estimate $E_{radio}$. The surface covered by the autonomous station array being much bigger than the surface covered by the scintillator array, most of the events fall outside the scintillator array and consequently their core location is not reliable and the energy estimate is underestimated (the reconstruction method trying to pull the core location inside the array limits). Using the core location provided by the radio signal reconstruction, the energy can be deduced again from the scintillators (Figure 8). The agreement is then clearly improved.

## 8. Conclusions

This preliminary analysis on a limited data set shows quite good agreements between the autonomous station signals compared to SELFAS simulations. First comparisons with the scintillator array are encouraging but still needs detailed work to improve the reconstruction in the particle detectors. These results suggest that the radio detection technique has the intrinsic capabilities to infer the main properties of high energy cosmic rays. This ongoing work clearly needs improvement on several aspects : a larger set of data can be quickly processed, careful analysis of the various source of error must be done in order to determine the resolution for the core location, $X_{max}$ and energy values. Probabilities must be computed for a particle identification estimate to study its variations over the energy range covered by CODALEMA. Extending this method to the low multiplicity and to the highly inclined events will increase the overall reconstruction efficiency and acceptance. A blind reconstruction can be undertaken to assess the cosmic ray detection efficiency and purity among the full set of recorded data. Finally, one can artificially lower the coincidence multiplicity (by ignoring some antennas) to estimate the resolution loss in a degraded mode which would be useful to anticipate the performances of a sparser antenna array.

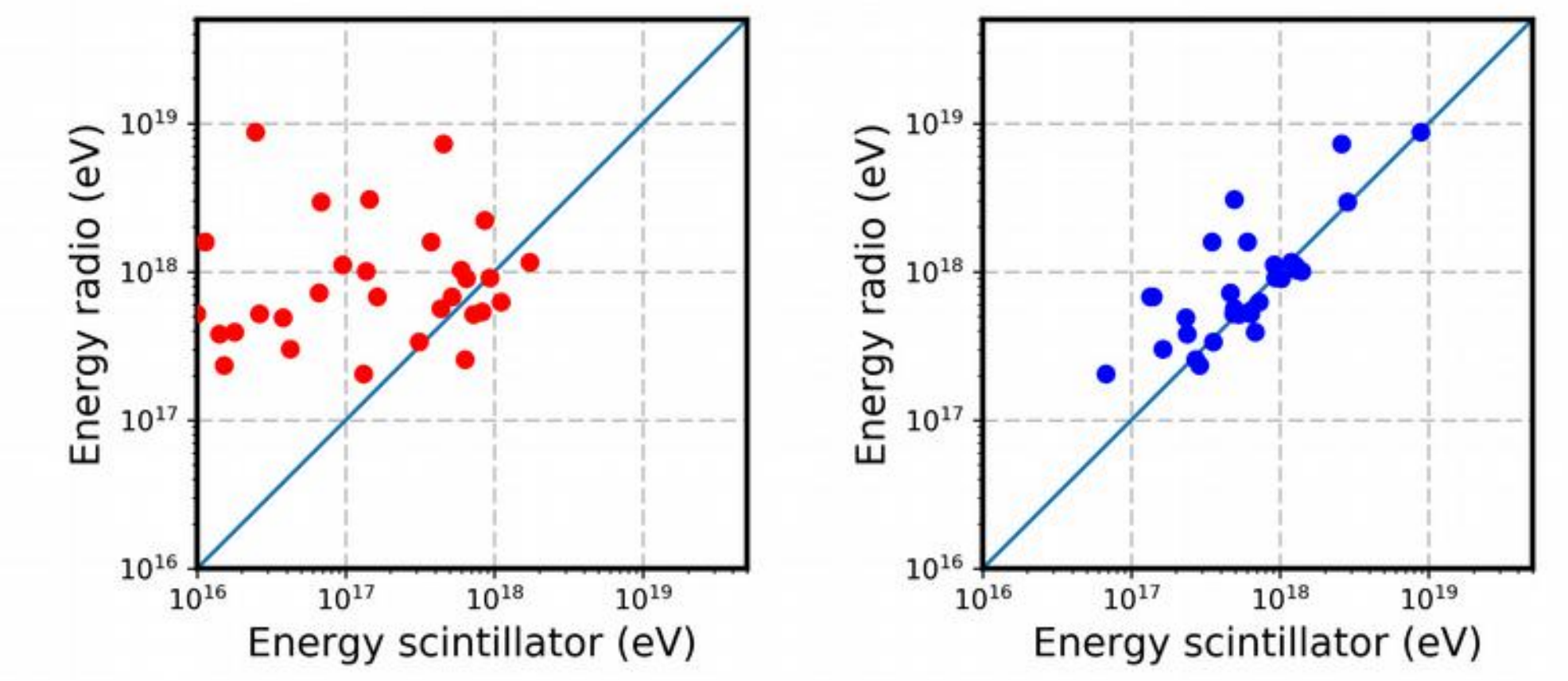

*Figure 8: CR energy deduced from the antennas versus the energy extracted from the scintillators (left). Energy deduced from the antennas versus the energy extracted from the scintillators with the core location taken from the radio signal analysis (right).*

35th International Cosmic Ray Conference
The Astroparticle Physics Conference

# On timing accuracy in observing radio impulses associated with Extensive Air Showers

**Alain Lecacheux**[†,1], **D. Charrier**[2], **R. Dallier**[2] , **A. Escudié**[2], **D. García-Fernández**[2], **L. Martin**[2] **and B. Revenu**[2].

[1]*CNRS-Observatoire de Paris, Meudon, France*

[2]*Subatech, CNRS-IN2P3, Nantes, France*

*E-mail:* alain.lecacheux@obspm.fr

A large subset of EAS radio observations by CODALEMA experiment in Nançay was used to measure the departure from a plane of the signal arrival times to the antennas. The study was done at two different scales in terms of range to the EAS axis. Below about 300m, by using a 10 antennas array triggered by neighbouring surface detectors, the average departure was found to be negligible (i.e. lower than the measurement incertitude of about 2-3 ns), leading to an equivalent "wave front curvature" radius larger than 30 km. Above the 300m range, and within the first kilometre, an extended array of several tens of self-triggered antennas was used. The measurements suggest the existence of significant time departures (of order 10 ns at 1 km range), which corresponds to an apparent curvature radius much larger than the atmosphere thickness. The main limitation of the method, in addition to the inferior time precision due to self-triggering, was the lack of accurate knowledge of corresponding EAS core locations, especially when those cores are located well outside the surface detector extent.

[†]Speaker

## 1. Introduction

Radio impulses associated with Extended Air Showers (EAS) produced in terrestrial atmosphere by High Energy Cosmic Rays (UHECR) of energy $10^{17}$eV and above, are now routinely observed by dedicated radio instruments on ground. Two basic, non-radiative processes [1] are able to produce significant, impulsive change in the electric field measured by a distant observer: i) a "charge excess" due to differential absorption of negatively and positively charged secondary particles along the EAS development; ii) a "charge displacement" of those particles, leading to a "transverse current" and its implied electric field by the Lorentz force exerted by Earth's magnetic field [2].

In both processes, two different versions of the resulting field can be produced, depending on whether the charge is moving slower of faster than the velocity of light in the medium [3]. In the first version, - the infra-luminous case -, only one instant in the charge's past history has a light cone which reaches any given location in space-time (e.g. an antenna). In the second version, - the supra-luminous case -, the field visibility is restricted to from inside the Čerenkov cone. Since the air index is variable with altitude and composition (in particular humidity), the two infra and supra luminous versions of each emission mechanism can coexist within a single shower observed by a spatially distributed antenna array.

Furthermore, in the aim to better explain radio observations, most of the recent numerical simulation models (e.g. CoREAS [4], ZHAireS [5], SELFAS2 [6]) do include an additional radiative term (acceleration), expressing the "time-variation" of both charges and transverse current.

In order to check and disentangle all those contributions, the real signals have to be carefully analysed. For instance, the accurate measurement of times of arrival on the individual antennas of a radio array is a way of sampling in space the EAS electric field distribution (i.e. the shape of the wave front in case of a purely radiative process), for comparison with corresponding model predictions.

In the following, a large subset of EAS radio observations by CODALEMA experiment in Nançay (France) is used to accurately measure the departure from a plane of the signal arrival times to the antennas.

## 2. Observations and analysis

### 2.1 Instrumentation

Three different CODALEMA sub-systems (Fig.1) have been used in this work (see [7] for a more detailed description of the CODALEMA experiment).

The surface detector (SD) is made of 13 scintillators covering a limited area of 400 m × 400 m, allowing unambiguous detection of high energy cosmic ray events. It provides a trigger signal sent to CA and to a dedicated GPS station for accurate time recording and off-line time coincidence with the stand-alone antenna stations (SA). Firing the trigger requires that five scintillators out of 13 can see a signal above the level corresponding to an energy detection threshold in the range $10^{15}$ to $10^{16}$eV.

The compact array (CA) is made of a cluster of 10 active antennas in dual linear horizontal polarisation, distributed over a square of about 150 m × 150 m (24 to 146 m spacing). The elementary antenna is a dual inverted V, fat dipole, similar to the ones used in the Long Wavelength Array interferometer (LWA1) in USA [8], but equipped with LONAMOS wideband, low noise amplifiers [9]. The output voltages of each pair of dipoles are digitized by a 400 MHz, 14 bits ADC, then down converted to base band and added in quadrature, in order to provide 10 circularly polarized output signals without too much dependence on azimuth. Each ADC, triggered by the surface detector (scintillators), runs for 6.5 µs, with a time resolution of 2.5 ns and a time (relative) accuracy much better than 1 ns.

Fifty seven self-triggered standalone stations (SA) are scattered over roughly 1 km$^2$ around the two former CODALEMA sub-systems. Each station is made of two crossed horizontal fat dipoles, fed to same LONAMOS low noise amplifiers as those used in the CA array. Same ADC boards are also used, but working with higher (1 GHz) sampling rate. A multi-level triggering system has been developed to select narrow transients specific to air shower induced electric field and to reject signals produced by anthropic noise sources. An internal GPS clock ensures common time scale with SD and CA, with an accuracy better than 5 ns.

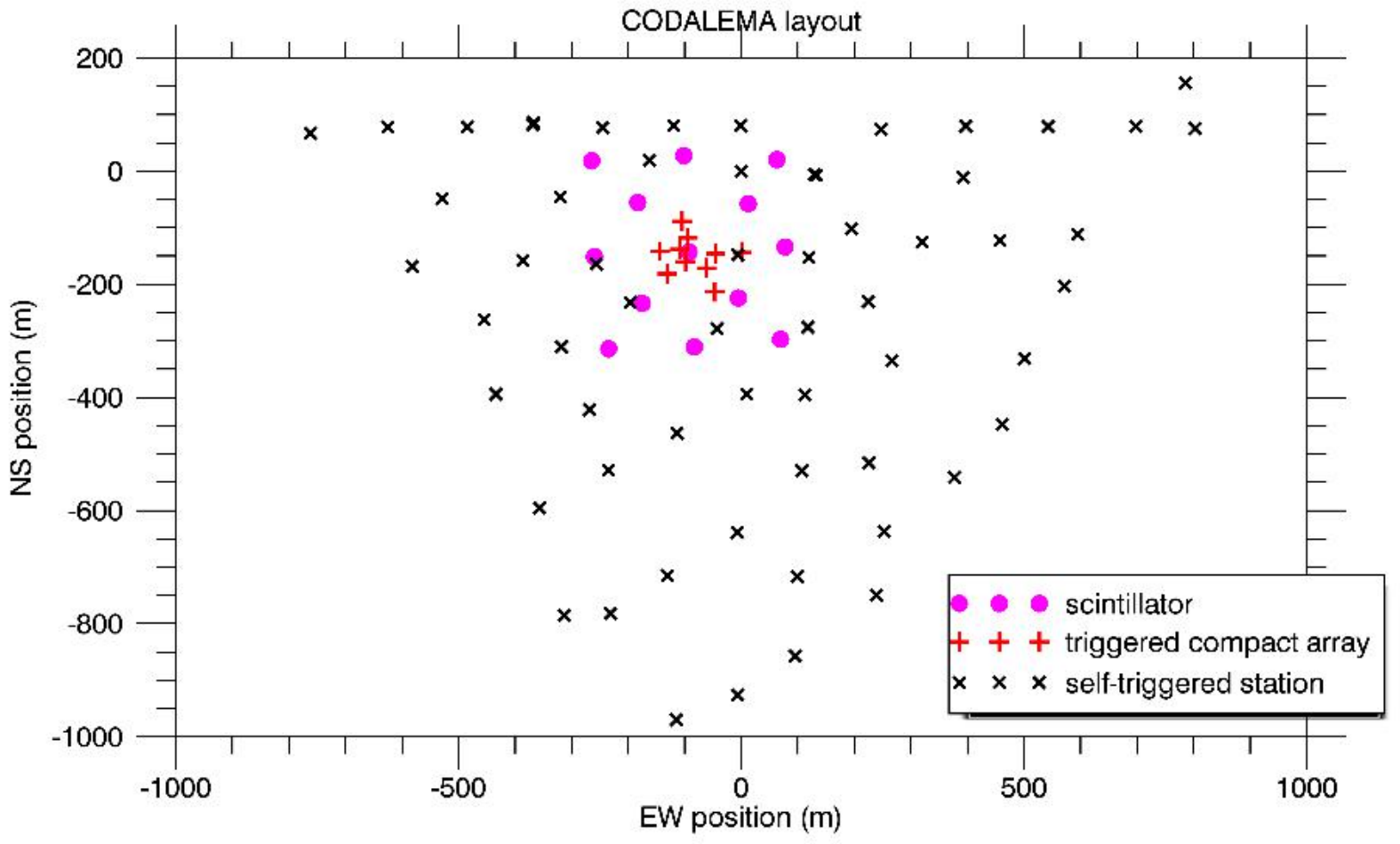

***Figure 1***: layout of the CODALEMA experiment

## 2.2 Method

Timing reconstruction of EAS radio signals is a special case of the well-studied TDoA (Time Difference of Arrival) problem, belonging to the general DoA (Direction of Arrival) problem when only signal timing is considered. Correctly solving this problem is conditioned by the accuracy of the sensor positions knowledge, the accuracy of the event dating and the relevance of the emitting source model.

Let $N$ sensors $\{\mathcal{A}_j\}_{j=1..N}$ located at positions $\vec{r}_j$ and measuring at times $t_j$ (or pseudo-ranges $p_j = ct_j$) the arrival of a perturbation propagating at the light velocity $c$ in direction $-\vec{u}$. Let $\vec{R}$ and $P$ the position and pseudo-range (time of arrival at ground) of the EAS radio core (by analogy with the EAS particles core). We can consider four models:

1) the perturbation source is at distance $\mathrm{C}^{-1}$ in direction $\vec{\mathrm{u}}$:

$$p_j = P - C^{-1} + \left\|C^{-1}\vec{u} - \vec{r}_j\right\| = P - C^{-1}\left(1 - \sqrt{1 - 2C\vec{u}\cdot\vec{r}_j + C^2\left\|\vec{r}_j\right\|^2}\right)$$

which reduces (in far field limit $\|\vec{r}_j\| C \ll 1$) to:

$$p_j \approx P - \vec{u}\cdot\vec{r}_j + \frac{C}{2}\left\|\vec{u}\times\vec{r}_j\right\|^2$$

2) the source is at infinite ($C = 0$):

$$p_j = P_\infty - \vec{u}_\infty \cdot \vec{r}_j$$

3) the isochrones are planar surfaces perpendicular to $\vec{u}$ :

same expression as in model 2

4) the isochrones are curved surfaces of revolution around the $\vec{u}$ axis:

$$p_j = P - \vec{u}\cdot\vec{r}_j + \frac{C}{2}\mathcal{F}\left(\vec{u}\times\left(\vec{r}_j - \vec{R}\right)\right)$$

where $\mathcal{F}$ is a function of the antenna distance to the EAS axis, satisfying:

$\mathcal{F}(0) = \mathcal{F}'(0) = 0$ and $\mathcal{F}''(0) = 1$.

The latter expression (model 4) can summarize all the other models. Here, the parameter $C$ is the curvature of the isochrone at its intersection with the EAS axis (i.e. at the EAS core location).
It might be tempting to use this expression for building, from the $\{r_j, p_j\}$ measurements alone, a least squares estimation of the $P, \vec{u}, C, \vec{R}$ unknowns as well as of some additional parameters describing $\mathcal{F}$ (as, for instance, in [10]). Without entering further mathematical discussion (not possible because of the allowed length of these Proceedings), it should be clear that such a calculation would lead to solve an ill-conditioned problem and likely be not conclusive, since $\vec{R}$ and $\mathcal{F}$ are of the second order with respect to $P, \vec{u}, C$.
Therefore, we have adopted a simplified two steps procedure: the first step is a fit of the model 2 (or 3) to the $\{r_j, p_j\}$ measurements, providing the $P_\infty, \vec{u}_\infty$ solution and a set of residuals $\delta p_j$; the second step consists in a fit of model 4 to $\{r_j, \delta p_j\}$ by choosing $\mathcal{F}(d) = d^2$ (parabolic isochrone), yielding the system of equations:

$$\left\{\delta p_j = \delta P - \delta\vec{u}\cdot\vec{r}_j + \frac{C}{2}\left\|\vec{u}_\infty\times\vec{r}_j\right\|^2\right\}_{j=1..N}$$

in which $\delta P, \delta\vec{u}$ and $C$ are new unknowns, assumed to be of the same order of magnitude. Finally, the amount of curvature can be tested upon magnitude of the $\chi^2$ decrease and significance of the C parameter departure from 0.

### 2.3 Results

All elements of the CODALEMA experiments were recently geolocalized with an accuracy of about 0.3 m (1 ns), including altitudes (the ground level indeed varies vertically by $\pm 8$ m over the Nançay radioastronomy observatory area).

The efficiency of our DoA reconstruction was tested on several thousands of transient aircraft signals (obtained during hundredths of air flights over Nançay during the two last years) as well as on a large amount of recorded EAS radio events (more than 3000 at time of this writing). These two kinds of signal are by far the most frequent ones when looking at high elevation. Regarding the aircraft signals, the reconstruction (model 1 of section 2.2) was accurate enough for allowing identification of commercial flights and successful check of altitude, speed and heading of the aircraft.

From the statistical analyses above, the distribution of pseudo-range residuals, for the CA antenna array, was found to be unimodal, with a 1.5 m upper boundary and a peak (modal value) located at 0.75 m, consistent with the used 2.5 ns sampling time (cf. Section 2.1). On the other hand, the pseudo-range residuals for SA antenna array display a similar distribution, but shifted towards higher values by a factor of about 3. The mode of the distribution of time residuals is located at about 2.4 m (or 8 ns), reflecting the loss of precision due to the GPS synchronisation scheme.

In order to answer the initial question about the shape of the EAS radio isochrones, we distinguished between two different ranges of distances to EAS axis.

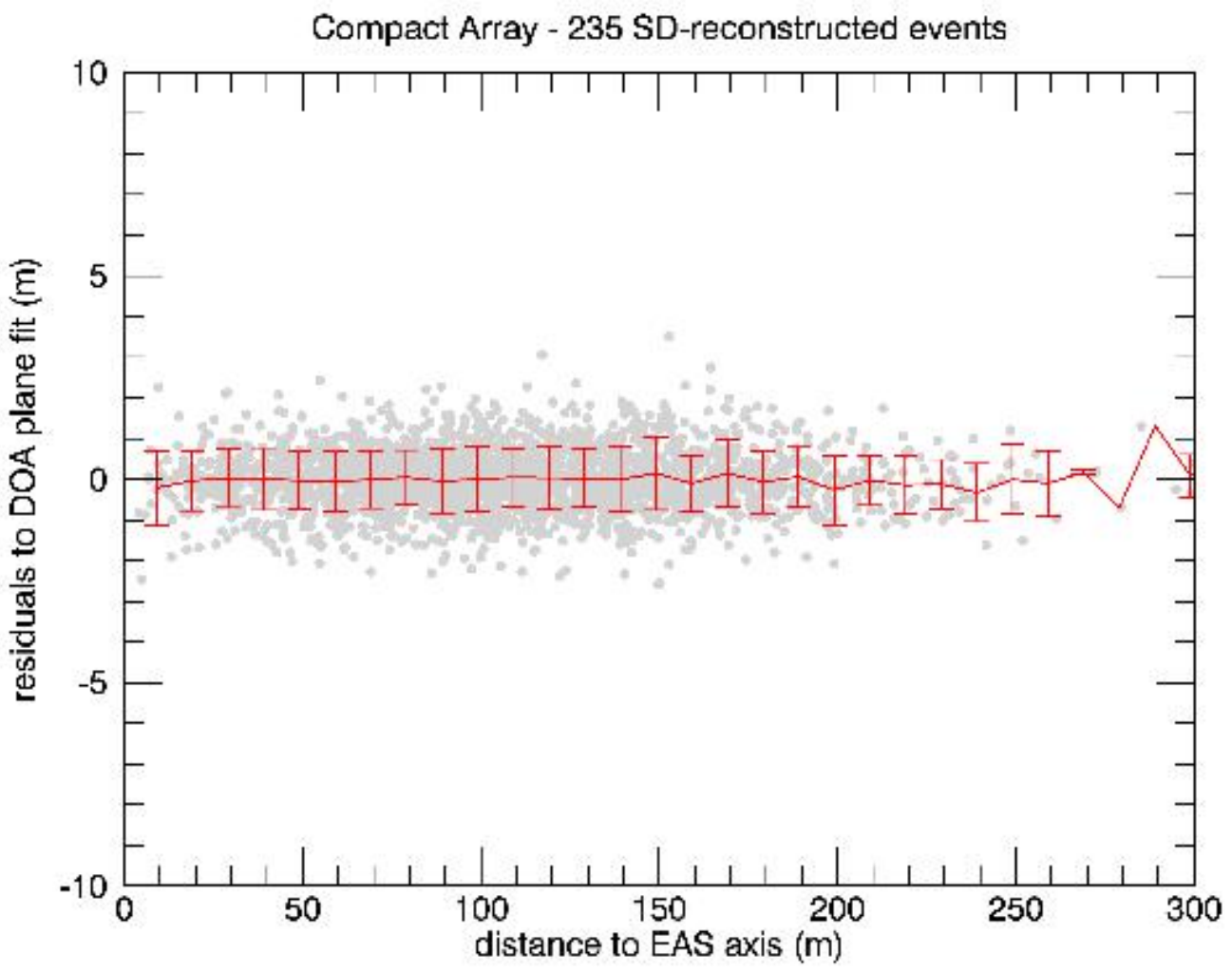

***Figure 2***: CA analysis of 235 EAS events. The $1\sigma$ error bar is calculated for each 10m range.

By using the CA instrument, we could explore the first 300 m, by selecting 235 high quality events, well reconstructed (energy, core location, etc...) from the corresponding SD data. Results are displayed in Fig.2. No obvious curvature is visible, leading to a curvature radius, if any, larger than 30 km.

By using the SA instrument, we could explore larger distances, up to about 900 m, by using 34 more EAS events, also seen by the SD, but, unfortunately, badly reconstructed (as usually when the EAS core is outside the area containing the scintillators). Results are displayed in Fig.3. As for CA, but with a larger scattering, the distribution of residuals below 300 m appears to be flat around zero. Beyond 300m, a slight tendency for increase of the residuals is barely visible, reaching about 10 ns at 1000 m, and leading to a curvature radius larger than 100 km.

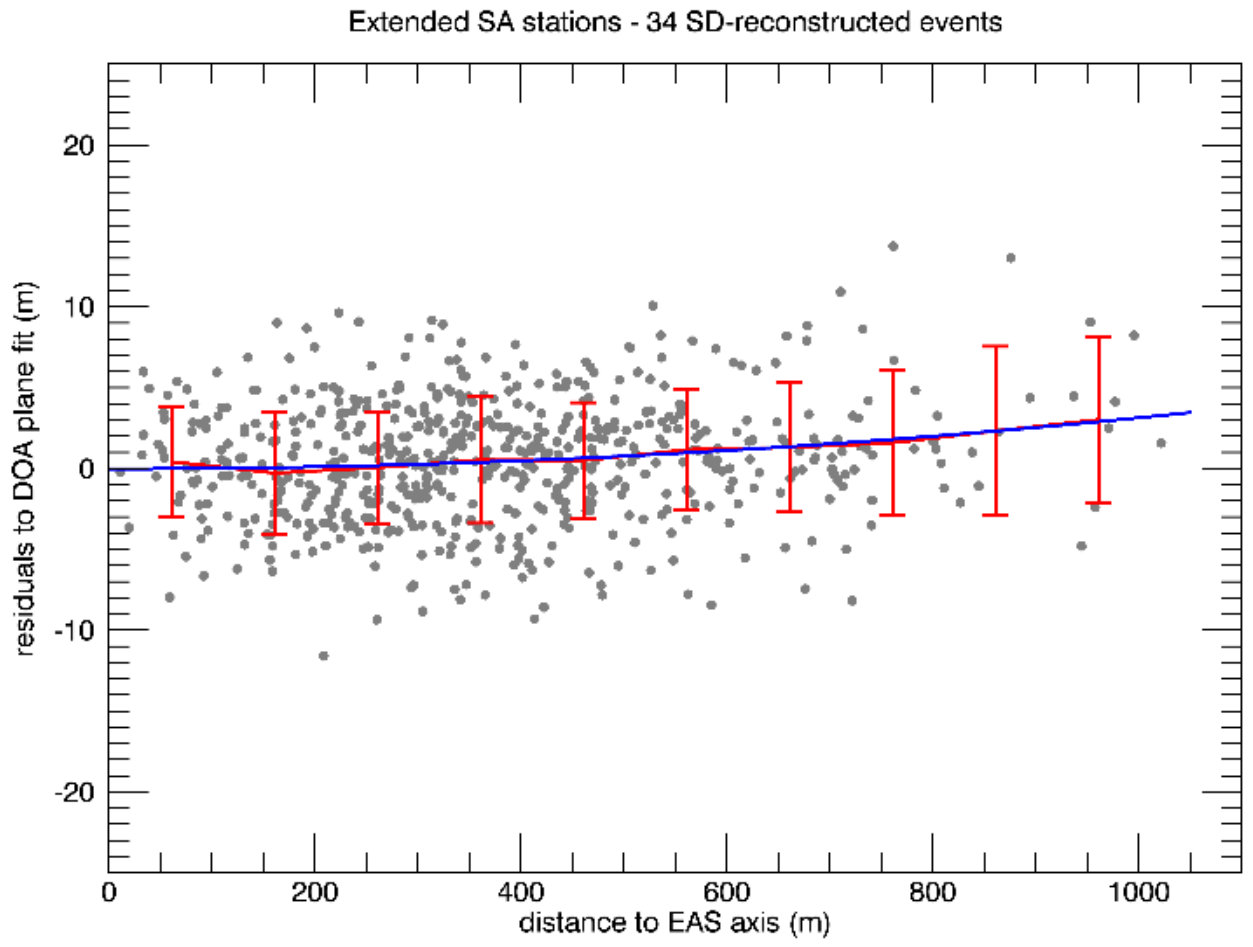

***Figure 3***: SA residual analysis for 34 external EAS events, for which core locations could only be estimated. The $1\sigma$ error bar is calculated for each 100 m range. The blue curve is the best parabolic fit corresponding to the average curvature $C = 0.0064\ \pm 0.0006\ km^{-1}$.

## 3. Discussion and Conclusion

Statistical analysis of the most recent CODALEMA data suggests that the isochrone surfaces of radio signals produced by Extensive Air Showers are weakly or not curved, with a radius of curvature larger than ~20 km, at the difference of the pancake of secondary particles which is known to show up a pronounced curvature (curvature radius of ~8 km for a shower with energy of $10^{18}$ eV [11]) linked to the altitude of the maximum development ($X_{MAX}$) of the shower.

Our result is in contradiction with previously published works [10,12] which, on the contrary, found a more pronounced curvature of the radio front, with typical time departure of ~20 ns (~7 m) at 300 m from EAS axis (cf. Fig.6 in [10] and compare with our Fig.2). Some further analysis is needed to resolve the contradiction.

Let us note that the existence of such a "radio curvature" is only legitimated by the assumption of a radiative part in the radio emission mechanism (implying that the radio antenna would be in the near field of an emitting distant source). In the opposite case, the relativistic effects are constraining impulses of electric field to occur when the observing antenna is just contained in a plane perpendicular to the particles propagation, implying a basically planar radio front.

# Direct measurement of the vertical component of the electric field from EAS

**R. Dallier**[*1,3], **H. Carduner**[1], **D. Charrier**[1,3], **L. Denis**[3], **A. Escudie**[1], **D. García-Fernàndez**[1], **A. Lecacheux**[2], **L. Martin**[1,3], **B. Revenu**[1,3]

[1] *SUBATECH, Institut Mines-Telecom Atlantique - CNRS - Université de Nantes, Nantes, France*
[2] *CNRS/Observatoire de Paris, Meudon, France*
[3] *Station de radioastronomie de Nançay, CNRS/Observatoire de Paris - PSL - Université d'Orléans/OSUC, Nançay, France*
*E-mail:* richard.dallier@subatech.in2p3.fr

A three-fold antenna system has been installed nearby the center of the CODALEMA particle detector. Its goal is to measure the complete electric field produced by air showers, i.e. along the 3 polarizations East-West (EW), North-South (NS) and vertical. Indeed on all currently operating radio detection arrays, only the horizontal NS and EW polarizations or their projections are measured. This allows the vertical electric field component to be reconstructed, provided that the far-field assumption is valid, but though strong hints based on the theories of air shower radio emission tend to validate this hypothesis, it has never been verified experimentally. We present the 3D antenna and its acquisition system, and the first results obtained.

*Speaker.

## 1. Objectives

Since the renewal of the radio detection method for the observation of high energy cosmic ray extensive air showers (EAS) at the beginning of the $21^{st}$ century, several experiments (well reviewed in [1] for example) have been set up throughout the world, and their results are widely exposed in this conference. Among them, one of the most ancient still in activity is CODALEMA, hosted by the Nançay Radio Observatory in France and described notably in [2]. Up to now, every currently operating experiment (for instance, AERA [3], LOFAR [4] or Tunka-Rex [5], and including CODALEMA) measures the electric field with antennas having two horizontal polarizations, along the East-West (EW) and North-South (NS) directions. The signal from each polarization is recorded as a voltage by an ADC, and knowing the complete acquisition chain characteristics (antenna transfert function, cable attenuation, amplifier gains, filter responses...), the electric field components can be derived through an unfolding procedure [6]. If most of the items composing the latter are only frequency dependent, the antenna transfert function also depends on the arrival direction of the signal and on the antenna environment, and is generally derived from simulations. To infer this antenna response, the "far field assumption" is made in the simulation codes, assuming that the signal received by the antenna has no longitudinal component along the shower axis, therefore the total electric field including the vertical, third component, can be reconstructed from the other two. Though current experimental results compared to EAS electric field simulations are very convincing (main component should be the transverse current due to the separation of charges in the geomagnetic field) and enhance the robustness of the radio detection method (see for example [7]), this far-field assumption has never been verified experimentally. It is worth noticing that, in the results of every EAS electric field simulation codes, based on the calculation of the contribution of each individual shower particle along its path to the global electric field, the longitudinal component is present, though most of the time negligible. By confirming or rejecting the far field hypothesis, measuring the complete electric field could help reaching a better accuracy, notably on the conversion of the electric field value to an energy. It could also help recovering the arrival direction of the signal with a single three-fold antenna, provided that accurate determination of the 3 polarization components is made. Several attempts have been made to measure the vertical polarization of the electric field, for example on LOPES in 2011 [8] and AERA in 2014 (internal communication). However, none has given convincing results, partly due to the mechanical difficulty to use as efficient antennas as in the horizontal plane. Hence, a revision of the concept of the three-dimensional measurement of the electric field should be done and is engaged in the EXTASIS project hosted by the CODALEMA experiment. The main goal of EXTASIS is to measure the low frequency ($\leq$ 10 MHz) contribution of EAS and possibly detect the so called "Sudden Death" contribution, the expected radiation electric field created by the particles that are coherently stopped upon arrival to the ground [9]. Regarding the large wavelengths associated to low frequencies with respect to the distances involved in an antenna array, it is also of primary importance to properly define and calculate the near field contribution of the electric field at all frequencies, as it is done in [10]. In that respect, EXTASIS aims at giving the most global and precise view on the properties of EAS electric field from 1 to 200 MHz and, associated to CODALEMA, at defining the experimental limits of the radio detection method.

## 2. Experimental setup and data set

The CODALEMA/EXTASIS experiment is described in [2]. It is composed of 5 arrays of detectors, among them 3 are of main interest for the work described in this paper: the particle detector array, the array of 57 standalone radio detectors and the cabled "compact array" in the middle of the particle detector (Fig. 1, left).

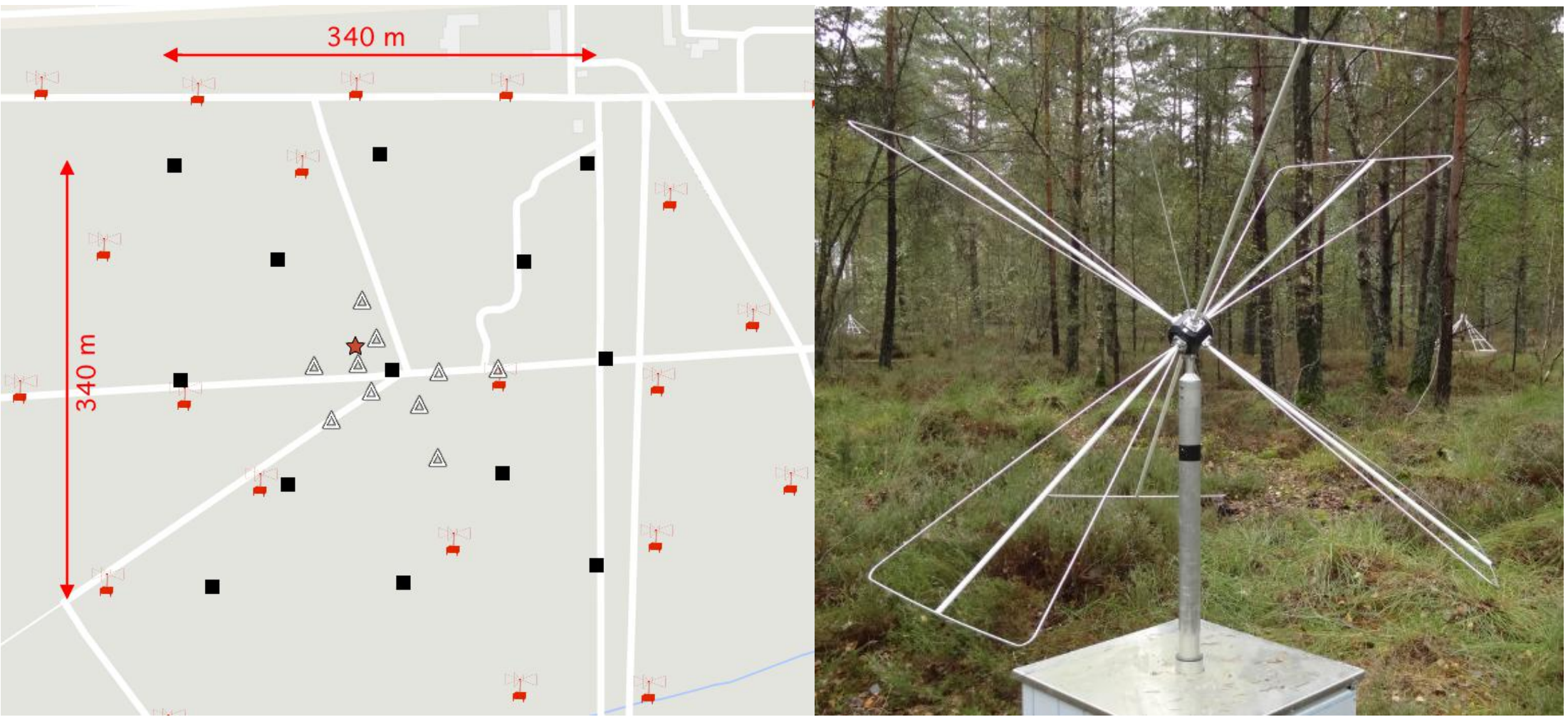

**Figure 1: Left:** zoom on the map of CODALEMA (North on top). Red symbols: some of the 57 CODALEMA autonomous radio detection stations; black squares: 13 scintillators; white triangles: 10 antennas of the compact array; orange star: "tripole" antenna. See [2] for more explanations. **Right:** the tripole antenna. Each Butterfly antenna is rotated twice to form a direct trihedron inclined by 45°.

Recently, a three-fold antenna system has been installed nearby the center of the particle detector. It is called "Tripole" antenna. In previously cited attempts to detect the vertical electric field, the antenna was either made of 3 simple dipoles (Lopes3D) or of Butterfly, double plane-polarization antennas to which a single monopolar antenna was added (AERA). In all cases, none of these vertical antennas had good enough performances and sensitivity. We chose to reconsider the problem and to use 3 regular Butterfly antennas equipped with the LONAMOS LNA, which have shown, notably on CODALEMA and AERA, their very good performances [6, 7]. The main difficulty lies in the mechanical holding of such a triple antenna system along the defined polarization axes. Therefore a special holding system has been developed, on which two rotations of 45° and 54.75° are applied around the $X$ (EW) and $Y$ (NS) axes respectively (Fig. 1, right). Triggered by the particle detector, the 3 signals are recorded and dated at a nanosecond accuracy, with the same type of electronics as for the scintillators (2.56 $\mu$s record length, 1 GS/s, 12 bits ADC). This ensures that the Tripole antenna signals, if any, are strongly correlated to particle EAS, though some accidental detection are possible, but extremely rare in the considered time window.

The operation mode of the standalone antenna (SA) and surface detector (scintillators, SD) arrays are described in [7] and [2], while the one of the compact array (CA) is more detailed in [11] (note that the "LWA" CA antennas described herein, though different in shape from the SA "Butterfly" ones, are equipped with the same LONAMOS Low-Noise Amplifier (LNA) described in [12] and have the same detection performances). We will focus on these 3 subsystems to build the data set for the analysis of the Tripole antenna data. The latter are obtained on each SD trigger, which are dated for further comparison with the SA events. The CA is also externally triggered by the SD, and its signals are recorded on a longer time base (6.4 $\mu$s at 400 MS/s) with the same type of ADC as for the other instruments. For the SA data analysis purposes of [7], we have selected a small subset of 64 events self-triggered by the SA with a minimal multiplicity of 4 and a maximal multiplicity of 48 (number of SA fired), and within $\pm 5$ $\mu$s from a SD trigger. These events are spread over the whole array, thus it is not expected that they all present a CA counterpart: indeed, among them, only 34 exhibit a clear signal in the ten antennas of the CA in at least one polarization, and finally 24 of these events with also a clear signal in at least two Tripole antenna polarizations are selected as our data set. The Tripole being located very close to the CA (surrounded by 4 of its antennas, see Fig. 1), these 24 events will be compared to the ones of the CA in order to derive some properties of the recorded signals.

## 3. Compact Array antenna signal unfolding

As mentioned before, it is possible to recover the vertical component of the electric field vector knowing its $X$ and $Y$ components, and assuming a far field propagation and the knowledge of the antenna gain pattern. Based on NEC-2 simulations of the LWA antenna including its environment, a vector equivalent length (VEL) matrix was calculated in the frequency domain for each direction of an incoming electric field $\vec{\mathbf{E}}_{\theta,\phi}$ in a hemisphere with a resolution angle of $1^\circ$ and a [10-199] MHz frequency range. This VEL, noted $\vec{\mathbf{L}}(\theta,\phi)$, is defined as the ratio of a vector induced voltage over a 50 $\Omega$ LNA terminal load and the vector value of the incoming electric field defined in a spherical coordinate system (Eq. 3.1) for a given $\theta$ and $\phi$ direction. The LWA antenna is centered on the $Z$-axis of a cartesian coordinate system. One dipole is aligned on the $X$-axis (EW) and the other on the $Y$-axis (NS). The following spherical coordinate system convention is adopted: $\theta$ is the zenith angle, counted from the $Z$-axis, and $\varphi$ is the azimuth angle, counted anticlockwise from the $X$-axis. The VEL fully characterizes the active antenna since it depends both on the antenna radiator directional pattern and on the LNA transfer function. It is worth noting that two VEL matrix are performed for any crossed-polarization antenna, one for the dipole along the $X$-axis $\vec{L}^X(\theta,\phi)$ and the other for the $Y$-axis $\vec{L}^Y(\theta,\phi)$. With assumed far field propagation conditions, the radial component of the incoming electric field $\vec{\mathbf{E}}_r$ is null and $\vec{\mathbf{E}}_{\theta,\phi}$ is written:

$$\vec{\mathbf{E}}(\theta,\phi) = E_\theta(\theta,\phi)\vec{\mathbf{e}}_\theta + E_\phi(\theta,\phi)\vec{\mathbf{e}}_\phi \tag{3.1}$$

The VEL of the $X$-axis antenna is written for the same spherical coordinate system as:

$$\vec{\mathbf{L}}^X(\theta,\phi) = L^X_\theta(\theta,\phi)\vec{\mathbf{e}}_\theta + L^X_\phi(\theta,\phi)\vec{\mathbf{e}}_\phi \tag{3.2}$$

The voltage $U^X$ induced on the terminal load of the LNA connected to the $X$-axis antenna is the scalar product of $\vec{\mathbf{E}}(\theta,\phi)$ by $\vec{\mathbf{L}}^X(\theta,\phi)$. With $L^X_\theta$, $L^X_\phi$ the VEL matrix data for $X$-axis, $U^X(\theta,\phi)$ can be calculated (Eq. 3.3) for any electric field in all directions for both frequency and time domain:

$$U^X(\theta,\phi) = \vec{\mathbf{E}}(\theta,\phi)\vec{\mathbf{L}}^X(\theta,\phi) = E_\theta(\theta,\phi)L^X_\theta(\theta,\phi) + E_\phi(\theta,\phi)L^X_\phi(\theta,\phi) \tag{3.3}$$

Eq. 3.2 and Eq. 3.3 are the same for the $Y$-axis antenna replacing $X$ by $Y$. In practice, the signal are recorded in both EW and NS polarizations in volts as a function of time. Applying the unfolding algorithm to these time series under the far-field hypothesis gives the $E_\theta$ and $E_\phi$ components of the total field $\vec{E}$ in $V.m^{-1}$, with $E_r$ fixed at 0. The cartesian components of the electric field along EW, NS and vertical axis ($X$, $Y$ and $Z$) are simply obtained with

$$\begin{pmatrix} E_X \\ E_Y \\ E_Z \end{pmatrix} = \begin{pmatrix} \sin\theta\cos\phi & \cos\theta\cos\phi & -\sin\phi \\ \sin\theta\sin\phi & \cos\theta\sin\phi & \cos\phi \\ \cos\theta & -\sin\theta & 0 \end{pmatrix} \begin{pmatrix} E_r \\ E_\theta \\ E_\phi \end{pmatrix} \tag{3.4}$$

where $\theta$ and $\phi$ are the zenith and azimut angles of the shower axis respectively.

## 4. Recovering Tripole antenna polarizations

The Tripole antenna dipoles are not oriented directly in the ($X$, $Y$, $Z$) directions, but experienced 2 rotations around $X$ and $Y$ axis. For sake of simplicity, we have chosen to align one of the dipoles along the EW direction, though not in the horizontal plane (see Fig. 1). The two other dipoles are thus oriented in NW-SE and NE-SW directions, and if we apply the inverse of the rotation matrix made of the combination of the three rotations of angles (45°, 54.75° and 0°) to this Tripole base (EW, NW-SE, NE-SW), we recover the Tripole signal in the ($X$, $Y$, $Z$) = (EW, NS, Vertical) base. This would be a satisfactory first step of an unfolding procedure if the three dipole antenna patterns were the same (provided that, as it is indeed the case, the LNA, cables, filters and ADC have also the same response for each dipole acquisition chain). Currently there are no NEC-2 simulations of the Tripole antenna, thus no possibility to recover the VEL of each dipole. As a first estimate, we will consider the following:

- same acquisition chain for each dipole of the Tripole, from the LNA down to the ADC;
- at low frequencies in the frequency range considered (*i.e* $\leq$ 80 MHz), the gain patterns of the dipoles are the same and slowly depend on the frequency and on the arrival direction of the shower. This is a reasonable hypothesis, based on previous simulations of Butterfly antennas, considering the small size of the antennas versus wavelength (short dipole hypothesis);
- above $\simeq$ 100 MHz, important side lobes appear in the gain pattern, which becomes highly frequency and direction dependent: the sensitivity to $(\theta, \phi)$ is strong and the 3 dipoles can not be considered as equivalent anymore;
- the simple basis transformation from (EW, NW-SE, NE-SW) to (EW, NS, Vertical) should thus be considered as valid in the low frequency regime, allowing reasonable comparison of the polarization patterns between the Tripole voltages and the recovered electric field components by the CA antennas on the same events.

## 5. Results and comparison

### 5.1 Example of Tripole event

An example of event recorded by the Tripole antenna is shown on Fig. 2. Here, the raw signals have been transformed in the (EW, NS, Vertical) base and later on filtered within the [24-82] MHz frequency window.

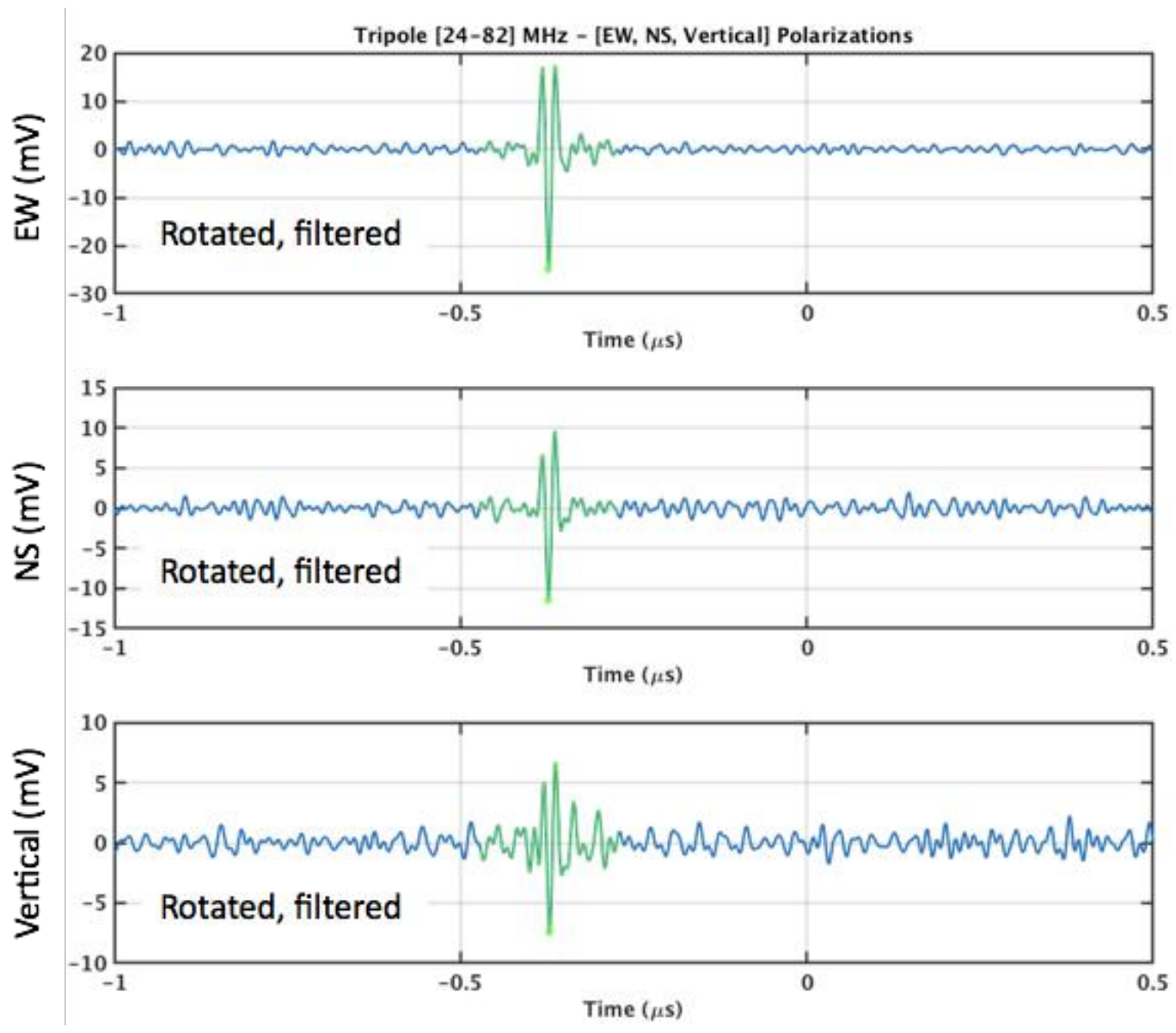

**Figure 2:** A Tripole event in the (EW, NS, Vertical) polarizations, filtered in [24-82] MHz. The signal is clearly visible on the three polarizations.

On Fig. 3, top, the polarization pattern for this event is presented for the three polarization combinations, together with the prediction in the case of a pure geomagnetic event (where charge excess contribution is expected to be null). Regarding the arrival direction of the event ($\theta \simeq 49^\circ$, $\phi \simeq 120^\circ$ counted counterclockwise from East) and the geomagnetic vector angle direction in Nançay ($\theta_B = 27.1^\circ, \phi_B \simeq 270^\circ$ (South)), such an event is not expected to exhibit a significant charge excess contribution and its polarization should follow the geomagnetic prediction, which seems indeed to be the case. On Fig. 3, bottom, the polarization pattern for the closest Compact Array antenna is traced for the same event, after applying the unfolding procedure. Both patterns are very similar for this example, which would support the far field hypothesis. It is worth noting that all the Compact Array antennas exhibit the same pattern for this event, as expected due to the small extent of the array and the event arrival direction.

### 5.2 Polarization pattern over the whole data set

The same study has been led over the whole data set of 24 events, from which 2 outliers were removed. As a summary, the average polarization angles of the ten antennas of the CA are plotted versus the ones of the Tripole for the NS/EW direction, and compared to the predicted geomagnetic angle for both instruments in the 3 directions (NS/EW, Vertical/NS and EW/Vertical). For this set of events, the arrival directions are spread over all the azimuts and zenith angles, without any cut thus no estimate of the possible charge excess contribution. At first glance, if one considers the NS/EW polarization only, a clear correlation is found between both instruments and also with the prediction (Fig. 4). The correlation is less clear - but still here - for the EW/Vertical polarization angles, even if an instrumental inversion when connecting the LNA of one of the Tripole dipoles leads to opposite trends. This is not essential, because the expected linear trend is still visible. However, it disappears almost completely for both instruments (CA and Tripole) when considering the Vertical/NS polarization, which should by construction be close to the $\theta_B$ component of the magnetic field vector (*i.e* $\simeq 27^\circ$ in Nançay). This unexpected result supports further investigations with a larger data set, cuts on the event arrival direction and different filtering bandwidth in order to verify the frequency dependence of the Tripole gain pattern.

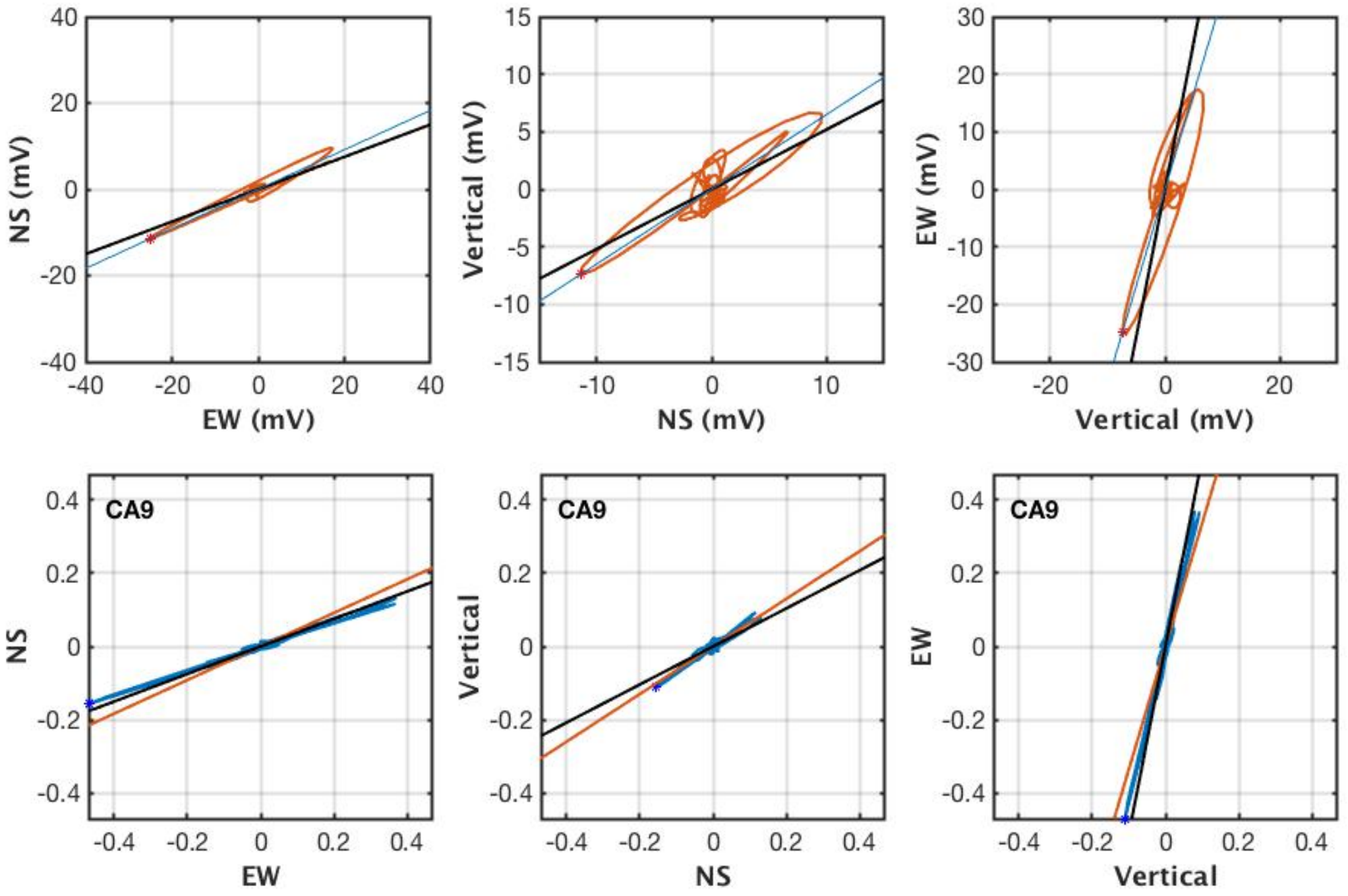

**Figure 3: Top:** the polarization pattern obtained for the Tripole event of Fig. 2. In orange, the polarization ellipses in the three directions. The black line indicates the geomagnetic prediction orientation for the event arrival direction, under the far-field emission hypothesis ($\vec{\mathbf{E}} = \vec{\mathbf{v}} \times \vec{\mathbf{B}}$). **Bottom:** in blue, the polarization pattern for the closest Compact Array antenna, once unfolded in the same frequency band ([24-82] MHz)). The black lines are the same as above, orange line stands for the polarization direction of the tripole.

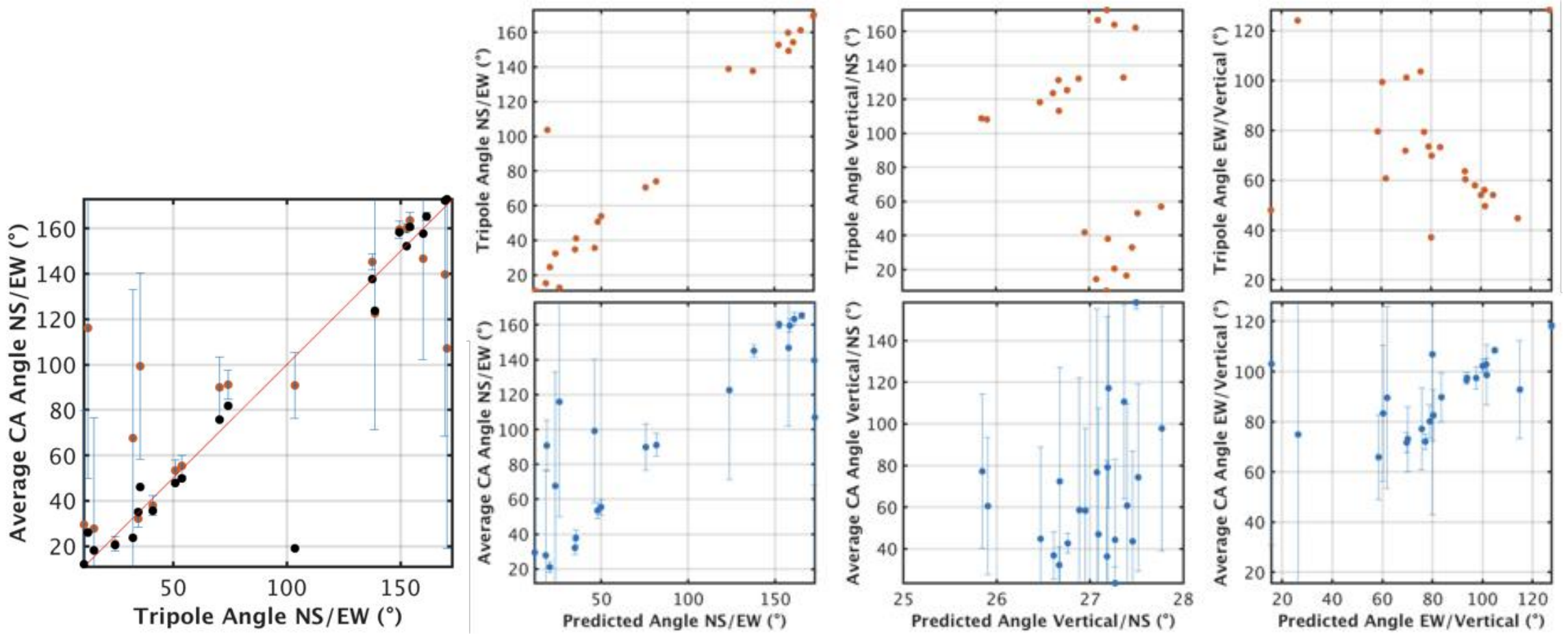

**Figure 4: Left:** the NS/EW polarization angles for the 22 event data set: average of the CA antennas angles vs the Tripole angle. The vertical error bars are the standard deviation of the angles over the 10 CA antennas. The black dots stand for the geomagnetic prediction. **Right:** Comparison to the predicted geomagnetic polarisation angles in the 3 directions for the 22 event data set. *Top:* Tripole vs prediction, *bottom:* Compact Array vs prediction. *See text for details.*

## 6. Conclusion and outlook

A three-fold polarization antenna has been installed in the middle of the particle detector of the CODALEMA experiment, and is surrounded by a small extent array of well known, externally triggered antennas. At the current stage of development, this so-called Tripole antenna already gives convincing results, being able to detect the EAS radio signal components in a complete, three axis orthogonal base. However, more investigations are required, notably on the Tripole antenna simulations, in order to properly unfold the voltage signals into electric field components. A larger study over thousands of events is in preparation, aiming at disentangling the various EAS electric field contributions and finally conclude about the validity - or not - of the widely used far field hypothesis in radio detection of cosmic ray air shower events.

*We thank the Région Pays de la Loire for its financial support to the Astroparticle Group of Subatech and in particular for its contribution to the EXTASIS experiment.*

# Near-field radio emission induced by extensive air showers

**Daniel García-Fernández*[a], Didier Charrier[a], Richard Dallier[a,c], Antony Escudie[a], Alain Lecacheux[b], Lilian Martin[a,c], Benoît Revenu[a,c], Matias Tueros[d]**
[a]*Subatech, Institut Mines-Télécom Atlantique, CNRS, Université de Nantes, Nantes, France*
[b]*CNRS-Observatoire de Paris, Meudon, France*
[c]*Unité Scientifique de Nançay, Observatoire de Paris, CNRS, PSL, UO/OSUC, Nançay, France*
[d]*Instituto de Física La Plata, CONICET, CCT-La Plata, La Plata, Argentina*
*E-mail:* daniel.garcia-fernandez@subatech.in2p3.fr

The measurement of the electric field created by cosmic-ray induced air showers is nowadays a well-established technique. Due to technical limitations, the low-end part of the frequency spectrum of the field has not been thoroughly exploited or understood, even though some experiments have indicated a large electric field at low frequencies. In this work, we present a new equation for the electric field of a particle track in time-domain valid for all frequencies, and therefore suited for the treatment of the near-field regime. After that, we calculate the low frequency radio signal of extensive air showers in the far-field and near-field regimes using a new version of the SELFAS Monte Carlo code that includes our equation.

*Speaker.

## 1. Introduction

The low flux of cosmic rays with energies above $\sim 10^{15}$ eV impedes their direct measurement, so we are forced to detect them indirectly via the extensive air showers (EAS) created upon the interaction of a primary cosmic ray with the atmosphere. The particles composing the EAS can give us some information on the properties of the primary cosmic ray. There are three main detection techniques – the detection of the particles arriving at ground level with Cherenkov detectors or scintillators, the detection of the fluorescence light emitted by the atmosphere after the passage of the charged particles of the EAS, and the detection using radio antennas.

Radio detection is nowadays a well-established technique [1] that allows us to infer the properties of the primary cosmic ray by measuring the electric field created by the charged particles in the EAS. The electric field is created, at first order, by the current induced by the deflection of the secondary particles of the EAS in the geomagnetic field, a mechanism known as the geomagnetic effect [2]. A subdominant mechanism, called the Askaryan effect, is also present, and it consists in the creation of a radiation electric field by the excess of negative particles in the shower [3, 4].

While every currently operating experiment (for instance, CODALEMA [6], AERA [7] or Tunka-Rex [8]) measures the electric field above 20 MHz, several experiments have detected an inequivocal low-frequency electric signal (below 20 MHz) coming from EAS, for instance [9, 10], although their properties have not been properly explained with a known mechanism. A review of many low-frequency experiments can be found in [5], and they all agree in the existence of a low-frequency counterpart with an amplitude higher than the usual emission above 20 MHz. Moreover, the coherent deceleration of the shower front, that is, the sudden stop of the shower particles when they arrive to the ground, has been proposed in [5], but also in [9], as the possible source of the low-frequency signal. We will call this the *sudden death* mechanism.

In order to measure the electric field of an EAS at low frequencies, the EXTASIS experiment has been set up at the Nançay radio observatory [11], with an effective frequency band spanning from 1.7 to 3.7 MHz. One of the main theoretical problems that arises is the accurate calculation of the electric field. Standard codes for the calculation of the electric field, such as SELFAS, ZHAireS or CoREAS, calculate the electric field of the shower in far-field, that is, when the wavenumber of the emission $k$ and the distance from emitter to observer $R$ verify $kR = 2\pi R/\lambda \gg 1$. If we expect an important emission from the lower part of the shower (as in the sudden death mechanism), and if the antennas are situated at a distance smaller than 100 m from the shower core, we have that at 1 MHz: $kR = 2\pi 100/300 \sim 2$, indicating that near-field effects could be important for antennas near the shower core.

In this work, we will present a formula for the field of a particle track that does not require the far-field approximation, compare it with a previous well-known formula, and then introduce it in the SELFAS Monte Carlo code to apply it to calculate the electric field of an EAS at low frequencies.

## 2. Electric field for a particle track

We define a particle track as the trajectory of a particle that begins at rest, gets suddenly accelerated up to a velocity $\mathbf{v}$ at $t = t_1$, travels in a straight line at constant speed and then is suddenly stopped at $t = t_2$. The track is the foundation of practically all the Monte Carlo codes that calculate the electric field of EAS in a microscopic way by discretising a particle trajectory in straight segments, an then calculate the total field as the superposition of the fields created by the tracks. Assuming an homogeneous, non-magnetic medium, we can write the electric field as [12]

$$\mathbf{E}(\mathbf{x},t) = \frac{1}{4\pi\varepsilon}\int d^3x' \left\{ \left[\frac{\rho(\mathbf{x}',t_{\rm ret})\mathbf{r}}{R^2(1-n\beta\cdot\mathbf{r})}\right]_{\rm ret} + \frac{n}{c}\frac{\partial}{\partial t}\left[\frac{\rho(\mathbf{x}',t_{\rm ret})\mathbf{r}}{R(1-n\beta\cdot\mathbf{r})}\right]_{\rm ret} - \frac{n^2}{c^2}\frac{\partial}{\partial t}\left[\frac{\mathbf{J}(\mathbf{x}',t_{\rm ret})}{R(1-n\beta\cdot\mathbf{r})}\right]_{\rm ret}\right\}, \tag{2.1}$$

where $n = \sqrt{\varepsilon}$ is the refractive index of the medium, $\mathbf{r}$ is the normalised vector of the line of sight, $R$ the distance between observer and emitter, $c$ is the speed of light in the vacuum, $\mathbf{v}$ is the velocity and $\beta = \mathbf{v}/c$ and the calculation is performed in retarded (or emission) time. No far-field frequency approximation has been assumed whatsoever, which means that Eq. (2.1) is suitable for obtaining the electric field at low frequencies and short distances as well, as long as the trajectory is accurately represented. The only caveat is that we must choose a charge density that verifies charge conservation, that is, the total charge in space must be conserved, otherwise the field will not be a solution of Maxwell's equations and the field will not be correct, in general. We can write such a charge density for a single track as

$$\rho(\mathbf{x},t) = q\delta^{(3)}(\mathbf{x}-\mathbf{x}_1)\Theta(t_1-t) + q\delta^{(3)}(\mathbf{x}-\mathbf{x}_1-\mathbf{v}(t-t_1))\Theta(t-t_1)\Theta(t_2-t_1) + q\delta^{(3)}(\mathbf{x}-\mathbf{x}_2)\Theta(t-t_2), \tag{2.2}$$

with $q$ the charge of the particle. The current is readily written as

$$\mathbf{J}(\mathbf{x},t) = q\delta^{(3)}\mathbf{v}\Theta(t-t_1)\Theta(t_2-t_1). \tag{2.3}$$

Contrary to what one could expect, the addition of the first and third terms in Eq. (2.2) does create a radiation electric field. Physically, this is due to the fact that an acceleration field is not the same as a charge creation field, the latter being physically impossible without an opposite charge being created at the same time. Mathematically, we have a field of the form

$$\mathbf{E}_{\rm cons,1}(\mathbf{x},t) = \frac{q}{4\pi\varepsilon}\frac{\mathbf{r}_1}{R_1}\frac{\partial}{\partial t}\Theta(t_{\rm ret}-t_1) \tag{2.4}$$

and an analogous term for the deceleration. The field in Eq. (2.4) represents an impulse field. We will give more details of this calculation in an upcoming paper.

### 2.1 Calculation of field and comparison with far-field (ZHS) formula in time domain

We will now consider an electron track and calculate numerically the electric field with the help of Eq. (2.1) using the densities in Eqs. (2.2) and (2.3), but adding a positive static charge at position $\mathbf{x}_1$ so that the total charge in space is zero, and the static electric field at large distances is zero (this does not affect the generality of our result). The numerical step in time is taken to be $\Delta t = 0.1$ ns and the numerical derivative is a two-point derivative using the values of the two adjacent points and calculating their difference.

Although our main interest is to calculate the field from EAS, formulas involving refractive indexes are better studied within dense media. Because of this, let us take a 1.2 m long electron track along the $z$ axis at $v \sim c$ in an homogeneous, non-magnetic medium with refractive index $n = 1.78$ (akin to that of deep Antarctic ice), and let us place an observer at a distance of 10 m from the centre of the track and at an angle of $\theta_C + 10°$, with $\theta_C = \mathrm{acos}\frac{1}{n\beta}$ the Cherenkov angle. We show in Fig. 1 (left) that the pulsed fields for the acceleration and deceleration of the track in this configuration using Eq. (2.1) are the same as those obtained with the well-known ZHS time formula [13] (radiation field calculated in the Coulomb gauge). Not taking into account the charge conservation results in an unphysical field (green line) that differs from the Coulomb gauge calculation. Placing an observer at 100 m yields a similar result (see Fig. 1, right), this time within the far-field zone, where no field is seen between the two acceleration and deceleration pulses. Again, not taking into account the charge conservation results in an incorrect radiation field.

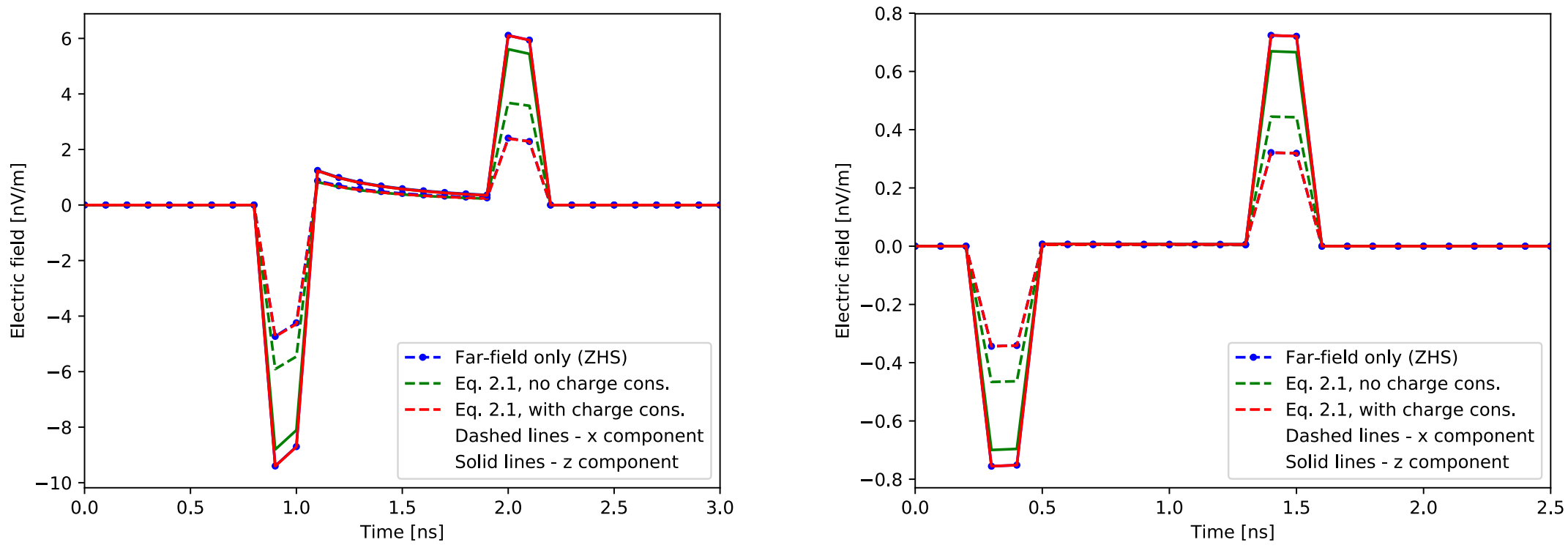

**Figure 1:** Electric field (in different approaches) created by a 1.2 m long electron track travelling along the $z$ axis, as a function of time. Observer is placed at $\theta_C + 10°$, in a medium with $n = 1.78$. Left: $R = 10$ m. Right: $R = 100$ m. Times have been arbitrarily offset. See text for details.

## 2.2 Calculation of field and comparison with an exact formula in frequency domain

Eq. (2.1) can be transformed to frequency domain. This transformation, besides allowing us to study our instrumental frequency range, it helps us to overcome the numerical divergences at the Cherenkov angle, since in frequency domain we can use the retarded time alone and so the divergent denominators in Eq. (2.1) are not present. The field for a track in the frequency domain can be written as

$$\mathbf{E}(\mathbf{x},\omega) = \frac{q}{4\pi\varepsilon}\left\{\int_{t_1}^{t_2} \mathrm{d}t\, e^{i\omega t}\frac{e^{ikR}}{R}\mathbf{r}\left[\frac{1}{R} - \frac{i\omega n}{c}\right] + \sum_{j=1,2}\left[(-1)^j e^{i\omega t_j} e^{ikR_j}\mathbf{r}_j\left(\frac{n}{cR_j} + \frac{i}{\omega R_j^2}\right)\right]\right\} + \frac{i\omega\mu_0 q}{4\pi}\int_{t_1}^{t_2} \mathrm{d}t\, e^{i\omega t}\frac{e^{ikR}}{R}\mathbf{v} \tag{2.5}$$

Eq. (2.5) looks similar to an equation for the exact field of a track presented in [14], which has been proven to be equivalent to the usual radiation field in the far-field regime and to reproduce the Cherenkov radiation for an infinite track. Eq. (2.5) and the equation in [14] can be derived from Maxwell's equations with the same charge and current densities, and are therefore equivalent. We show an example of this equivalence in Fig. 2 (left). Since Eq. (2.5) is equivalent to an exact formula in frequency domain, its time domain counterpart Eq. (2.1) is then a correct implementation of the exact formula of a track in time domain.

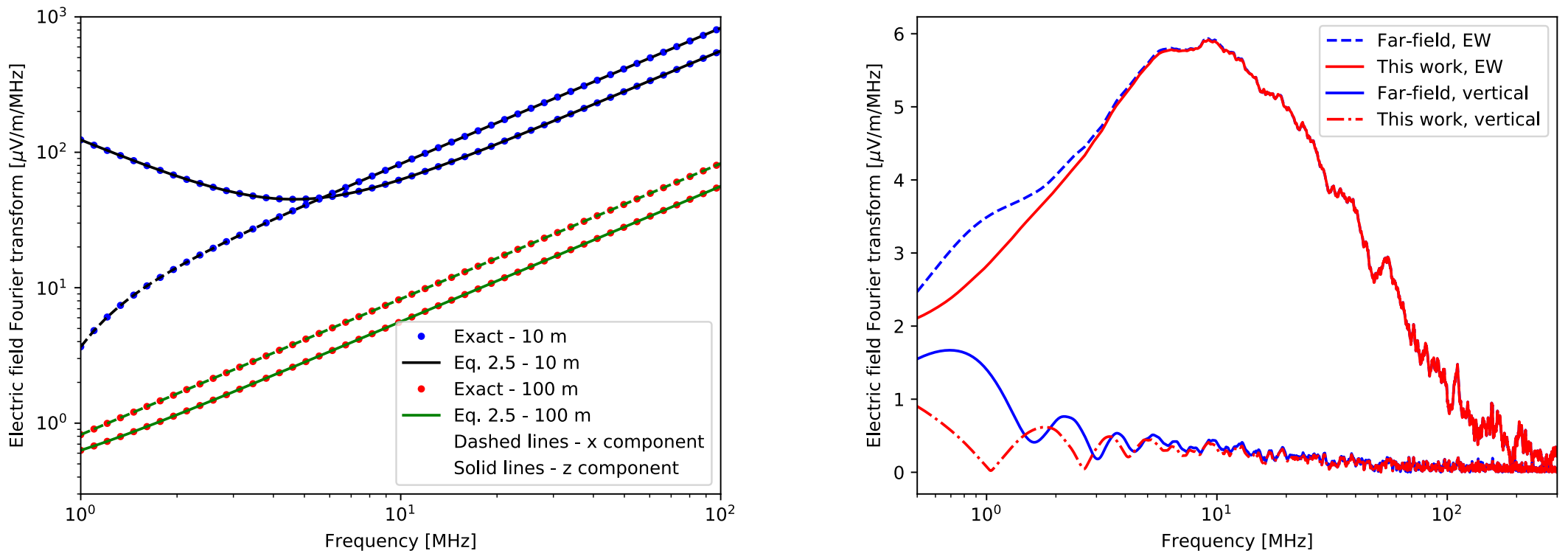

**Figure 2:** Left: Fourier transform of the electric field as a function of frequency for the same track as in Fig. 1. The observer is placed at the Cherenkov angle ($n = 1.78$). The field has been calculated with Eq. (2.5) and the exact calculation in [14]. Right: Fourier transform amplitude of the electric field created by a 1 EeV proton-induced vertical EAS. The antenna is placed 200 m East of the shower core. See text for details.

## 3. Near-field emission of EAS using the code SELFAS

### 3.1 Implementation in the SELFAS Monte Carlo code

SELFAS [12] is a code for the calculation of the electric field of an EAS that operates by sampling the particles of the shower, simulating their propagation in the atmosphere for a given depth, and then calculating the field created by the trajectory modelled as a collection of particle tracks. It includes all the state-of-the-art mechanisms for particle propagation and field emission, and it has been recently upgraded with the most realistic available description of the atmosphere using the Global Data Assimilation System (GDAS) [15, 16].

In order to embed Eq. (2.1) into SELFAS, we made the following assumptions:

- The static Coulomb field created after the particles have been stopped is not detectable, since the media (atmosphere and soil) will tend to eliminate all local charge excess and regain electrostatic equilibrium. We will not include static fields, as a consequence.
- When a track reaches ground, we can calculate its field as the sum of the field of one track that stops at the boundary and another one that immediatly accelerates at nearly the same place. This approach can explain transition radiation [17]. In this case there are three contributions to the electric field:
  - The direct contribution (as if there was no boundary) that we are able to calculate at all frequencies with Eq. (2.1).
  - The reflected (far-field) and surface wave contribution (can be near or far-field) to the electric field, that we will not calculate. The effect of reflection and far-field surface waves in a detector can be taken into account knowing the emission pattern of our antenna with the ground present, by the theorem of electromagnetic reciprocity, without the need of calculating the reflected EAS field at the antenna.

– The underground transmitted component. Given the large attenuation of radio waves inside soil, we will ignore this contribution.

In other words, we will calculate the direct, near-field regime electric field of an EAS.

### 3.2 Results for the simulated showers

We have simulated a vertical proton-induced 1 EeV shower at the Nançay site (altitude 180 m) and placed antennas east of the shower core. We show in Fig. 2 (right), the comparison between the far-field spectrum and the spectrum calculated with Eq. (2.1) (same shower), for the East-West and vertical components of an antenna located 200 m East of the shower core. It is clearly seen that for the East-West component, both calculations agree for frequencies above $\sim$ 10 MHz, which was expected ($kR \gg 1$). However, below 10 MHz both approaches differ for both polarisations and we must use Eq. (2.1) if we want to accurately depict the electric field. In this case, the near-field effects tend to decrease the amplitude of the electric field.

We can find in Fig. 3 (left) the filtered ($<$ 5 MHz, sixth order low-pass Butterworth) vertical pulses for two antennas located East of the shower core. We find that the second pulse seen in Fig. 3, related to the sudden death of the shower [5] is always present regardless of whether we take into account or not the near-field effects, albeit the pulse seems to be a bit smaller in the near-field case (5% in this example). Since it is related to the deceleration of the shower front, the arrival time of this pulse to the detector should be proportional to the distance $d$ between antenna and shower core ($t \sim d/c$, if we take the arrival of the shower at ground level to be $t = 0$), something evidenced in Fig. 3 where the antenna at 500 m sees this pulse after the antenna at 200 m (with the correct relative delay of $(500-200)/c = 1000$ ns. In contrast, what is drastically changed at low frequencies is the vertical component of the principal pulse (geomagnetic and charge excess). Even the filtered EW polarisation exhibits non-negligeable differences, see Fig. 3 (right).

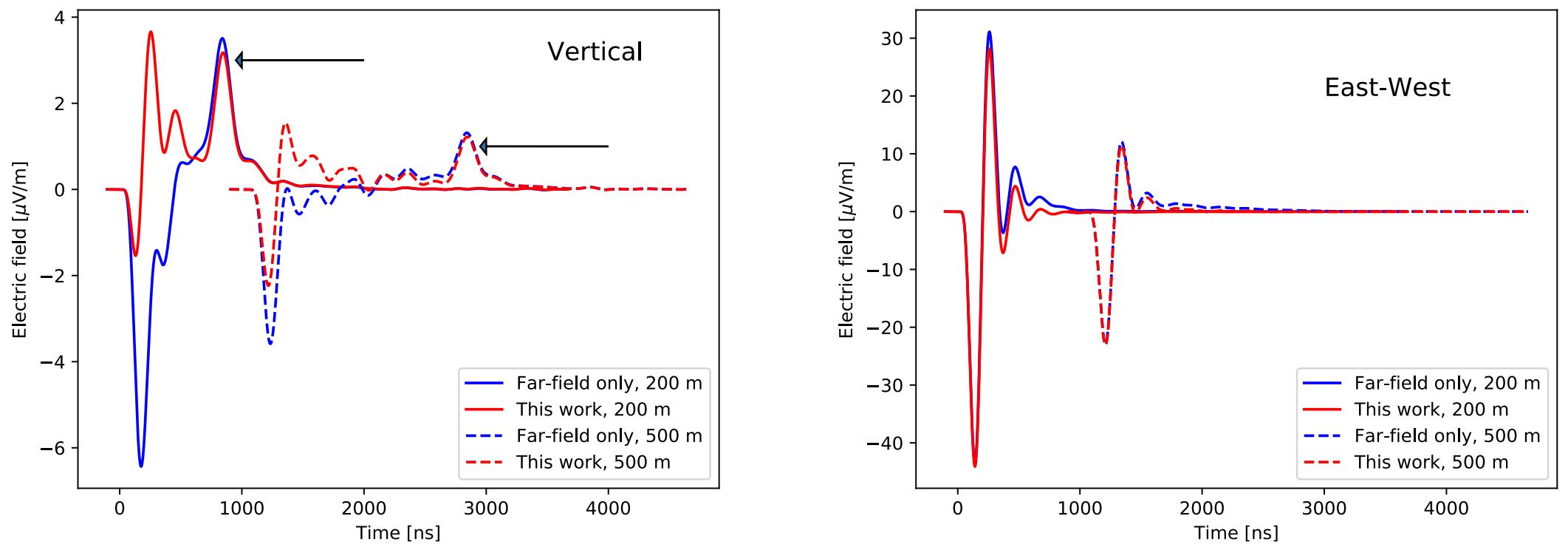

**Figure 3:** Electric field created by a 1 EeV proton-induced vertical shower for antennas East of the shower core, as a function of time. Times have been arbitrarily offset. The traces have been numerically transformed to frequency domain, filtered using a low-pass sixth order Butterworth filter with a cut frequency of 5 MHz and then transformed back to time domain. The sudden death field (indicated by the arrows) is visible after the principal pulse in each trace. See text for details. Left: Vertical component. Right: East-West component.

In order to know what order of magnitude should be expected from the so far undetected sudden death pulse (SDP), we calculate its amplitude for a simulated shower. We show in Fig. 4 the filtered amplitudes for proton-induced vertical showers at several distances from the core, as a function of the energy of the primary proton. At 300 m and 1 EeV, the SDP is $\sim 1$ $\mu$V/m. These results are just $\sim$ 10-20% lower than those obtained with the standard far-field formula in [5]. The SDP amplitude is proportional to the primary energy.

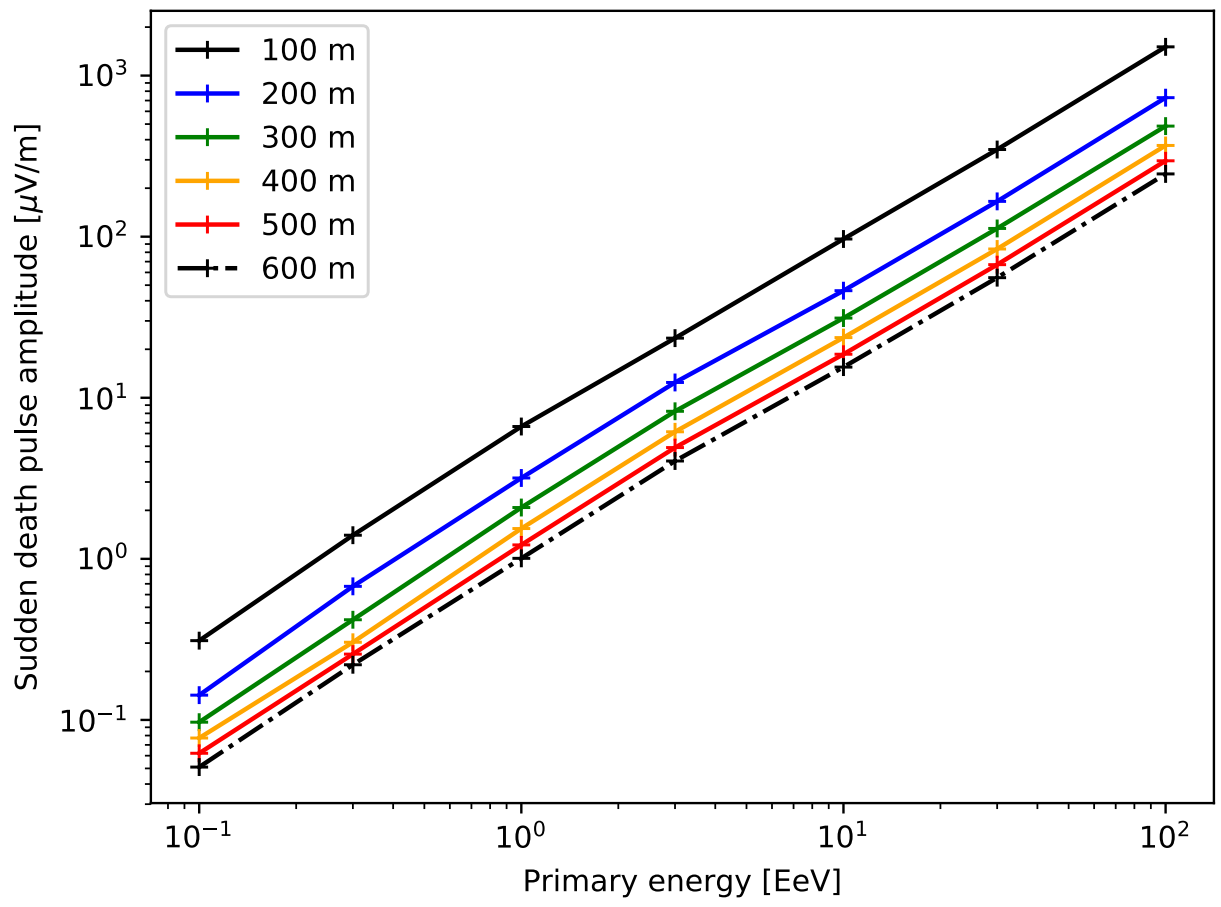

**Figure 4:** SDP amplitude (filtered as in Fig. 3) for the vertical polarisation, created by vertical proton showers, as a function of the primary energy for different core distances. The antennas are located East of the shower core at the distances specified in the legend.

The sudden death pulse is more prominent at sites with higher ground altitude, where the ground is closer to the shower maximum. This is precisely what we show in Fig. 5 for a 5 EeV, 30° shower at 180 (left) and 2650 (centre) m of altitude (as in GRAND [18]), where the SDP is larger for the latter case. We also show in Fig. 5 (centre) that, for an antenna at 200 m, the vertical component filtered in the $[5,50]$ MHz band presents similar results for the far-field approximation and for Eq. (2.1). It is also patent that the SDP is less evident at high frequency, therefore low frequency is preferable in order to detect it.

In Fig. 5 (right) from 20 MHz upwards, both formulas give very similar amplitudes for the electric field. However, at low frequencies, this is no longer the case. Both formulas predict a higher amplitude between 1 and 10 MHz than in the rest of the spectrum, something that is in agreement with the low-frequency experiments carried out so far [5, 9, 10].

## 4. Outlook and conclusions

We have shown that Eq. (2.1), with the appropriate charge and current density included, reproduces the electric field created by a particle track in the near-field regime. We have shown that its Fourier transform (Eq. (2.5)) is equivalent to an exact calculation of the field of a track [14]. Eq. (2.1) is useful if we want to calculate the electric field of a track at any frequency, in particular below 5 MHz, corresponding to the near-field regime, as it is the case for the EXTASIS experiment [11].

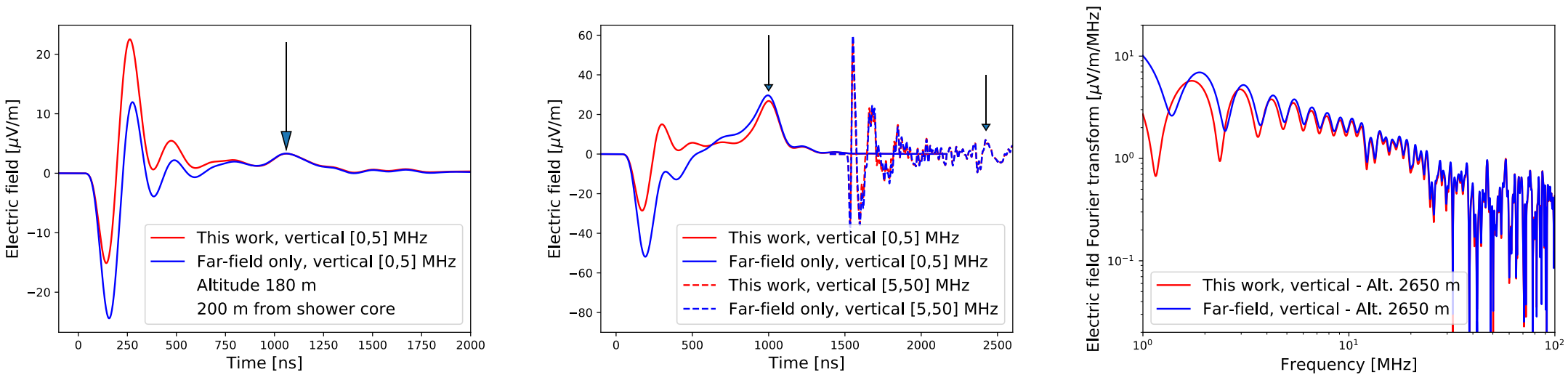

**Figure 5:** Left and centre: electric field (vertical polarisation) created by a 5 EeV proton-induced 30° shower for an antenna at 200 m East of the shower core. The SDP is visible after the principal pulse (indicated by the arrows) at $\sim$ 1000 ns after the first, negative pulse. Left: 180 m of altitude. Centre: 2650 m of altitude. A trace filtered between 5 and 50 MHz is plotted, with time arbitrarily offset. Right: Fourier transform amplitude of the electric field (vertical).

We have implemented Eq. (2.1) in the SELFAS code and used it to calculate the electric field in the near-field regime created by an EAS. Differences with the far-field case are quite important for frequencies below 5 MHz and more for the vertical component than the EW component. The SDP introduced in [5] is still present when calculating the near-field effects.

Nevertheless, since most of the difference between Eq. (2.1) and the far-field approach come from the existence of a longitudinally polarised component (**r** in Eq. (2.1)) and since antenna responses are known usually for far-field only (without longitudinal component), further studies are needed in order to translate the electric field at the antenna into a measured voltage. A careful analysis of the antenna response, by means of simulations or measurements is needed for understanding our final data. A study of the influence of the ground on the electric field of an EAS is underway.

## 5. Acknowledgements

We thank Région Pays de la Loire for its financial support of the Groupe Astro of Subatech and in particular for their contribution to the EXTASIS experiment.

# Low frequency observation of cosmic ray air shower radio emission by CODALEMA/EXTASIS

**Antony ESCUDIE**[*1], **D. Charrier**[1,3], **R. Dallier**[1,3], **D. García-Fernández**[1], **A. Lecacheux**[2], **L. Martin**[1,3] **and B. Revenu**[1,3]

[1] *SUBATECH, Institut Mines-Telecom Atlantique – CNRS/IN2P3 – Universite de Nantes, Nantes, France*
[2] *CNRS-Observatoire de Paris, Meudon, France*
[3] *Unite Scientifique de Nancay, Observatoire de Paris, CNRS, PSL, UO/OSUC, Nancay, France*
*E-mail:* antony.escudie@subatech.in2p3.fr

Over the years, significant efforts have been devoted to the understanding of the radio emission of extensive air shower (EAS) in the range [20-200] MHz. Despite some studies led until the early nineties, the [1-10] MHz band has remained unused for 20 years. However, it has been measured by these pioneering experiments and suggested by theoretical calculations that EAS emit a strong electric field in this band and that there is evidence of a large increase in the amplitude of the radio pulse at lower frequencies. The EXTASIS project, located within the radio astronomy observatory of Nançay and supported by the CODALEMA instrument, aims to study the [1-10] MHz band, and especially the so-called "Sudden Death" contribution, the expected radiation electric field created by the particles that are stopped upon arrival to the ground. We present our instrumental setup, the objectives of the EXTASIS project and our first results.

[*] Speaker.

## 1. Introduction

It is a well known fact that the coherent radio emission during the development of air shower has two main origins: geomagnetic and charge excess mechanisms [1, 2, 3]. The resulting emission appears as a fast electric field transient lasting few tens of ns, and this can be detected by large bandwidth (∼100 MHz) antennas and acquisition systems. Man-made broadcasting at low and medium frequencies constrains the common observation in the range $[30-80]$ MHz, but several detections at low frequencies have been realized in the 70's and up to the 90's, and a conclusion can be drawn from these observations: extensive air showers (EAS) undoubtedly emit a strong electric field at low frequencies. At that time, several experiments [4, 5, 6, 7] have shown that when frequency decreases, there is a clear evidence of a strong increase of the radio pulse amplitude. We have therefore chosen to reinvestigate low frequency measurements by implementing in the CODALEMA experiment low frequency antennas and using the SELFAS simulation code [8] to highlight a possible new emission mechanism. Figure 1 shows the simulation with SELFAS code of the electric field strength as a function of time, recorded by an antenna in vertical polarization located 300 m to the shower core (in $\mu V \cdot m^{-1}$) for a vertical shower initiated by a proton at 0.1 EeV. On the full-band trace (blue line), the negative peak at 0 ns is due to is due to the shower development in the air. The second, positive peak around 1 µs has been interpreted as the effect of the coherent deceleration of the shower front hitting the ground: the sudden death pulse (SDP) [9, 10]. The sudden death pulse arrives 1 µs after the normal pulse, which is consistent with the propagation time to antenna (300 m). By filtering in different frequency bands, we show that both signals are still detectable for frequencies below <10 MHz, suggesting the use of low frequency antennas: this is the goal of the EXTASIS (EXTinction of Air-Shower Induced Signal) project. In the following, we present the objectives of the EXTASIS project, our instrumental setup and our first results.

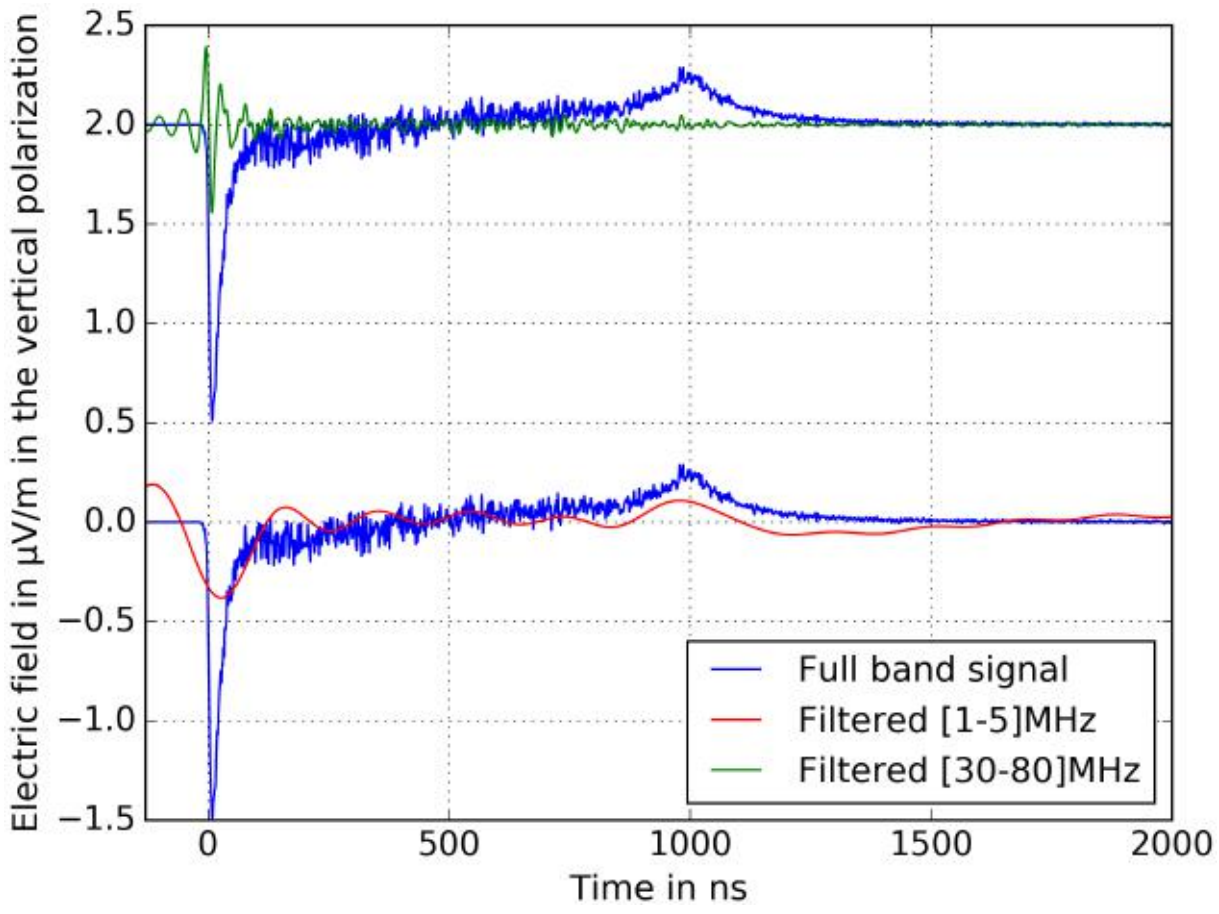

Figure 1: Electric field as a function of time obtained with a SELFAS simulation for a vertical shower at 0.1 EeV: the associated filtered response in different bands showing that the two pulses can be seen in the range $[1-5]$ MHz.

## 2. Instrumental setup

Currently, the EXTASIS project is based on low frequency antennas triggered by scintillators: two low frequency antennas directly triggered via cables, and five antennas for which the trigger is distributed over a network (see figure 2), and is supported by the CODALEMA experiment [11, 12].We know that the particle detector (scintillator) detects only cosmic ray events, because it is sensitive only to the particles arriving at ground. When at least 5 out of the 13 scintillators are fired in a given time window, a master trigger is built by a multiplicity card and sent to a dedicated datation station which dates by GPS the event, to the compact array and to the low frequency antennas. For the two cabled antennas, the timing is made by this dedicated GPS station, and data are acquired thanks to a digital oscilloscope (PicoScope). The devices are classical Butterfly antennas, a dual polarization (East-West and Vertical, since the SDP is expected to be also vertically polarized) active antenna using a bow-tie shaped rod as radiating element, with an adapted low noise amplifier to the band $[2-6]$ MHz. They are disposed at equal distance from the scintillators center on a 9 m high mast. As suggested by NEC simulations, the losses in ground are minimized and the signal to noise ratio in the range $[2-6]$ MHz is upper than 10 dB at 9 m height. The 2.10 m long dipoles composing the Butterfly antennas are considered as short when compared to the measured wavelengths. Each dipole feeds a low noise amplifier (LNA) whose input impedance is adapted to the dipole. The output signal of the LNA is transported by a cable to a radio frequency chain, followed by an analogical chain composed of filters. The signals of both polarizations feed the 2 inputs of a digital oscilloscope serving as a digitizer and driven by a dedicated acquisition software hosted by a distant PC. The oscilloscope sampling time bin is 2 ns and the traces contain 20,000 time bins, so the trace is recorded over a total time of 40 µs. Moreover, the two cabled low frequency antennas are supplemented by 5 network-connected low frequency antennas. These antennas are physically the same as the cabled antennas, and are also externally triggered by the scintillators. They are network-connected due to technical limitations, and the trigger is thus distributed over an Ethernet network. When the trigger is produced by the particle detectors, it is sent to the GPS datation station which duplicates the trigger signal and sends it to an emission chassis which distributes the trigger signal over Ethernet connexion to the five concerned low frequency antennas composed of a reception chassis. First, the 5 network-connected low frequency antennas had an East-West (EW) and a vertical polarization, but to provide the detection of the low frequency counterpart, we have decided to replace the vertical polarization by an EW high frequency polarization: thus, the actual EXTASIS apparatus is composed of one Butterfly antenna in EW polarization at 9 m high (EXTASIS Low Frequency (LF)) working in the range $[1-6]$ MHz and one Butterfly antenna in EW polarization at 1.5 m (EXTASIS High Frequency (HF)) working in the range $[20-200]$ MHz, and digitized by the same oscilloscope. The electronic modules of the reception chassis is located below the EXTASIS HF antenna. The two EXTASIS antennas are supplemented by one standalone antenna (SA, CODALEMA HF) working in the range $[20-200]$ MHz. The objective is to observe a pulse in the associated CODALEMA HF, observed the same pulse in the EXTASIS HF and use the timing of the pulse to find it in the EXTASIS LF, allowing us to affirm that the transients seen in LF are evidence of the low frequency counterpart of the radio signal in cosmic ray induced air showers.

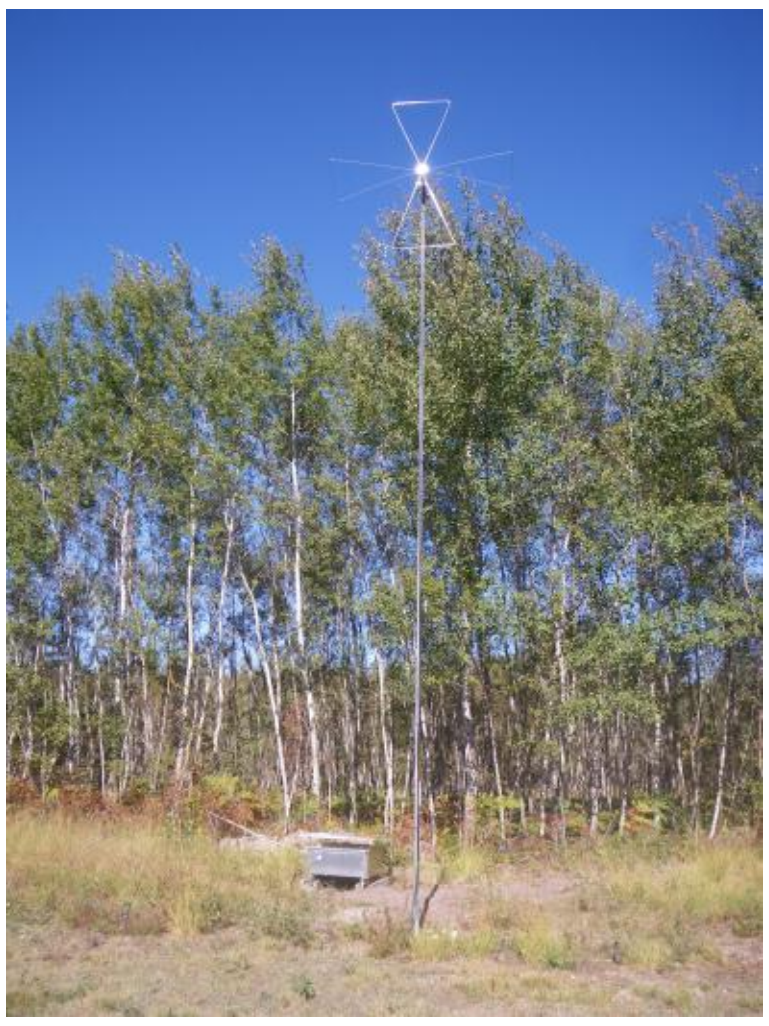

Figure 2: Photography of a low frequency antenna.

## 3. First results

### 3.1 Low frequency sky

Figure 3.(a) presents the atmospheric noise temperature in function of the frequency for the site of Nançay. In our case, the studied frequency range is dominated by the atmospheric noise due to the high brightness temperature of the atmosphere at these frequencies, which grows considerably during the night, but also by man-made noise. Figure 3.(b) shows a time-frequency diagram, which is a daytime power density spectrum for one low frequency antenna, and highlights the ambient noise variation during one day, and a day/night dependence can be seen, confirming that our low frequency antennas are sensitive to the day/night variation of atmospheric noise temperature as predicted and shown on figure 3.(a). The vertical black-dashed lines represent the sunrise and sunset, and delimit two regions: the "night region" where our instrument is not efficient, and the "day region", where our instrument is efficient, and where we expect to detect a low frequency signal due to a cosmic ray air shower. We can also see a clear man-made emitter at around 7.5 MHz. From

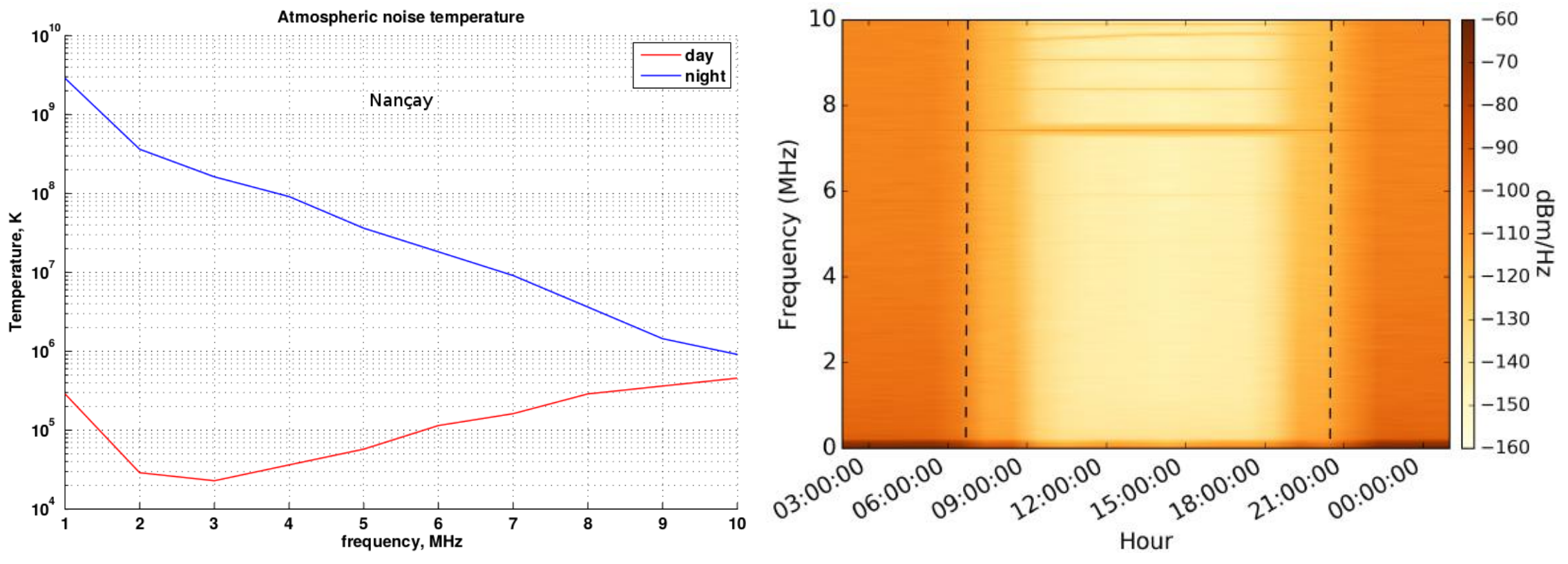

(a) (b)

Figure 3: Left: atmospheric noise temperature in function of the frequency for the site of Nançay. Right: time-frequency diagram of a low frequency antenna. See text for more details.

this result, we have decided to focus our study on daytime, and particularly in the $[1.7-3.7]$ MHz frequency domain, which is the cleanest band below 10 MHz.

### 3.2 Low frequency signal detections

First, the selection of low frequency events with a significant signal was done with an amplitude threshold. With this method, we have detected one low frequency event involving 2 LF antennas. After a comparative study between three detection methods (amplitude threshold, wavelet analysis and LPC), we have chosen to use the LPC (linear predictive coding) method [13, 14]. With this method, we are more sensitive to events with a small signal to noise ratio, as shown in figure 4. This event is seen in coincidence by CODALEMA and EXTASIS, and the arrival direction reconstructions are in good agreement: $\theta_{LF}$ =31.1°, $\Phi_{LF}$ =146.1°, $\theta_{SA}$ =40.6°, $\Phi_{SA}$ =145.2° and $\theta_{SC}$ =32.4°, $\Phi_{SC}$ =144.1°. One can see that the transients are hardly visible on the filtered traces (figure 4.(a) and (b)), but that they begin to appear after the LPC processing (figure 4.(c) and (d)).

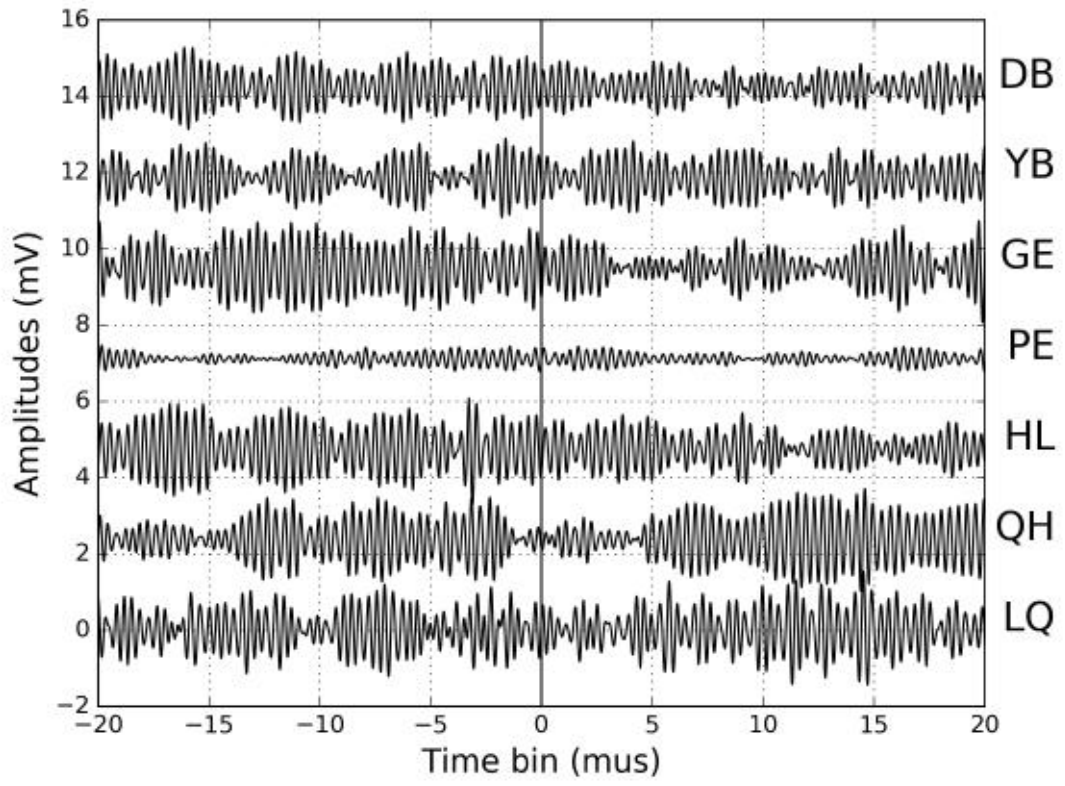

(a) Filtered $[1.7-3.7]$ MHz LF signals.

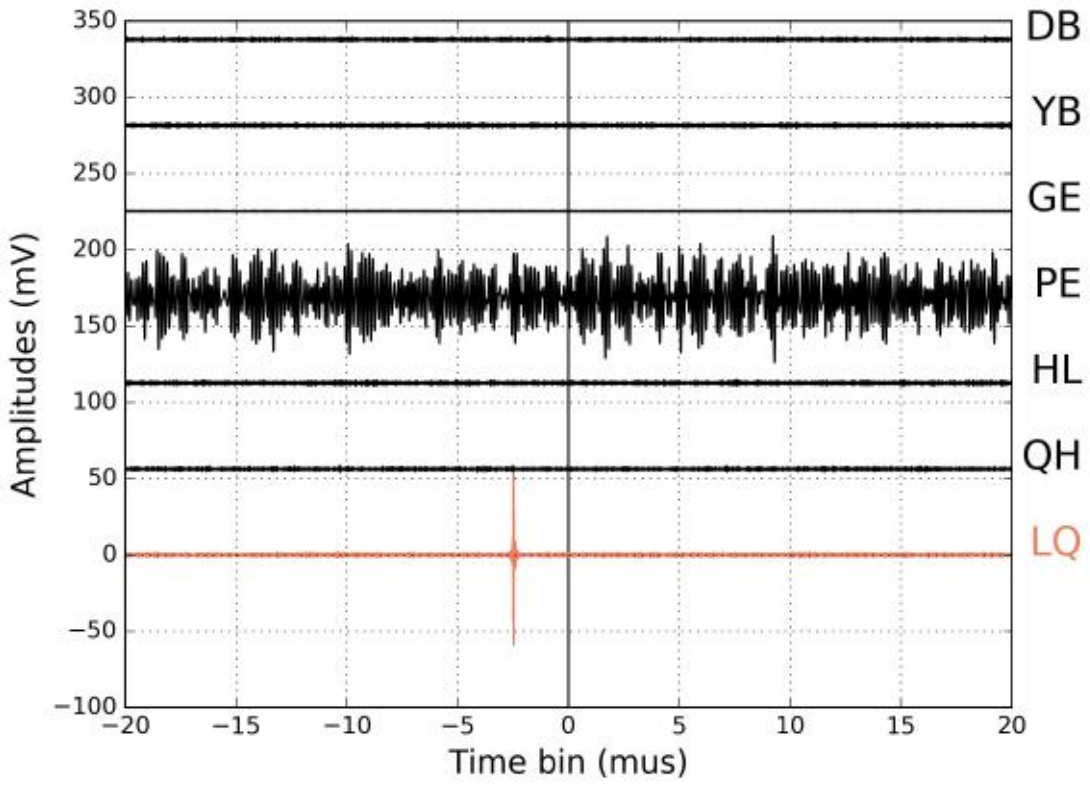

(b) Filtered $[30-80]$ MHz HF signals.

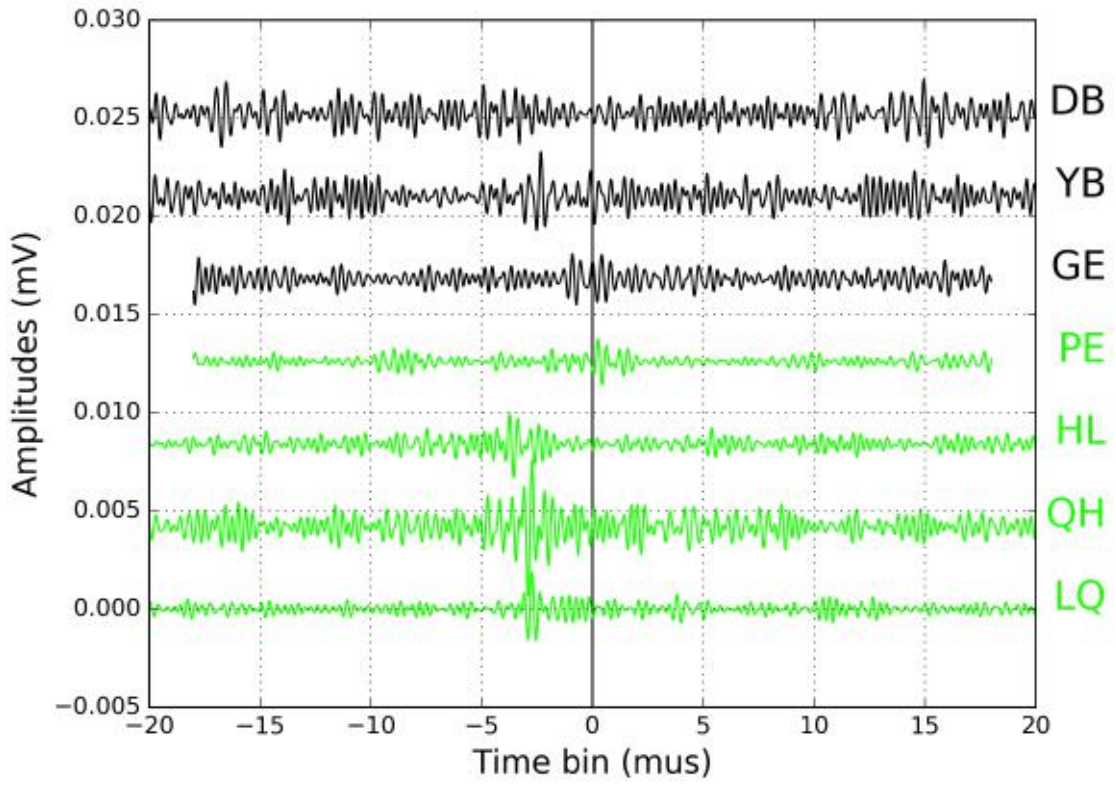

(c) Error of prediction on LF signals.

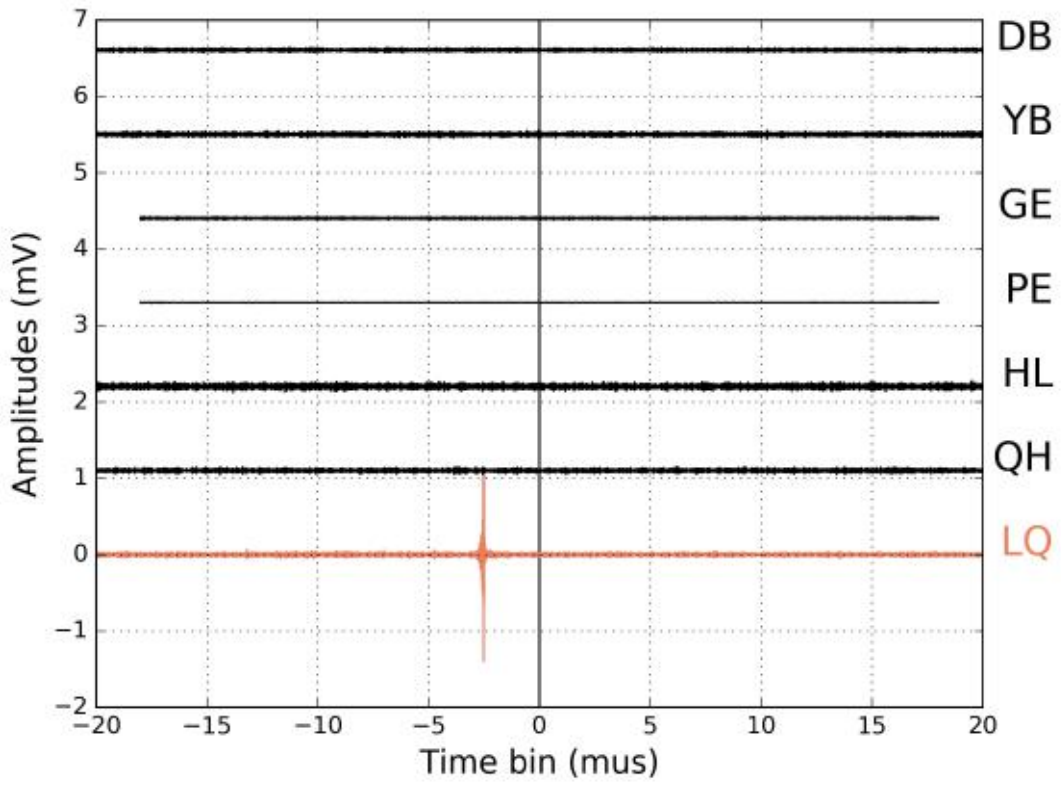

(d) Error of prediction on HF signals.

Figure 4: LF event seen with the EXTASIS instrument. From the top, left: Filtered LF signals, filtered HF signals, error of prediction on LF signals and on HF signals.

With SELFAS, we simulated this event, and made the radio core reconstruction, see [15] for more explanation on the method. The best core position is in $x=$259 m and $y=-$809 m, and represented by a magenta square in figure 5. Figure 5 shows the map of a part of facilities in Nançay. Grey dots represent the standalone antennas, and the stars the low frequency antennas. The involved standalone antennas in the coincidence are represented by coloured circles indicating the order in which the signal is seen by the antennas, and the involved low frequency antennas are represented by green crosses. The low frequency antenna named LQ (see figures 4.(c) and (d)) presents a signal in both polarizations at the same position in the trace, particularly in the HF polarization, while the 3 other ones have only a signal in the LF polarization. A possible explanation is that the LQ antenna is close to the estimated shower core, as one can see on figure 5, and that the 3 other ones are very far away from it, hinting that the detection range at low frequency is higher than at high frequency. Making a precise simulation of the event thanks to the information from the radio reconstruction [15], it is shown in figure 6 which presents the simulated lateral distribution of the electric field depending on the frequency band that at low frequency (figure 6 right), the LF antennas very far away from the shower core begin to be in the detection zone, explaining why the PE antenna situated at 850 m presents a signal in the low frequency band, and not in the high frequency band. Moreover, figure 7 presents the simulated power spectrum density as a function of frequency for the four involved LF antennas. We see that the PSD quickly drop in the classical band with the shower axis distances, while they decrease much more slowly in the LF band. This can explain also why LQ presents a signal in both polarizations while the three others present a signal only in the LF polarization. This result indicates that the detection range is larger at low frequency than at high frequency.

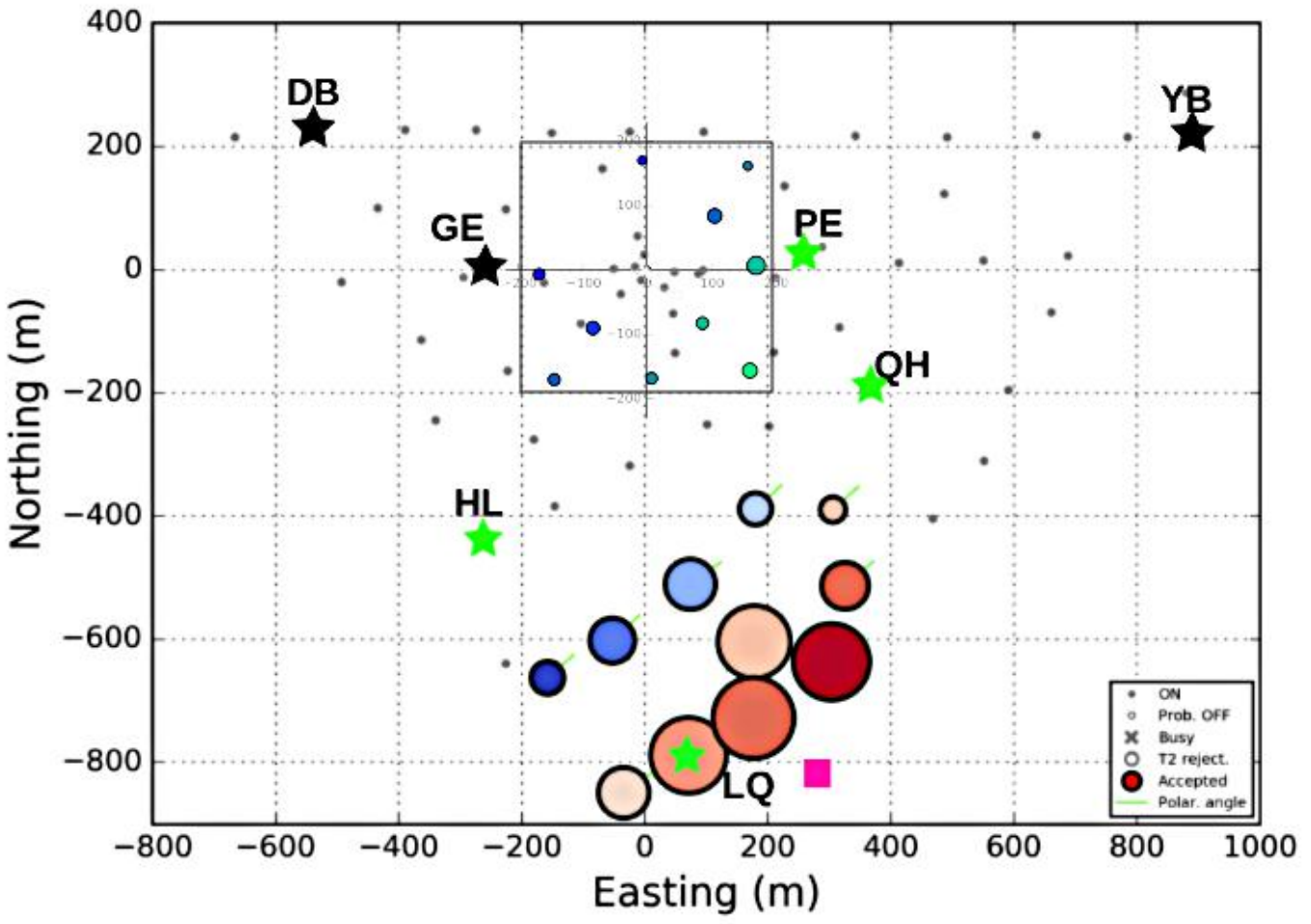

Figure 5: Map of a part of facilities in Nançay. Grey dots represent the standalone antennas, the square area with 13 circles represents the scintillators and the stars the low frequency antennas. The involved standalone antennas are represented by coloured circles indicating the order in which the signal is seen by the antennas, and the involved low frequency antennas are represented by green crosses. The estimated shower core is represented by a red square.

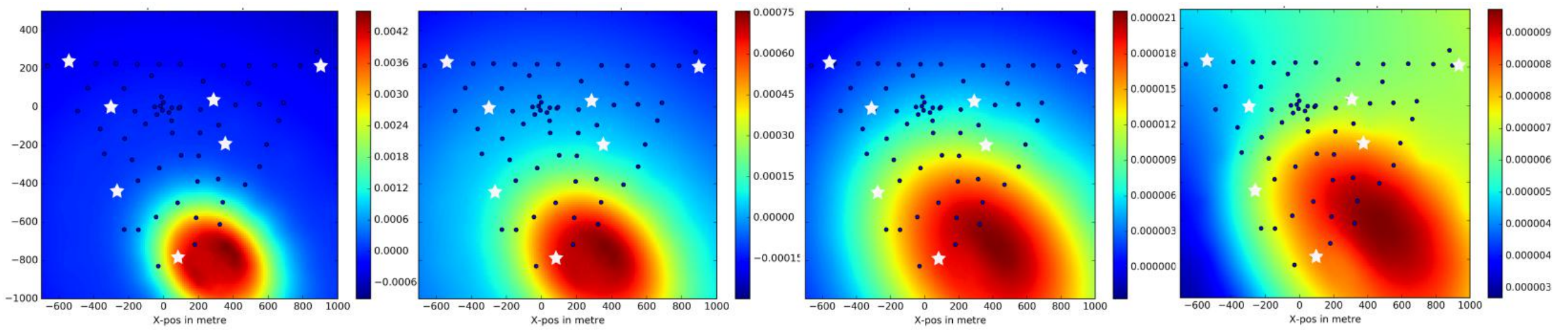

Figure 6: Lateral distribution of the electric field depending on the frequency range. From left to right: $[30-80]$ MHz, $[10-30]$ MHz, $[5-10]$ MHz and $[1-5]$ MHz. White stars represent the low frequency antennas. At low frequency, the LF antennas very far away from the shower core begin to be in the detection zone and thus sensitive.

This finding is confirmed by another event figure 8, seen by four LF antennas. Three of them have a signal in both polarizations, while the farthest one from the shower core presents a signal only in the LF polarization. This result endorses our argument that detection range is better at low frequency than at high frequency. The arrival direction reconstructions are again in good agreement: $\theta_{LF}$ =66.8°, $\Phi_{LF}$ =155.0°, $\theta_{SA}$ =60.1°, $\Phi_{SA}$ =153.8° and $\theta_{SC}$ =61.6°, $\Phi_{SC}$ =154.6°.

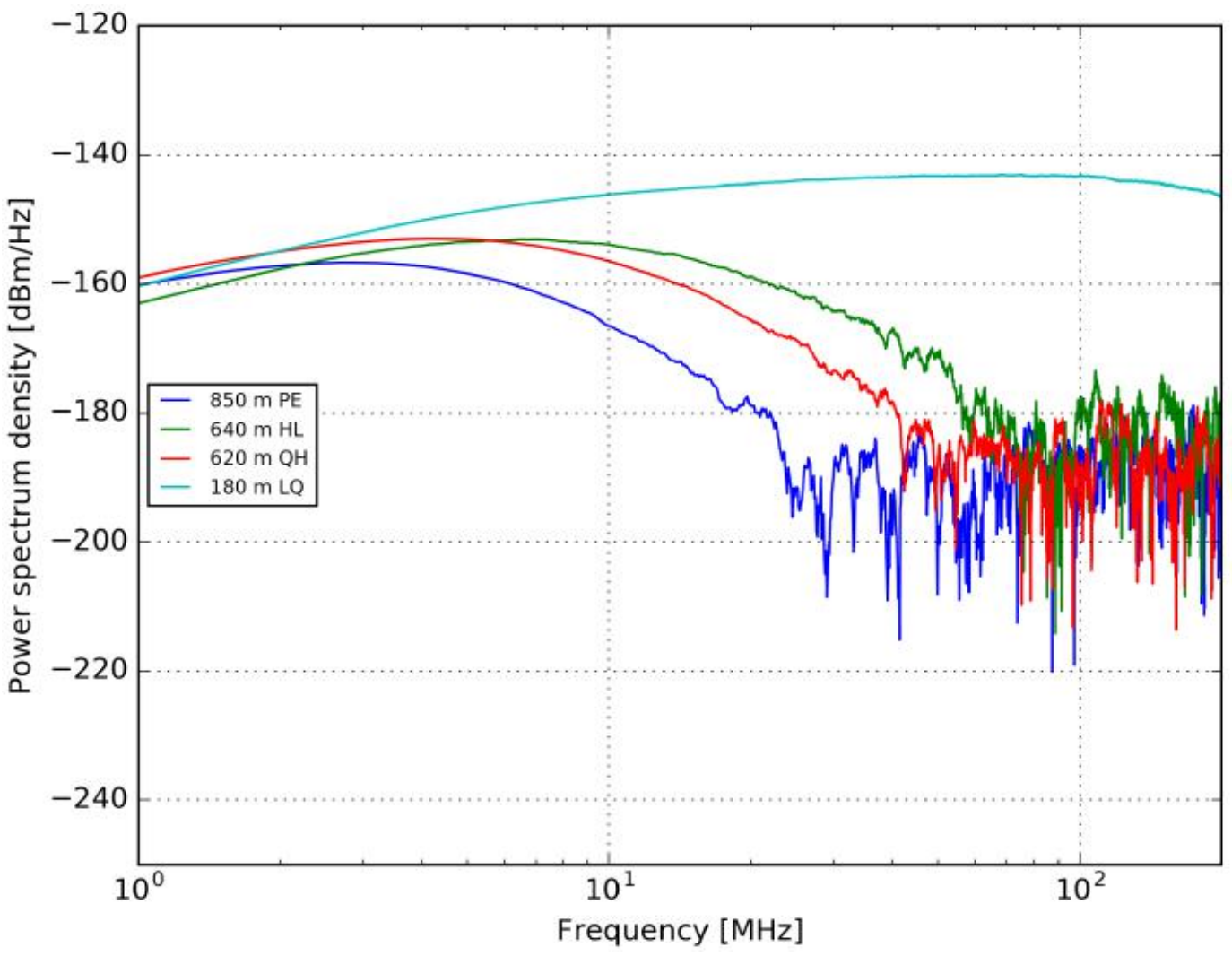

Figure 7: Simulated power spectrum density as a function of frequency for the four involved LF antennas.

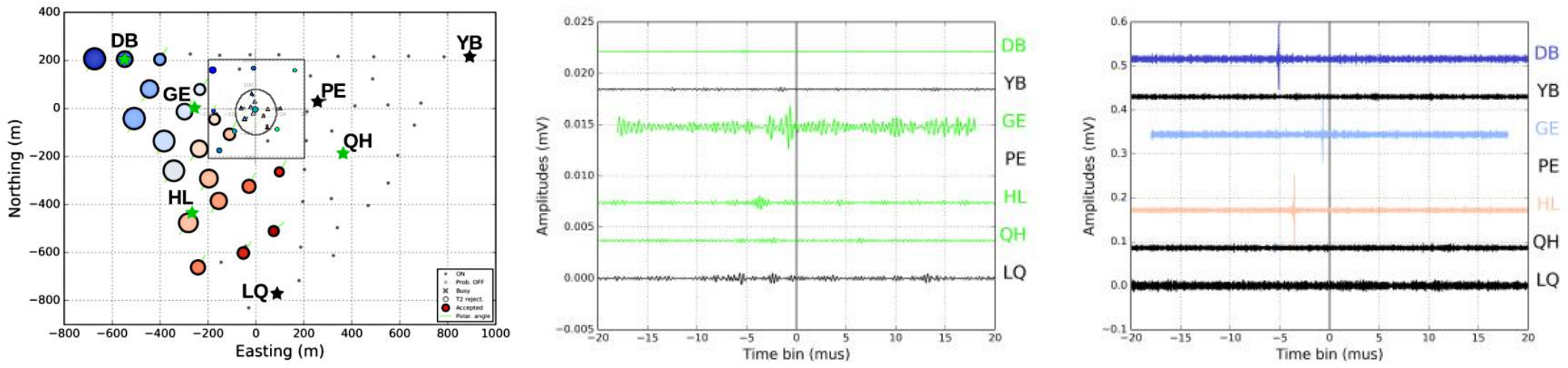

Figure 8: LF event seen with the EXTASIS instrument. From left: map of event (green crosses represent the involved low frequency antennas), the square area with 13 circles represents the scintillators and the triangles in the circular area represent the compact array, error of prediction on LF signals and error of prediction on HF signals.

## 4. Conclusion and outlooks

Since the installation of the instrumental setup of EXTASIS experiment, the low frequency sky in Nançay has been explored, allowing us to understand the low frequency antenna environment, to put in evidence its behaviour and therefore to convince us that our low frequency instrument is functional.

In this proceeding, we have shown two low frequency events in coincidence with the particle detectors and the other radio instruments installed in Nançay, what is a clear evidence of a low frequency counterpart of the radio signal emitted by cosmic ray induced air showers. Moreover, these results seem to confirm the expectations, particularly that the detection range at low frequency is better than at high frequency.
Now that the timing of signal is verified thanks to the high frequency polarizations installed on the low frequency antennas, we need to switch to the previous instrumental setup version, and replace the high frequency polarization by a vertical polarization, in order to be sensitive to the sudden death signal which is expected also in this polarization.

## Acknowledgements

We thank the Région Pays de la Loire for its financial support of the Astroparticle group of Subatech and in particular for its contribution to the EXTASIS experiment.

# Computing the electric field from Extensive Air Showers using a realistic description of the atmosphere

**B. Revenu[1,3], F. Gaté[1], V. Marin[2], R. Dallier[1,3], A. Escudie[1], D.García-Fernández[1], L. Martin[1,3]**
[1] *Subatech, Institut Mines-Telecom Atlantique, CNRS, Universite de Nantes, Nantes, France*
[2] *Nantes, France*
[3] *Unité Scientifique de Nançay, Observatoire de Paris, CNRS, PSL, UO/OSUC, Nançay, France*
*E-mail:* revenu@in2p3.fr

The composition of ultra-high energy cosmic rays is still poorly known and this is an very important topic in the field of high-energy astrophysics. We detect them through the extensive air showers they create after interacting with the atmosphere constituents. The secondary electrons and positrons of the showers emit an electric field in the kHz-GHz range. It is possible to use this radio signal in 20-80 MHz for the estimation of the atmospheric depth of maximal development of the showers $X_{\text{max}}$, with a good accuracy and a duty cycle close to 100%. This value of $X_{\text{max}}$ is strongly correlated to the nature of the primary cosmic ray that initiated the shower. We present the importance of using a realistic atmospheric model in order to correct for systematic errors that can prevent a correct and unbiased estimation of $X_{\text{max}}$.

## 1. Introduction

In the last years, most of the air shower detection experiments run arrays of radio detectors in order to measure the electric field emitted by showers. Antennas are commonly used in the band 30-80 MHz where the electric field is coherently emitted by all secondary electrons and positrons. Detecting this field is now routinely achieved but more difficult is extracting the primary cosmic ray characteristics using this radio signal. This is done with simulations codes such as [1, 2, 3]. They compute the expected electric field as a function of time and of the observer's location with respect to the shower axis. They rely on the choice of the nature of the primary cosmic ray (light or heavy nucleus), its energy and the shower geometry (zenith and azimuthal angles). Then, secondary particles are created and tracked; they evolve in the atmosphere, generally described by its density $\rho_{\text{air}}$, as a function of the altitude $z$. The atmospheric depth $X = \int \rho_{\text{air}}(z)\text{d}z$ is a critical quantity as it describes the shower development in the air and represents the amount of matter crossed by the particles. Taking the example of the code SELFAS, we compute the atmospheric depth by numerical integration, taking into account the curved shape of the Earth and its atmosphere; we don't use anymore the flat approximation which is valid up to 60°. The (new) SELFAS electric field formula is (see [4]):

$$\mathbf{E}(\mathbf{x},t) = \frac{1}{4\pi\varepsilon}\int \text{d}^3x' \left\{ \left[\frac{\rho(\mathbf{x}',t_{\text{ret}})\mathbf{r}}{R^2(1-n\beta\cdot\mathbf{r})}\right]_{\text{ret}} + \frac{n}{c}\frac{\partial}{\partial t}\left[\frac{\rho(\mathbf{x}',t_{\text{ret}})\mathbf{r}}{R(1-n\beta\cdot\mathbf{r})}\right]_{\text{ret}} - \frac{n^2}{c^2}\frac{\partial}{\partial t}\left[\frac{\mathbf{J}(\mathbf{x}',t_{\text{ret}})}{R(1-n\beta\cdot\mathbf{r})}\right]_{\text{ret}} \right\}. \tag{1.1}$$

This is the electric field expression at the observer's location $\mathbf{x}$ at time $t$ from the charge density $\rho$, current $\mathbf{J}$, $\mathbf{r}$ being the normalized vector particle-observer and $\mathbf{v} = c\beta$. $R$ is the distance particle-observer. Integration is performed for retarded time $t_{\text{ret}} = t - nR/c$. The air index $n$ also plays an important role as it drives the electric field amplitude (in particular close to the Cherenkov angle) and the arrival time of the signal at the observer's location, i.e. the pulse shape. For those reasons, it is very important to describe the atmosphere accurately enough.

## 2. Atmospheric model

The atmosphere is a layer of gas around the Earth. The properties of this gas are more or less stable according to the time of the day (daily effect) and the time of the year (seasonal effect). A rough description assumes the atmosphere is in a steady state at fixed altitude; this is the case for the US Standard model [5]. A much more detailed, and updated model is provided by the Global Data Assimilation System (GDAS) [6].

### 2.1 The US Standard atmosphere

This model exists since 1976 and is an idealized representation of the atmosphere from the sea level up to 1000 km of altitude, in period of moderate solar activity. Values are estimated from yearly averages under the assumption of hydrostatic equilibrium; air is considered as an homogeneous mixture of several gases. The US Standard air density profile can be retrieved easily from [7] as large table; analytic formulas are also available, using 4 ou 5 atmospheric layers with continuity conditions, as it is the case for the Linsley parameterization. The US Standard atmosphere provides for instance, $\rho_{\text{air}}(z)$ as an average value for all locations on the Earth, be it in winter or summer, day or night. We can expect quite large deviations with respect to specific weather conditions that can be very different from the US Standard values: it should also have an impact on the electric field emitted by showers during those specific weather conditions. The SELFAS code used the Linsley parameterization up to December, 2016. It can also use a much more refined model, based on the GDAS.

## 2.2 The GDAS model

The GDAS is the system used by US official agencies (such as the National Center for Environmental Prediction, the Global Forecast System) "to place observations into a gridded model space for the purpose of starting, or initializing, weather forecasts with observed data" (see [8]). The resulting 3D model space uses various ground observations, balloon data, wind profiler data, aircraft reports, buoy, radar and satellite data. This model can be retrieved on grid of various size in terms on longitude and latitude ($1° \times 1°$, $0.5° \times 0.5°$ or even $0.25° \times 0.25°$) starting from year 2001. The timestep for these data is 3 hours. It is possible to get the 3D model for the actual atmospheric conditions for any location on Earth at any time, up to an altitude of $z_{\rm max}^{\rm GDAS} = 26$ km. Many variables are available in the GDAS model. Four our needs, we will focus on the variables driving the values of interest ($\rho_{\rm air}$ and $n$) for the electric field computation: the relative humidity ($R_h$), the temperature ($T$) and the total pressure ($P$). Using such a refined model is a good approach as the data show large deviations, see for instance in FIG. 1 the relative humidity as a function of the altitude for the location of Nançay, France on March 18, 2014. At fixed altitude, the variations during the day are very important; the relative humidity has a role in the value of the air refractive index so that we also expect an influence on the electric field from air showers compared to a "standard" and constant relative humidity.

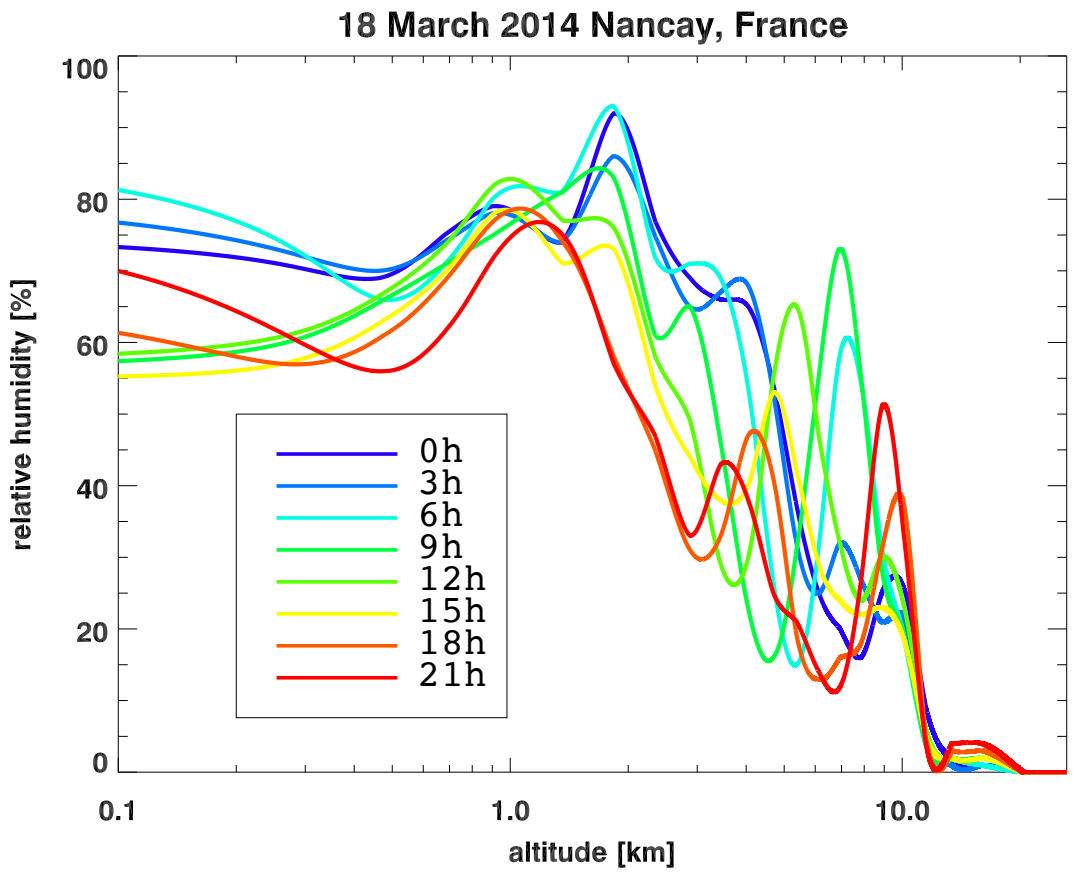

**Figure 1:** Daily variations of the relative humidity as a function of the altitude, using the GDAS data at Nançay on March 18, 2014.

The idea is to compute the atmosphere density $\rho_{\rm air}$ and air index $n$ at the time and location an event is detected. Then, the shower can be simulated with SELFAS, running with the corresponding atmosphere at the time of detection. The simulation will be done with the best astmospheric model possible, on a case-by-case basis. In order to compute accurately $\rho_{\rm air}$ and $n$, the procedure consists in using the pressure (in hPa), the temperature (in K) and the relative humidity (in %) as a function of the geopotential height $G_h$ in geopotential meters (gpm). We convert these meters into altitude above sea level. Then, the air density $\rho_{\rm air}$ is computed from the ideal gas law:

$$\rho_{\rm air}(z) = \frac{p_d(z(Z_g,\phi))M_d + p_v(z(Z_g,\phi))M_v}{RT(z(Z_g,\phi))}, \tag{2.1}$$

where $z(Z_g,\phi)$ is the altitude above sea level corresponding to the geopotential height $Z_g$ at a latitude $\phi$; $p_d$ and $p_v$ are the partial pressures of dry air and water vapor and $M_d$ and $M_v$ the molar masses. The water vapor partial pressure is given by $p_v = R_h p_{\rm sat}$ where $p_{\rm sat}$ is given by (see [9, 10]):

$$p_{\rm sat} = 6.1121 \exp\left[\left(18.678 - \frac{T}{234.5}\right)\left(\frac{T}{257.14+T}\right)\right] \quad (T \text{ in } ^\circ\text{C}). \tag{2.2}$$

Then, $p_d = P - p_v$. This equation can be used in the range $[-80;+50]^\circ$C, which is our range of interest.

We have the recipe to compute all relevant quantities to simulate accurately the electric field emitted by air showers using the GDAS data, provided only up to $z_{\rm max}^{\rm GDAS} = 26$ km above sea level. As air showers can initiate at much larger altitudes, we use the US Standard model with a scaling factor to ensure continuity with the GDAS model below $z_{\rm max}^{\rm GDAS}$, up to an altitude of $\sim 110$ km, which is considered as the limit of the atmosphere.

Finally, we get an accurate atmosphere model for the time of the detection of an event, at any place in the world, using both the GDAS data below $z_{\rm max}^{\rm GDAS}$ and the rescaled US Standard model between $z_{\rm max}^{\rm GDAS}$ and 110 km. FIG. 2 presents the maximum relative differences $(\rho_{\rm GDAS} - \rho_{\rm US})/\rho_{\rm US}$ as a function of the altitude, between the GDAS model and the US Standard model during the year 2014 in Nançay. We also show the relative difference for the sample day March 18, 2014. We see that the relative differences can reach $\pm 15\%$, which will have a large influence of the atmospherical depths.

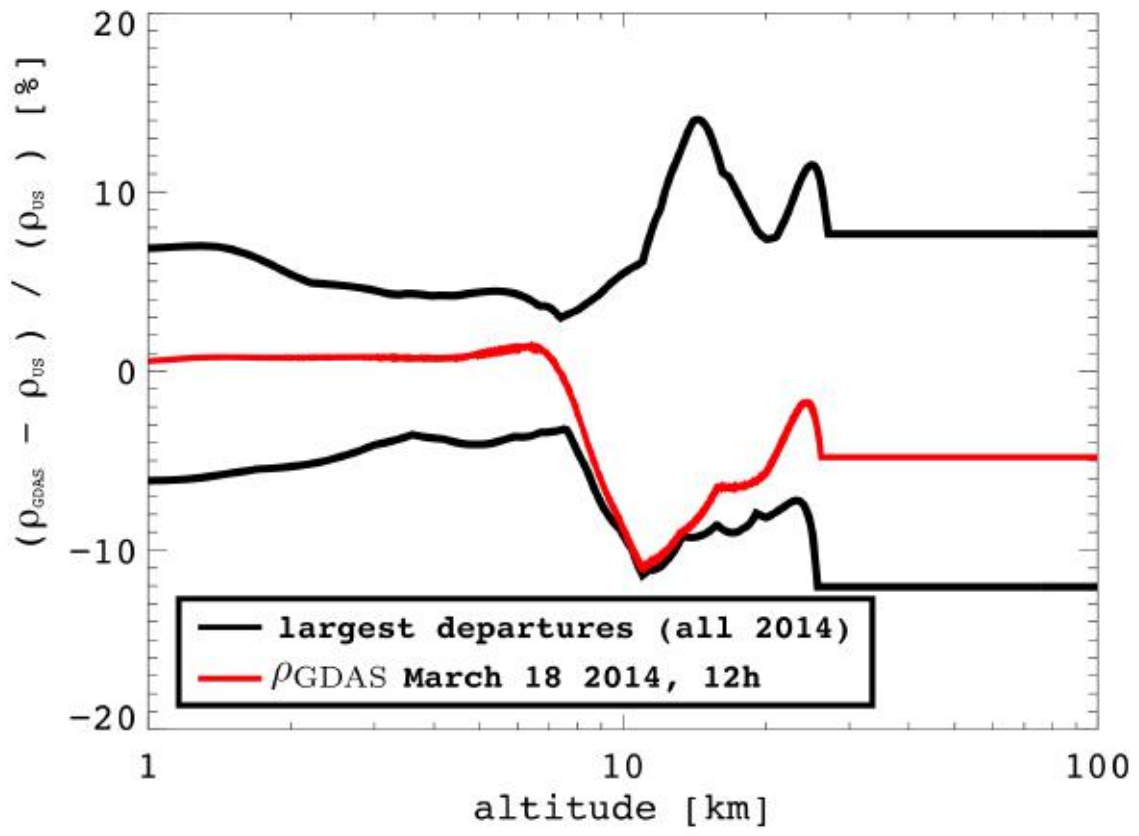

**Figure 2:** In black: extrema of the differences between the US Standard model air density profile and all the GDAS profiles along the year 2014, as a function of altitude. In red: the air density computed from the GDAS model on March 18, 2014.

## 3. Effect on the electric field computation

In order to check directly the influence of the chosen atmospheric model, we simulated a shower with SELFAS, initiated by a 1 EeV proton with a first interaction depth $X_1 = 100$ g/cm$^2$, a zenith angle $\theta = 30°$ and azimuth $\phi = 90°$ (coming from the North). We compute the total electric field amplitude as a function of the axis distance and in the direction of $\mathbf{v} \times \mathbf{B}$, where $\mathbf{v}$ is the shower axis direction and $\mathbf{B}$ the geomagnetic field at the observer's location. We did this simulation considering the US Standard atmosphere and taking the actual atmosphere on March 18, 2014, in Nançay. The electric field amplitudes are shown in FIG. 3. We observe that the electric field profile is larger when considering the US Standard model which is a source of systematic error when trying to reconstruct the $X_{max}$ from the radio signal. We can imagine a reversed situation where the simulated profile is larger in the GDAS case. This emphasize the importance of considering the actual atmosphere at the time of detection of an event.

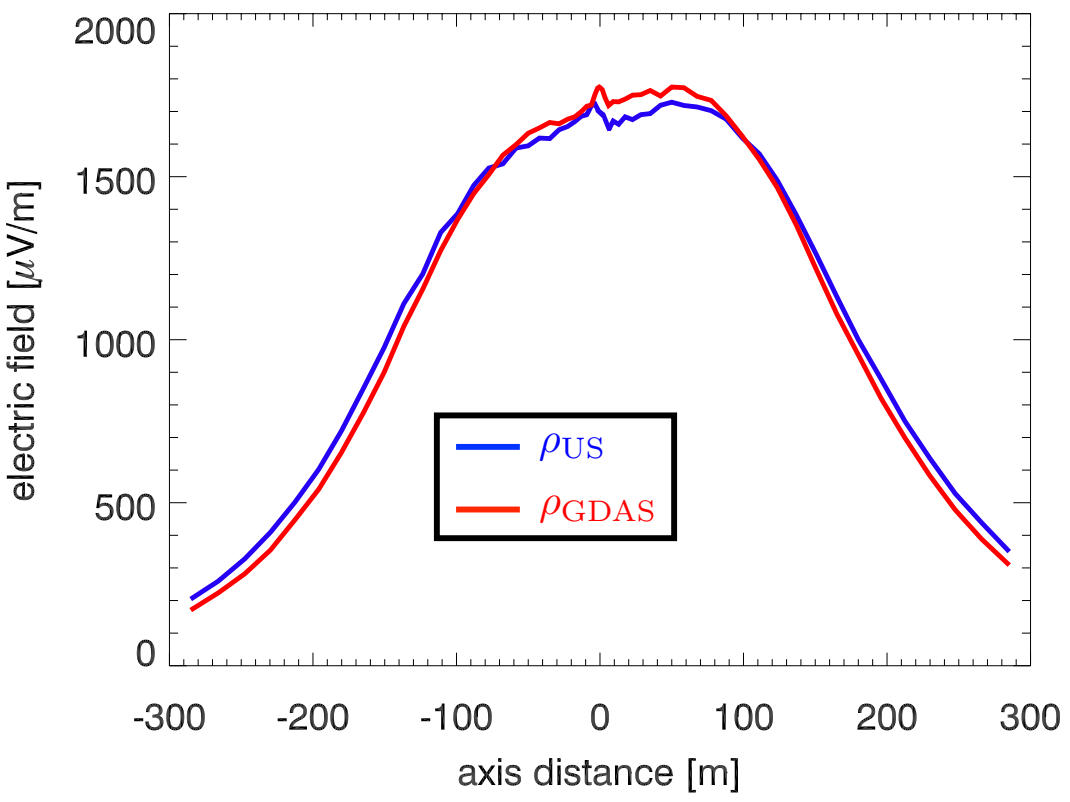

**Figure 3:** Total electric field amplitude in the shower front reference frame as a function of the axis distance in the direction $\mathbf{V} \times \mathbf{B}$, where $\mathbf{V}$ is the shower axis and $\mathbf{B}$ the geomagnetic field. This corresponds to a the same shower initiated by a proton at 1 EeV with a first interaction depth of 100 g/cm$^2$, $\theta = 30°$ and $\phi = 90°$, developing once in US Standard model (blue curve) and once in the atmospheric model based on the GDAS data on March 18, 2014 at noon (red curve).

## 4. Air index computation

The other fundamental parameter to properly estimate for the electric field computation is the air index. Let's consider a secondary charged particle of the shower at an altitude $z$ and a distance $R$ from the observer. The electric field amplitude depends directly on $n$ as explicited in EQ. 1.1. Then, the arrival time $t$ of the wave at the observer's location is $t = t_e + <n> R/c$ where $t_e$ is the emission time and $<n>$ is the average value of the air index on the line particle-observer. Usually, the air index is provided by the Gladstone and Dale law:

$$n(z(l)) = 1 + \kappa\rho(z(\ell)) \text{ with } \kappa = 0.226 \text{ cm}^3/\text{g}. \tag{4.1}$$

The corresponding average air index $< n >$ is given by:

$$< n(z(\ell)) >= 1 + \frac{\kappa}{\ell} \int_0^{\ell} \rho(z(\ell'))\, \mathrm{d}\ell'$$

But the Gladstone and Dale constant $\kappa$ depends on the medium and on the considered wavelength. We used in SELFAS (up to December 2016) the same value than that in use in CoREAS: the constant $\kappa = 0.226$ cm$^3$/g corresponds to optical wavelengths ($\lambda \sim 400$ nm). This is not suited for our MHz frequency range: $\lambda = 7.5$ m at 40 MHz. A more suited approach is to properly take into account the humidity fraction $R_h$ which plays an important role in the value of the air index according to [11]:

$$n = 1 + 10^{-6} N \quad \text{with} \quad N = \frac{77.6}{T}\left(P + 4810\,\frac{p_v}{T}\right) \qquad T \text{ in K}, \tag{4.2}$$

where $N$ is the refractivity. This equation is proposed for the MHz to GHz domain. The air density is hidden in $T$ and $P$ as we assume the ideal gas law approximation is valid. It means that the refractivity will be different when considering a GDAS model or the US Standard model. The water vapor partial pressure explicitly appears and can be the dominant term: the air refractive index should be computed with this parameter. As in section 2.2, we can use the accurate formula using the GDAS data up to $z_{\rm max}^{\rm GDAS}$. Beyond this altitude, we have no data for temperature and relative humidity. Hopefully, the air relative humidity beyond $z_{\rm max}^{\rm GDAS}$ can be considered as null as usually no clouds are observed above 12 km. We can therefore use a simpler formula above $z_{\rm max}^{\rm GDAS}$:

$$p_v = 0,\; P = p_d,\; T = \frac{P_d M_d}{R\rho} \quad \text{so that} \quad N = 77.6\,\frac{R\rho}{M_d} \quad \text{with} \quad \rho = \rho_{\rm US},$$

the $\rho_{\rm US}$ being the rescaled US Standard density ensuring continuity with $\rho_{\rm GDAS}$ at $z_{\rm max}^{\rm GDAS}$. The difference between the air refractivity $N$ in the US Standard atmosphere and in the GDAS model is shown in FIG. 4. We see that the relative difference can reach 35% close to the ground (where the humidity fraction can be very large) and around 15% at altitudes of interest for the shower development (10-20 km). As the refractivity appears with a $10^{-6}$ factor in the air index, the effect on the electric field is not so important, as shown in FIG. 5. In this figure, we show the electric field in the three polarizations East-West, North-South and vertical as a function of time for a shower initiated by a 1 EeV proton with $X_1 = 100$ g/cm$^2$ and $\theta = 30°$, $\phi = 90°$, for observers located at different positions around the shower axis. The only difference is the value of the refractivity: we consider the value using the US Standard atmosphere with the Gladstone and Dale law in one case and the GDAS model on March 18, 2014 with the high frequency formula (EQ. 4.2) in the other case. We see that the North-South amplitude can be modified by 30% and the arrival time of the wave at the observer location can be shifted by 5 ns. In this example, the East-West amplitude is dominant so that the total electric field will not be strongly affected by the choice of the air index model. But if the considered experiment does not measure all three polarizations then the choice of the model is primordial.

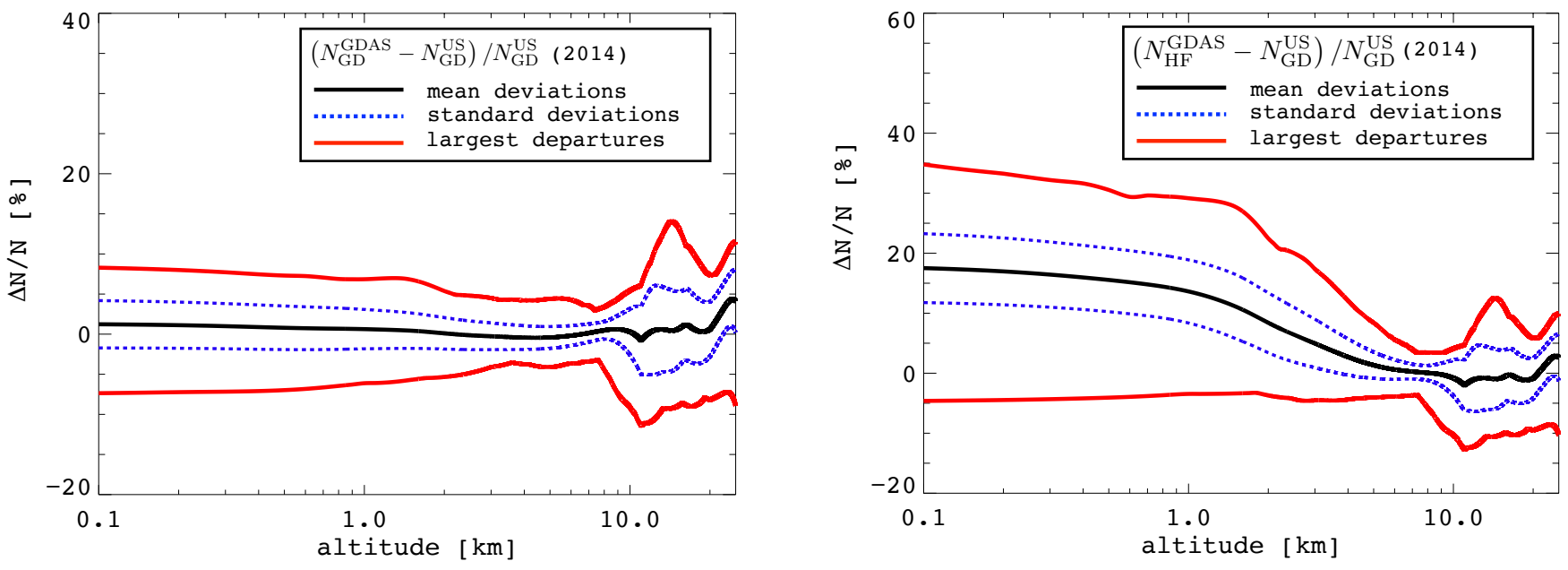

**Figure 4:** Relative difference in refractivity as a function of the altitude, with respect to the case $N_{\rm GD}^{\rm US}$, for $N_{\rm GD}^{\rm GDAS}$ (left) and $N_{\rm HF}^{\rm GDAS}$ (right). The black line corresponds to the mean values for the year 2014 and the red lines correspond to the maximum deviations. The largest deviations in the $N_{\rm HF}^{\rm GDAS}$ case (right) correspond to the high level of humidity below $5-6$ km of altitude. The blue dashed line indicates the standard deviation of the relative difference.

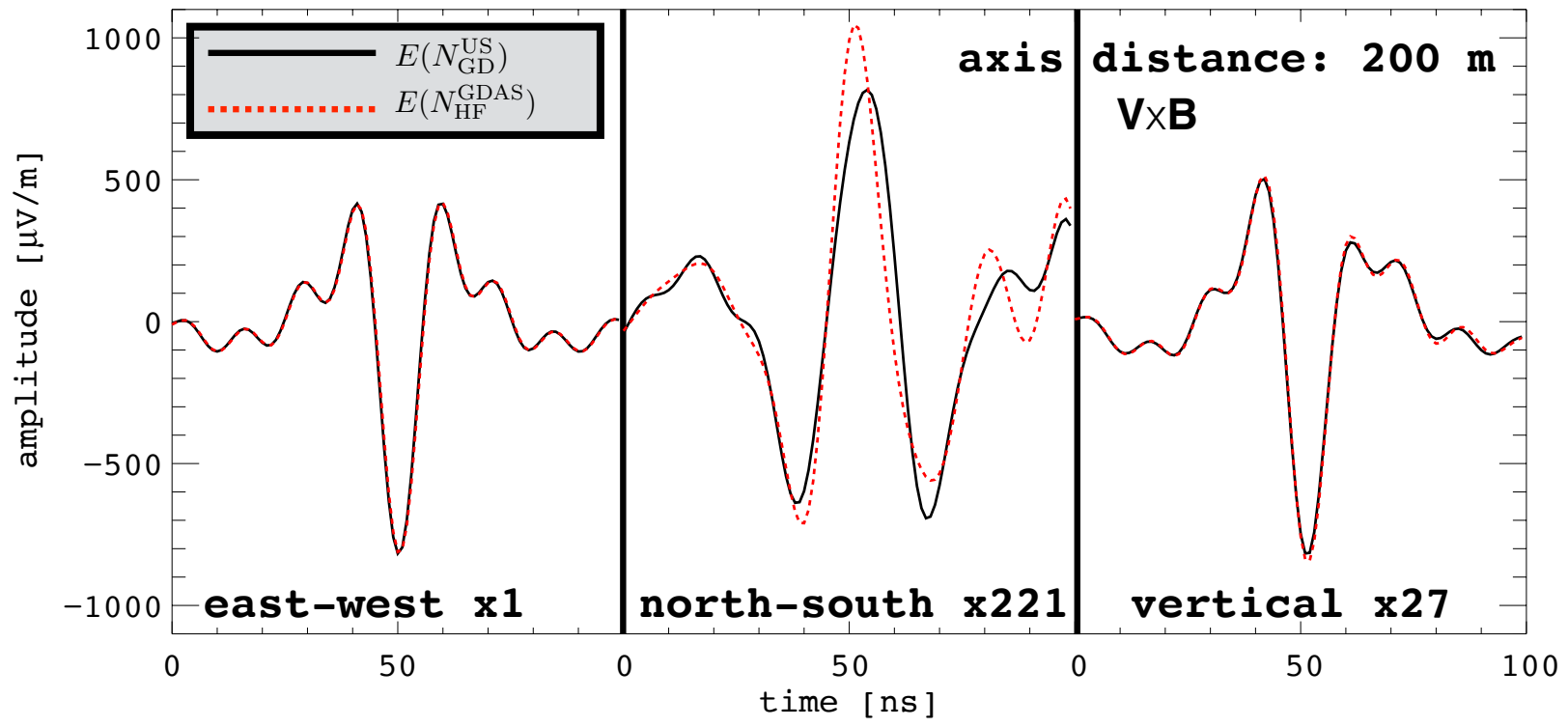

**Figure 5:** Time series of the electric field simulated with SELFAS using $N_{HF}^{GDAS}$ in red and $N_{GD}^{US}$ in black, with the same air density profile, at 200 m from the shower axis in the polarization parallel to $\mathbf{v} \times \mathbf{B}$. The shower is induced by a proton at 1 EeV with a first interaction depth of 100 g/cm$^2$, $\theta = 30°$, $\phi = 90°$; the electric field is filtered in the band [20; 80] MHz using the three polarizations that are indicated at the bottom of each plot together with the scale factor applied for better visibility.

## 5. Unbiased $X_{max}$ reconstruction

In this section we check the estimation of the $X_{max}$ in the context of the choice of the atmospheric model. For this, we simulate a shower on March 18, 2014 initiated by a 1 EeV proton with $X_{max} = 720$ g/cm$^2$ and $\theta = 30°$, $\phi = 90°$. This is our reference event, the one we want to reconstruct. We then apply the same procedure as we do with actual cosmic events. We simulate showers initiated by protons and iron nuclei having the same arrival direction and energy but random $X_{max}$. One set of protons and iron nuclei showers uses the US Standard air density and the corresponding refractivity $N_{GD}^{US}$. The other set uses the same atmosphere than our reference event (on March 18, 2014) with the refractivity $N_{HF}^{GDAS}$. As described in [12], the agreement between each simulated electric field and the mock data of the test event is tested with a $\chi^2$ test. The preferred $X_{max}$ value is 750 g/cm$^2$ when using the US Standard atmosphere. It is 718 g/cm$^2$ when using the GDAS atmosphere at the time of detection of the event. The true value is 720 g/cm$^2$. The discrepancy between the two models is due to the choice of the atmospherical model. It is very important to note that when converting atmospherical depths to distances to the point of maximum emission, both models leads to the same value of 4380 m from the shower core. It means that the electric field distribution is governed by the geometrical distance to the observer and not directly by the atmospherical depth. It also means that the preferred electric field distribution in the two data sets corresponds to showers having their maximum emission at the same distance to the observer and not at the same $X_{\mathrm{max}}$. In other words, using different atmospheric models allow to reconstruct directly, with no bias, the correct $X_{\mathrm{max}}$. This is not possible using the US Standard model which gives in this example a value shifted by 30 $\mathrm{g/cm^2}$ with respect to the true value.

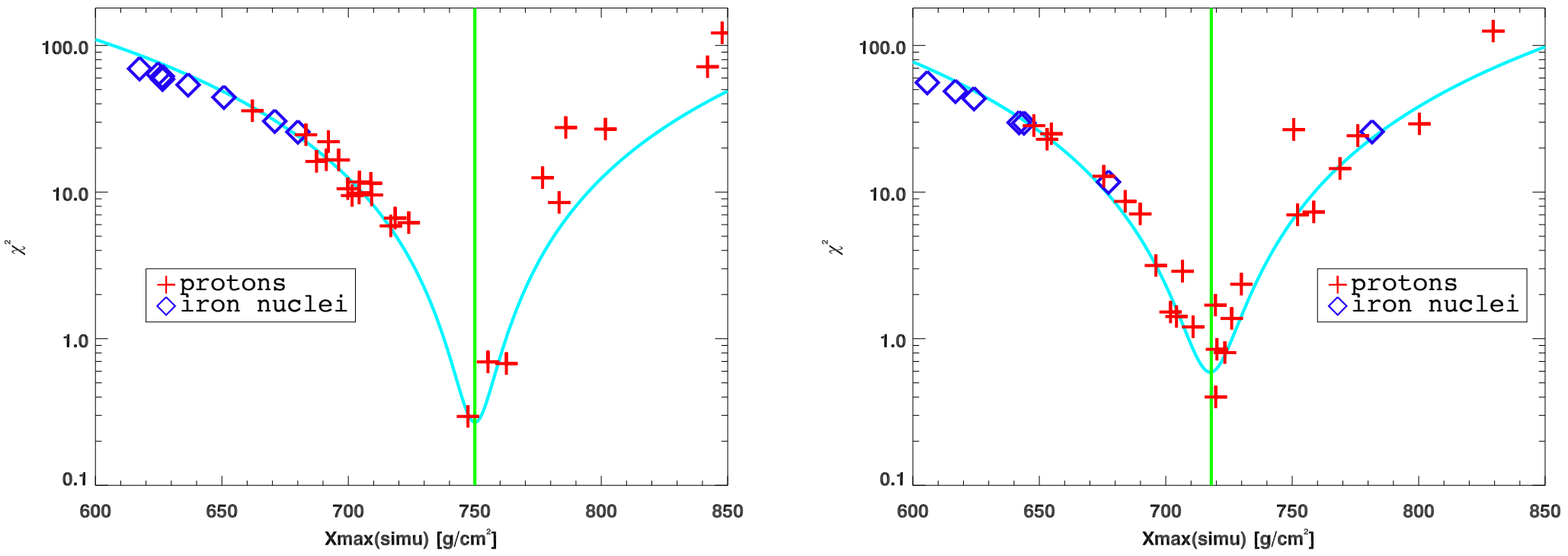

**Figure 6:** Value of the $\chi^2$ test as a function of the simulated $X_{\mathrm{max}}$ depths for the set of showers simulated using $\rho_{\mathrm{US}}$ and $N_{\mathrm{GD}}^{\mathrm{US}}$ (top) and the set of showers using $\rho_{\mathrm{GDAS}}$ and $N_{\mathrm{HF}}^{\mathrm{GDAS}}$ (bottom).

## 6. Conclusion

We have studied the importance of using a precise model of the atmosphere in the context of the computation of the electric field emitted by air showers during their development. The air density $\rho_{\mathrm{air}}$ and air index $n$ are influenced by the weather conditions. Two models are available: the US Standard description which provides average atmosphere characteristics; these are the same for all locations on the Earth and any time (day/night, winter/summer). The other one is based on the GDAS model which allows an accurate atmosphere estimation at the time an event is detected. We have shown that the relative difference in the air density can reach $\pm 15\%$ between both models. This can lead to differences up to some tens of $\mathrm{g/cm^2}$ in terms of atmospheric depths. This is clearly not negligible as the uncertainty on the $X_{\mathrm{max}}$ using the radio technique is of the order of 20 $\mathrm{g/cm^2}$. The air index value is also driven by the atmospheric model: its density but also the humidity fraction which can have a dominant influence on the refractivity. The Gladstone and Dale law is commonly used but it is not suitable for our frequency range and does not take into account the humidity fraction. Considering a basic refractivity model (with the US Standard atmosphere and the Gladstone and Dale law) and an accurate refractivity model (properly using both the GDAS model and the humidity fraction), we have shown that the electric field is altered by the model: amplitudes can be modified by some tens of % and the arrival time of the wave at the observer location can be shifted by some ns. This should be of relative importance according to the specific design of the considered experiment. Finally, we compared the $X_{\max}$ estimation using a reference event simulated using the atmosphere of a specific day, with a true value of $X_{\max} = 720$ g/cm$^2$. We used two simulated sets of showers initiated by protons and iron nuclei of the same energy and arrival direction than the reference event but with random $X_{\max}$. One set uses the US Standard atmosphere and the Gladstone and Dale law; the other one uses the same atmosphere than that of the reference event, together with the refined air index model. The first set leads to a biased $X_{\max}$ estimation of 750 g/cm$^2$ and the second one to a correct value of 718 g/cm$^2$. Using a crude atmosphere model leads to biased $X_{\max}$ values, up to some tens of g/cm$^2$. Nevertheless, their is no bias on the distance to the shower maximum emission point: US Standard and $X_{\max} = 750$ g/cm$^2$ correspond to the same distance than GDAS and $X_{\max} = 718$ g/cm$^2$. The conversion can be done afterwards but it is clearly better to run the shower simulation directly in the best atmospheric model in order to properly consider all effects into account (correct atmospheric depths and correct air index).